\documentclass[aps, reprint, prx, twocolumn, superscriptaddress,amsmath,amssymb, longbibliography,noeprint]{revtex4-2}

\usepackage{times}
\usepackage{amsmath}
\usepackage{amssymb}
\usepackage{xfrac}
\usepackage{dsfont}
\usepackage{graphicx} 
\usepackage{float}
\usepackage{dcolumn} 
\usepackage{bm} 
\usepackage{braket}
\usepackage{mathtools}
\usepackage{xcolor}
\usepackage{hyperref}
\hypersetup{
    colorlinks=true,
    linkcolor=blue,
    filecolor=blue,
    citecolor=blue,
    urlcolor=blue
}
\usepackage[all]{hypcap}

\usepackage{etoolbox}

\providecommand{\bibfield}[2]{#2}
\renewcommand{\bibfield}[2]{%
    \ifstrequal{#1}{title}
        {\textit{#2}}
        {#2}%
}

\begin{document}

\preprint{APS/123-QED}

\title{Static-Field Shielding of Bosonic Molecules: Evaporation to Degeneracy and Self-Bound Droplets}

\author{Jongheum Jung}
\thanks{These authors contributed equally to this work.}
\affiliation{Department of Physics, Princeton University, Princeton, New Jersey 08544, USA}

\author{Laura Futamura}
\thanks{These authors contributed equally to this work.}
\affiliation{Department of Physics, Princeton University, Princeton, New Jersey 08544, USA}

\author{Matteo Ciardi}
\affiliation{Institute for Theoretical Physics, TU Wien, Wiedner Hauptstra{\ss}e 8-10/136, 1040 Vienna, Austria}

\author{Jason Rosenberg}
\affiliation{Department of Physics, Princeton University, Princeton, New Jersey 08544, USA}

\author{Youssef Aziz Alaoui}
\altaffiliation{Present address: Laboratoire Kastler Brossel, Coll\`{e}ge de France, CNRS, ENS-PSL University, Sorbonne Universit\'{e}, 11 Place Marcelin Berthelot, 75005 Paris, France.}
\affiliation{Department of Physics, Princeton University, Princeton, New Jersey 08544, USA}

\author{Yukai Lu}
\affiliation{Department of Physics, Princeton University, Princeton, New Jersey 08544, USA}

\author{Thomas Pohl}
\affiliation{Institute for Theoretical Physics, TU Wien, Wiedner Hauptstra{\ss}e 8-10/136, 1040 Vienna, Austria}

\author{Waseem S. Bakr}
\email{Email: wbakr@princeton.edu}
\affiliation{Department of Physics, Princeton University, Princeton, New Jersey 08544, USA}

\date{\today}

\begin{abstract}
The strong, tunable dipolar interactions of ultracold molecules make them a powerful platform for quantum many-body physics, but reaching degeneracy by evaporative cooling requires suppressing inelastic collisions. Microwave collisional shielding has enabled the preparation of degenerate Fermi gases and Bose-Einstein condensates of polar molecules, whereas static electric field shielding has been limited to fermionic species, which are less prone to inelastic loss. We demonstrate static-field F\"{o}rster shielding of bosonic $^{23}$Na$^{87}$Rb molecules, suppressing two-body loss by up to four orders of magnitude. Within a bound-state-free electric-field window, three-body loss is also strongly suppressed, enabling efficient evaporation. Evaporating with an efficiency of 2.09(9), we increase the phase-space density of the gas by two orders of magnitude, reaching degeneracy with 7200(1000) molecules. We observe self-bound droplets at the end of evaporation over a wide range of field strengths, emerging from either degenerate or non-degenerate parent gases. We perform Path Integral Monte Carlo simulations, which suggest that the observed droplets are filamentary in nature, and find good agreement with the experimentally observed droplet formation temperatures. Our results establish F\"{o}rster shielding as a single-field route to prepare degenerate gases and self-bound droplets of bosonic polar molecules.

\end{abstract}

\maketitle

Ultracold polar molecules have emerged as a promising platform for quantum science and many-body physics~\cite{micheli2006toolbox, Barnett2006Magnetism, carr2009cold, bohn2017cold}. Their long-lived rotational states provide internal degrees of freedom for quantum control, while electric dipole--dipole interactions generate anisotropic and externally tunable couplings between molecules~\cite{yan2013observation,Burchesky2021CaFCoherence,gregory2024second,Holland2023CaFEntanglement,Bao2023SE, Picard2025SE, Ruttley2025SE}. Together, these ingredients enable the study of spin models in pinned arrays~\cite{christakis2023probing, Miller2024XYZ, lu2026probing} as well as strongly interacting itinerant systems~\cite{li2023tunable, Carroll2025tJ}. Over the past two decades, rapid advances in assembling molecules from ultracold atoms and in direct laser cooling have produced ultracold gases and arrays of molecules with increasing control over their internal and motional states~\cite{langen2024quantum, ni2008high, shuman2010laser, anderegg2019optical}. For bulk molecular gases, an important step is to reach quantum degeneracy while retaining control over the dipolar interactions, thereby accessing many-body phenomena governed by quantum statistics and long-range interactions~\cite{DeMarco2019KRbDegenerate,schindewolf2022evaporation, bigagli2024bec,shi2026bec,zhang2026droplet,biswas2026controlled,tang2018thermalization,chomaz2019long,su2023dipolar,he2025exploring,chandrashekara2026competing}. For example, strong dipolar interactions in bosonic systems can give rise to self-bound droplets, which are held together by interactions alone and were observed in magnetic atoms~\cite{kadau2016observing,ferrier-barbut2016droplets, Schmitt2016Droplet,Chomaz2016QuantumFluctuation}, atomic Bose mixtures~\cite{cabrera2018quantum,semeghini2018self}, and recently in polar molecules~\cite{zhang2026droplet,shi2026bec}.

Evaporative cooling is the standard route to quantum degeneracy, but it requires elastic collisions to rethermalize the gas much faster than molecules are lost. Satisfying this requirement is particularly challenging because molecules that reach short range can react or form collision complexes, resulting in rapid loss~\cite{ospelkaus2010quantum,Hu2019K2RB2,gregory2019sticky}. Collisional shielding addresses this challenge by creating repulsive barriers that suppress short-range encounters and yield high elastic-to-inelastic collision ratios~\cite{schindewolf2025few}. Two parallel approaches have emerged based on microwave dressing~\cite{Karman2018MWShielding} and static electric fields~\cite{Avdeenkov2006EfieldShielding}. Microwave shielding, first demonstrated with directly laser-cooled molecules in optical tweezers~\cite{anderegg2021observation}, has enabled evaporative cooling of assembled polar molecules to Fermi degeneracy~\cite{schindewolf2022evaporation} and to Bose--Einstein condensation~\cite{bigagli2024bec,shi2026bec}.

Static electric fields provide a different pathway to collisional stability through two mechanisms. In quasi-two-dimensional geometries, a strong electric field polarizes molecules perpendicular to the plane, favoring repulsive side-by-side collisions~\cite{de2011controlling,quemener2011dynamics}. Alternatively, in three dimensions, the field can tune rotational pair states near a F\"orster resonance, where resonant dipole--dipole coupling produces a repulsive barrier against short-range approach~\cite{gonzalez2017adimensional}. Both mechanisms have been demonstrated with fermionic KRb molecules, enabling strong suppression of inelastic two-body loss and evaporative cooling to quantum degeneracy~\cite{valtolina2020dipolar,matsuda2020resonant,li2021KRbEvap,lin2026KRbPauli}.

For bosonic molecules, however, three-body loss can become the limiting process once two-body loss is suppressed~\cite{Stevenson2024ThreeBody}. An attractive well in the shielded intermolecular potential can support field-linked bound states that can mediate three-body recombination~\cite{Avdeenkov2003FieldLink, Chen2023FieldLinked, Chen2024Tetramer}. Under single-microwave shielding, this process limited evaporation before condensation~\cite{bigagli2023collisionally, lin2023SingleMWShielding}. Dual-microwave shielding removed the bound state and enabled Bose--Einstein condensation, but requires two independently controlled dressing fields and introduces additional Floquet inelastic loss~\cite{bigagli2024bec, shi2026bec, Karman2025DoubleMWShielding}. A distinct possibility arises with dc F\"orster shielding: the shielding barrier can be strong even when the long-range attractive well is too shallow to support a bound state~\cite{mukherjee2024controlling}. Such a bound-state-free regime could provide the simultaneous two- and three-body collisional stability required to evaporatively cool a bosonic molecular gas to quantum degeneracy.

\begin{figure*}[t]
    \centering
    \includegraphics[width=15cm]{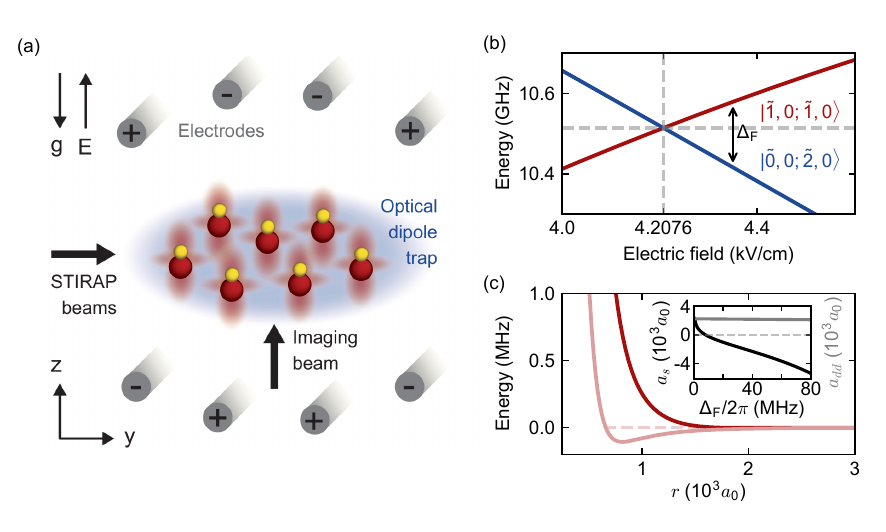}
\caption{Static electric field shielding above a F\"orster resonance. (a) Experimental setup. An eight-rod electrode configuration in vacuum enables precise tuning of the electric field at the molecular gas while supporting a numerical aperture NA$~=0.5$ imaging system~\cite{christakis2023probing}. The red shaded regions represent the repulsive shielding barriers experienced by the molecules. (b) Energy levels of the $\vert \tilde{1},0; \tilde{1},0\rangle$ (red) and $\vert \tilde{0},0; \tilde{2},0\rangle$ (blue) pair states near the F\"orster resonance at $4.2076$~kV/cm. The vertical dashed line indicates the field corresponding to the F\"orster resonance. (c) Adiabatic interaction potentials for the asymptotic incoming channel $\ket{\tilde{1},0;\tilde{1},0}$ with partial wave $L=0$. Results are shown for F\"orster defects of $\Delta_{\textrm{F}}=2\pi\times17$ MHz (dark red) and $2\pi\times314$~MHz (light red). For $\Delta_{\textrm{F}}=2\pi\times314$~MHz, the potential supports a bound state indicated by a light red dashed line. The inset shows the $s$-wave scattering length ($a_s$, left axis, black) and dipolar length ($a_{dd}$, right axis, gray) as functions of $\Delta_{\textrm{F}}$.}
    \label{fig:theory}
\end{figure*}

Here, we realize this regime and use it to produce a quantum-degenerate gas of bosonic NaRb molecules. Applying an electric field above the F\"orster resonance at $4.2~\mathrm{kV/cm}$ suppresses two-body loss by up to four orders of magnitude. By varying the field, we identify a bound-state-free region in which both two- and three-body losses are sufficiently weak for efficient evaporation. At an optimal field in this region, we demonstrate evaporation with efficiency of $2.09(9)$ to reach a phase-space density of $1.7(6)$, after which a compact droplet containing $230(80)$ molecules emerges. Its lack of measurable expansion during time of flight provides evidence for self-bound behavior. More generally, we observe self-bound droplets over a broad range of F\"orster defects, emerging from both quantum-degenerate and non-degenerate gases. Together, our results demonstrate an efficient route to quantum-degenerate molecular gases and self-bound dipolar matter. 

\section{F\"orster resonance shielding}

We prepare ultracold bosonic $^{23}\mathrm{Na}^{87}\mathrm{Rb}$ molecules in the rotational state $\ket{\tilde{1},0}$, confined in a crossed optical dipole trap (ODT). Here, $\ket{\tilde{N},m_N}$ denotes the field-dressed state adiabatically connected to the zero-field rotational state $\ket{N,m_N}$. An eight-rod electrode assembly surrounding the molecular cloud generates a homogeneous static electric field while preserving optical access for high-resolution imaging [Fig.~\ref{fig:theory}(a)]. By varying the applied field, we tune the molecular interactions near a F\"orster resonance at $4.2076~\mathrm{kV/cm}$.

The F\"orster resonance occurs when the asymptotic pair states $\ket{\tilde{1},0;\tilde{1},0}$ and $\ket{\tilde{0},0;\tilde{2},0}$ become degenerate [Fig.~\ref{fig:theory}(b)] where $\ket{\tilde{0},0;\tilde{2},0}$ denotes the exchange-symmetric pair channel. We define the F\"orster defect at infinite intermolecular separation as $\hbar\Delta_{\textrm{F}}=E_{\tilde{1},\tilde{1}}-E_{\tilde{0},\tilde{2}}$, such that $\Delta_{\textrm{F}}>0$ above the resonance. At finite intermolecular separation, dipole--dipole coupling between the two channels produces an induced repulsive interaction that suppresses access to short range. In the perturbative regime, the effective interaction can be written as~\cite{lassabliere2022model, Mukherjee2025EffectivePotential}
\begin{align}
V(\mathbf r)
&=V_{\textrm{short}}(\mathbf r)+V_{\textrm{long}}(\mathbf r), \nonumber\\
V_{\textrm{short}}(\mathbf r)
&=\left(
\frac{d_{\tilde{1}\tilde{2}}d_{\tilde{1}\tilde{0}}}
{4\pi\epsilon_0 r^3}
\right)^2
\frac{2(1-3\cos^2\theta)^2}
{\hbar\Delta_{\textrm{F}}}, \nonumber\\
V_{\textrm{long}}(\mathbf r)
&=\frac{d_{\tilde{1}\tilde{1}}^2}
{4\pi\epsilon_0}
\frac{1-3\cos^2\theta}{r^3},
\label{EffectivePotential}
\end{align}
where $\mathbf r=(r,\theta,\phi)$ is the relative coordinate, with $\theta$ measured from the electric-field axis, and
$d_{\tilde{i}\tilde{j}}
=\langle\tilde{i},0\vert d_z\vert\tilde{j},0\rangle$
is a dipole matrix element. The induced $1/r^6$ term $V_{\text{short}}(\bm{r})$ is non-negative for $\Delta_{\textrm{F}}>0$ and describes the repulsive shielding interaction, whereas the $1/r^3$ term $V_{\text{long}}(\bm{r})$ describes the anisotropic interaction between the field-induced molecular dipoles. Together they produce a repulsive core at short range and, along the field axis, an attractive well at larger separation. The full interaction potentials and bound-state spectrum are obtained from coupled-channel calculations (see Appendix~\ref{sec:interactionpotentials} and~\ref{sec:coupled-channel}). 

\begin{figure}[t]
    \centering
    \includegraphics[width=8.6cm]{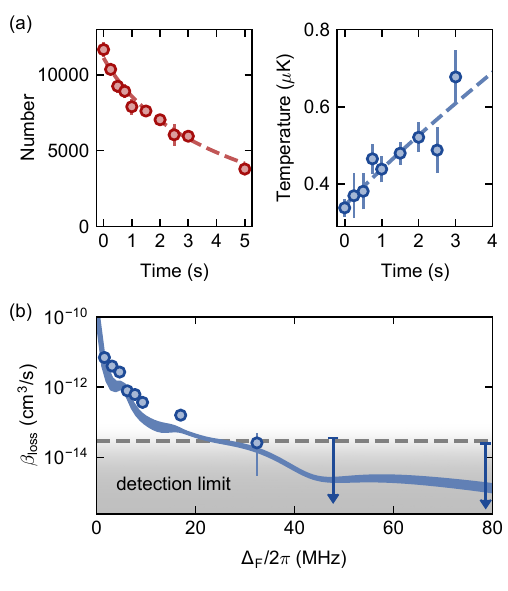}
    \caption{Two-body loss rate coefficients $\beta_{\textrm{loss}}$. (a) Number (left) and temperature (right) of molecules as functions of holding time for $\Delta_{\textrm{F}}=2\pi\times17$ MHz. Each point is obtained by averaging over 15 shots. Error bars of the number of molecules show 1$\sigma$ standard error of the mean. Error bars of the temperature of molecules show the 1$\sigma$ error from the fit of the time-of-flight expansion. (b) $\beta_{\textrm{loss}}$ at different F\"{o}rster defects $\Delta_{\textrm{F}}$. The blue shaded region represents the range of theoretical two-body loss-rate coefficients obtained from coupled-channel calculations after thermal averaging at temperatures between 0.4 $\mu$K and 1 $\mu$K, corresponding to the temperature range sampled during our two-body-loss measurements. The gray-shaded region denotes two-body loss-rate coefficients below the experimental detection limit, with the gray dashed line marking the detection threshold. For measurements whose fitted mean loss-rate coefficient is less than 10$\%$ of the detection threshold, downward arrows indicate one-sided upper confidence limits; the top of each arrow marks the calculated confidence bound. We use bootstrapping to estimate the 1$\sigma$ error of $\beta_{\textrm{loss}}$.}
    \label{fig:twobodyloss}
\end{figure}

Figure~\ref{fig:theory}(c) shows the calculated adiabatic interaction potentials for the asymptotic incoming channel $\ket{\tilde{1},0;\tilde{1},0}$ with partial wave $L=0$. The inset shows the corresponding $s$-wave scattering length $a_s$ and dipolar length $a_{dd}=m_{\textrm{NaRb}} d_{\tilde{1}\tilde{1}}^2/12\pi\epsilon_0\hbar^2$,
where $m_{\textrm{NaRb}}$ is the mass of NaRb molecules. At $\Delta_{\textrm{F}}=2\pi\times314~\mathrm{MHz}$, the attractive well supports a field-linked bound state that can mediate three-body recombination, whereas no bound state is present at $\Delta_{\textrm{F}}=2\pi\times17~\mathrm{MHz}$. The $s$-wave scattering length also strongly depends on the F\"orster defect and crosses zero near $\Delta_{\textrm{F}}=2\pi\times8~\mathrm{MHz}$. These calculations indicate qualitatively different bound-state and scattering regimes within the shielding region. We therefore measure two- and three-body loss as functions of the F\"orster defect to identify conditions suitable for evaporative cooling.

\section{Two- and three-body losses}
We first characterize the suppression of inelastic two-body loss across the dc F\"orster-shielding region. By evaporatively cooling the atomic gases in the ODT, we prepare a partially Bose-condensed gas of $1.60(8)\times10^6$ Na atoms in the $\ket{F=1,m_F=1}$ state at $390(60)~\mathrm{nK}$ (condensate fraction $\sim0.07$) and a thermal gas of $1.10(1)\times10^6$ Rb atoms in the $\ket{F=1,m_F=1}$ state at $650(30)~\mathrm{nK}$. The Rb temperature is slightly higher than the Na temperature because there is a thermalization lag in the sympathetic cooling in the ODT. We next associate NaRb Feshbach molecules by ramping the magnetic field from $350~\mathrm{G}$ to $346.7~\mathrm{G}$ in $0.4~\mathrm{ms}$ across the interspecies Feshbach resonance at $347.6~\mathrm{G}$. A subsequent stimulated Raman adiabatic passage (STIRAP) pulse transfers the molecules to the field-dressed rotational state $\ket{\tilde{1},0}$ in the electronic and vibrational ground-state manifold. The one-way transfer efficiency is $94(2)\%$. Because of the limited bandwidth of the electric-field control, we perform STIRAP in the static electric field used for each measurement, ensuring that the molecules are shielded immediately after transfer to the ground state.

To characterize two-body loss under well-defined conditions, we first partially evaporate the molecular cloud and then recompress the ODT to a depth $U/k_B\simeq9~\mu\mathrm{K}$ with trapping frequencies $(\omega_x,\omega_y,\omega_z)=2\pi\times[143.8(1.2),106(2),139.7(1.1)]$~Hz. The resulting samples have temperatures between $400$ and $600~\mathrm{nK}$ and a truncation parameter $\eta=U/k_BT>10$, such that evaporation loss during the subsequent hold time is negligible. To determine temperature, we release the ground-state molecules and allow them to expand in time of flight (ToF) before applying a bound-to-free reverse STIRAP pulse immediately prior to imaging. This protocol avoids the short lifetime of Feshbach molecules during expansion and eliminates the need for magnetodissociation before ToF, whose dynamics can perturb the momentum distribution used for thermometry~\cite{Lam2022NaCsFeshbach}. The bound-to-free reverse-STIRAP efficiency is $87(2)\%$. We measure the molecule number and temperature as functions of hold time and fit both observables simultaneously using a kinetic model that includes one-body decay and two-body loss (see Appendix~\ref{sec:kinetic model}). The one-body lifetime, primarily limited by off-resonant scattering from the trapping light, is measured independently to be $7.07(12)~\mathrm{s}$ and is held fixed when extracting the two-body loss coefficient at different F\"orster defects. 

Figure~\ref{fig:twobodyloss}(a) shows representative number and temperature dynamics at $\Delta_{\mathrm F}=2\pi\times17~\mathrm{MHz}$, together with the kinetic-model fit. The fit yields a two-body loss rate coefficient of $\beta_{\mathrm{loss}}=1.6(3)\times10^{-13}~\mathrm{cm^3/s}$, corresponding to a suppression by more than three orders of magnitude relative to the unshielded value $\beta_{\mathrm{loss}}=3.8(5)\times10^{-10}~\mathrm{cm^3/s}$~\cite{lin2023SingleMWShielding}. Figure~\ref{fig:twobodyloss}(b) shows the measured two-body loss rate coefficient as a function of the F\"orster defect, together with coupled-channel calculations. At small defects, $\Delta_{\mathrm F}<2\pi\times35~\mathrm{MHz}$, where two-body decay remains sufficiently rapid to be resolved, the measurements agree well with the calculation. For larger defects, the extracted loss rate coefficients reach the experimental sensitivity floor and are consistent with $\beta_{\mathrm{loss}}\lesssim3\times10^{-14}~\mathrm{cm^3/s}$, corresponding to a suppression of up to four orders of magnitude. The detection limit is set primarily by the finite one-body lifetime and the uncertainties in the measured molecule number and temperature (see Appendix~\ref{sec:detection limit}), which prevents us from resolving smaller two-body loss-rate coefficients. The measured two-body loss rate coefficients are comparable to those reported for dual microwave shielding~\cite{yuan2025extreme, shi2026bec}.

\begin{figure}[t]
    \centering
    \includegraphics[width=8.6cm]{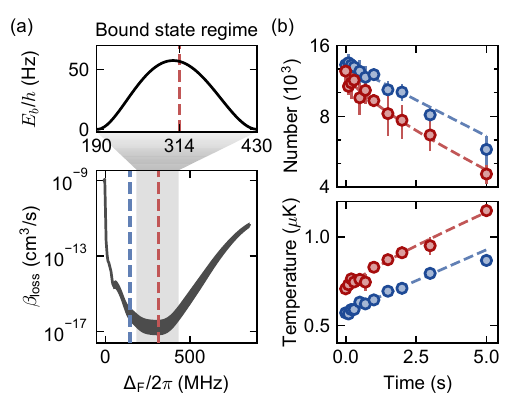}
    \caption{Three-body loss. (a) Two-body loss-rate coefficient $\beta_{\textrm{loss}}$ (lower) and bound-state energy $E_b$ (upper) as functions of the F\"orster defect $\Delta_{\textrm{F}}$. In the lower plot, the dark gray shaded region shows the calculated $\beta_{\textrm{loss}}$ for collision energies ranging from $0.1~\mu\mathrm{K}$ to $1~\mu\mathrm{K}$. The light gray shaded region indicates the range of $\Delta_{\textrm{F}}$ over which a bound state exists, which is expanded in the upper plot. We choose two electric fields: one outside the bound-state regime, $\Delta_{\textrm{F}}/2\pi=146~\mathrm{MHz}$ (blue), and one within the bound-state regime, $\Delta_{\textrm{F}}/2\pi=314~\mathrm{MHz}$ (red), where the bound-state energy is $E_b/h=60~\mathrm{Hz}$. (b) Fits to the molecule number and temperature as functions of hold time in the ODT. At $\Delta_{\textrm{F}}/2\pi=146~\mathrm{MHz}$ (blue), we obtain a three-body loss-rate coefficient of $L_3=3^{+8}_{-3}\times10^{-26}~\mathrm{cm}^6/\mathrm{s}$. At $\Delta_{\textrm{F}}/2\pi=314~\mathrm{MHz}$ (red), we obtain $L_3=1.0(3)\times10^{-24}~\mathrm{cm}^6/\mathrm{s}$.}
    \label{fig:threebodyloss}
\end{figure}

Having established strong suppression of two-body loss by F\"orster shielding, we next characterize three-body recombination. At large F\"orster defects ($\Delta_{\textrm{F}}>2\pi\times140$~MHz), coupled-channel calculations predict two-body loss rate coefficients well below $10^{-14}~\mathrm{cm^3/s}$. However, for $190~\textrm{MHz}\lesssim\Delta_{\textrm{F}}/2\pi\lesssim430~\textrm{MHz}$, the attractive well at large separations supports a field-linked bound state in this strongly shielded regime, providing a final-state channel for three-body recombination [Fig.~\ref{fig:threebodyloss}(a)].

To characterize this loss process, we compare two F\"orster defects for which the two-body loss rate coefficient lies below our detection limit, but only one supports a field-linked bound state. At $\Delta_{\textrm{F}}=2\pi\times146~\mathrm{MHz}$, the calculated interaction potential supports no bound state. At $\Delta_{\textrm{F}}=2\pi\times314~\mathrm{MHz}$, it supports a shallow field-linked bound state with binding energy $E_b/h\simeq60~\mathrm{Hz}$. In the bound-state region, where $a_s$ is large, we obtain the binding energy using the universal relation $E_b = \hbar^2/m_{\textrm{NaRb}}a_s^2$, consistent with Ref.~\cite{mukherjee2024controlling}. We measure the molecule number and temperature as functions of hold time at both defects and fit both observables simultaneously using a kinetic model that includes three-body loss [Fig.~\ref{fig:threebodyloss}(b)].

At $\Delta_{\textrm{F}}=2\pi\times314~\mathrm{MHz}$, we extract a three-body loss rate coefficient of $L_3=1.0(3)\times10^{-24}~\textrm{cm}^6/\textrm{s}$. In contrast, at the bound-state-free point $\Delta_{\textrm{F}}=2\pi\times146~\mathrm{MHz}$, we obtain $L_3=3^{+8}_{-3}\times10^{-26}~\textrm{cm}^6/\textrm{s}$, below our detection threshold of $L_{\textrm{det}}=7\times10^{-26}~\textrm{cm}^6/\textrm{s}$. At the densities relevant to these measurements, $n_0L_{\textrm{det}}/\beta_{\textrm{loss}}>100$ throughout the collision-energy range $E/k_B=100$~nK--$1~\mu$K, where $\beta_{\textrm{loss}}$ is obtained from coupled-channel calculations. Thus, even at the three-body detection threshold, the predicted two-body-loss contribution is negligible, justifying its omission from the three-body kinetic model.

The substantially larger three-body loss rate coefficient observed in the bound-state regime shows that three-body recombination mediated by the bound state imposes an additional constraint on the collisional stability of the shielded molecular gas. Nevertheless, even in the presence of the bound state, the three-body loss rate coefficient remains relatively low, consistent with the small binding energy of the field-linked state. 

\section{Evaporative cooling to quantum degeneracy}


\begin{figure}[t]
    \centering
    \includegraphics[width=8.6cm]{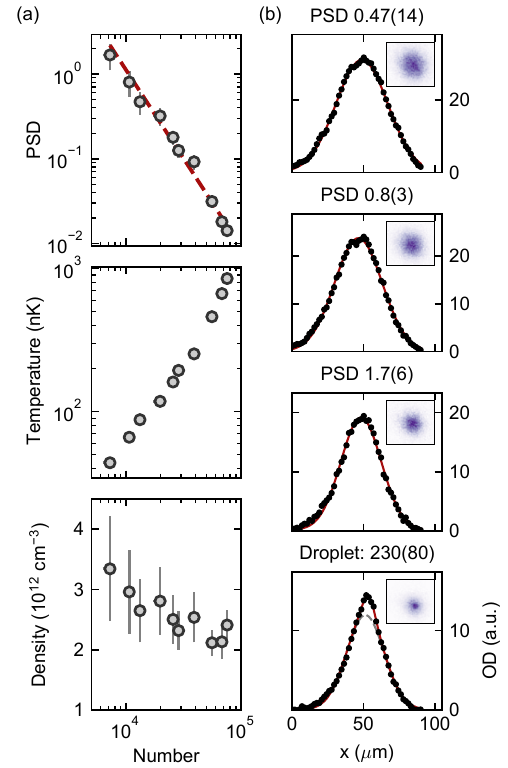}
    \caption{Evaporation curve of molecules at $\Delta_{\textrm{F}}/2\pi=17$ MHz. (a) Phase-space density (PSD, upper), temperature (center), and peak density (lower) versus molecule number. The red dashed line is a linear fit on log--log scales used to extract the evaporation efficiency. Each data point is an average of three to five consecutive shots. Error bars are 1$\sigma$ standard error of the mean. (b) One-dimensional optical-density (OD) profiles at different stages of evaporation, obtained by integrating the two-dimensional OD distributions along the $y$ axis. Images are taken after 6 ms time of flight. The corresponding PSD is indicated above each profile, with averaged two-dimensional OD images shown in the insets. Above PSD $~=1.7(6)$, a droplet containing $230(80)$ molecules emerges, shown in the bottom panel with a bimodal fit. Each inset color scale is independently normalized.
}
    \label{fig:psdcurve}
\end{figure}

Having demonstrated strong loss suppression in a bound-state-free regime under dc F\"orster shielding, we next investigate evaporative cooling of the shielded molecular gas. In choosing the F\"{o}rster defect at which to evaporate, there are two important considerations. First, the ratio of elastic to inelastic collisions needs to be high for efficient evaporation. Second, the relative strengths of the $s$-wave scattering length $a_s$ and the dipolar length $a_{dd}$ strongly influence the state reached at the end of evaporation. Evaporative cooling with a positive $a_s$ that exceeds $a_{dd}$ has been used to reach gas phase condensates~\cite{bigagli2024bec}. In the absence of a bound state, this regime is only accessible in our system very close to the F\"{o}rster resonance, where two-body loss becomes large. Two- and three-body losses are strongly suppressed throughout the explored window of F\"orster defects $\Delta_{\textrm{F}}\geq 2\pi\times17$~MHz, so we investigate evaporation in this range of defects. In this section, we report on the evaporation trajectory at $\Delta_{\textrm{F}}=2\pi\times17$~MHz, where we reach quantum degeneracy. At this defect, $a_s=-800a_0$ and $a_{dd}=2200a_0$, and the calculated elastic to inelastic collision ratio exceeds $10^4$ at the beginning of the evaporation (see Appendix~\ref{sec:coupled-channel}). 

\begin{figure*}[t]
    \centering
    \includegraphics[width=17.5cm]{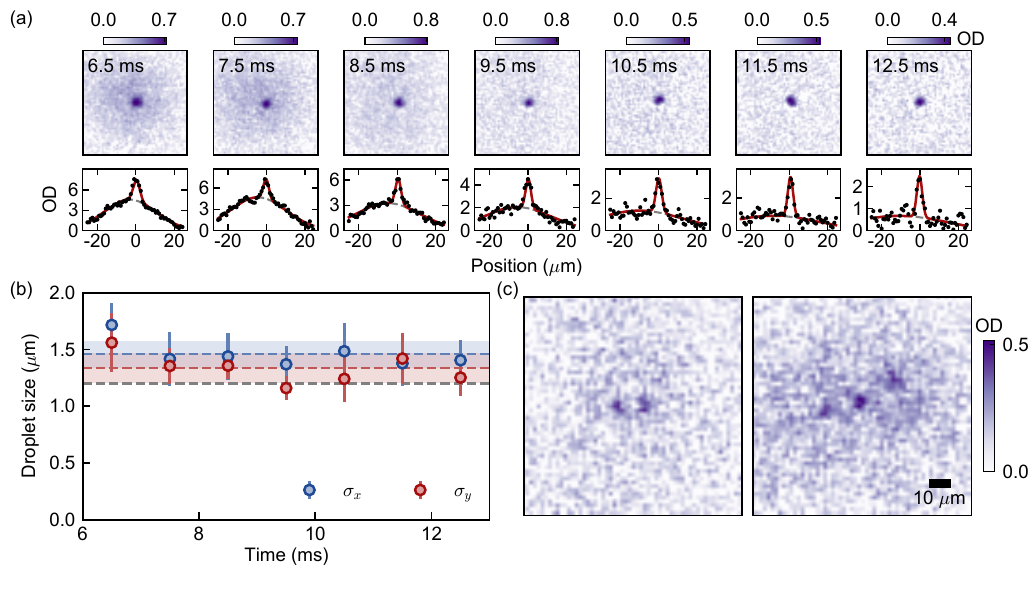}
    \caption{Cloud size of a self-bound droplet at different time-of-flight (ToF). (a) Each column shows an absorption image of the molecular cloud (top) and a 1D profile (integrated along the $y$ axis) of absorption images (bottom) after different ToF. The ToF is shown in the upper-left corner of each image. Bimodal fits are shown as solid red lines, whereas Gaussian fits to the thermal cloud are shown as dashed gray lines. We fit each two-dimensional density distribution to a sum of two Gaussian functions. Each image is averaged over $10$ shots. Each color bar is self-normalized. (b) Gaussian $1/e^{1/2}$ radii of the droplet, $\sigma_x$ and $\sigma_y$, as functions of ToF. Red ($\sigma_x$) and blue ($\sigma_y$) dashed lines indicate the mean radii, while the corresponding shaded regions denote one standard deviation over all ToFs. The gray dashed line indicates the imaging resolution. (c) Dual droplets (left) and triple droplets (right) after $10.5$~ms ToF. Each image is a single-shot absorption image.}
    \label{fig:dropletproperties}
\end{figure*}

We prepare $7.5(3)\times10^4$ molecules at an initial temperature of $850(30)~\mathrm{nK}$. The ODT trap frequencies are $(\omega_x,\omega_y,\omega_z)=2\pi\times[102(1),98(1),100(8)]~\mathrm{Hz}$. We evaporate the molecules by lowering the ODT depth over $1.2~\mathrm{s}$ at an electric field of $4.223~\mathrm{kV/cm}$ ($\Delta_{\mathrm{F}}=2\pi\times17$~MHz). Throughout the evaporation ramp, we determine the phase-space density
$\textrm{PSD}=N\omega_x\omega_y\omega_z(\hbar/k_BT)^3$
and peak density
$n_0=N\omega_x\omega_y\omega_z(m_{\textrm{NaRb}}/2\pi k_BT)^{3/2}$
from the measured molecule number $N$, temperature $T$, and corresponding trap frequencies, as shown in Fig.~\ref{fig:psdcurve}(a). In the thermal gas, $n_0a_{dd}^3\ll1$ and $n_0\vert a_s\vert^3\ll1$, so these ideal-gas expressions provide good estimates of the phase-space density and peak density.

Starting from an initial phase-space density of $\textrm{PSD}=0.014(2)$, we achieve an evaporation efficiency of
$-d\ln\textrm{PSD}/d\ln N=2.09(9)$. Figure~\ref{fig:psdcurve}(b) shows the corresponding evolution of the density profile during evaporation at $\mathrm{ToF}=6$~ms. As the trap depth is lowered, the density distribution evolves from a broad thermal cloud into a bimodal distribution, with a compact central component emerging at the final trap depth. At this final trap depth, the trapping frequencies are $(\omega_x,\omega_y,\omega_z)=2\pi\times[69(1),71(1),26(8)]$~Hz. At the last measured point before the compact component appears, the thermal molecular gas reaches $\textrm{PSD}=1.7(6)$ with $N=7200(1000)$ molecules at a temperature of $44(4)~\mathrm{nK}$. This indicates that the molecular sample has entered the quantum-degenerate regime.

A bimodal fit yields $230(80)$ molecules in the compact component. Its position fluctuates between experimental realizations, as also observed in Ref.~\cite{zhang2026droplet}. To account for these fluctuations, we recenter each absorption image on the compact component before averaging $7$--$10$ realizations.

\section{Evidence for self-bound droplets}
To characterize the compact component that emerges at the end of evaporation, we measure its expansion after release from the trap. Because the shallow depth of focus of our $\mathrm{NA}=0.5$ imaging system limits direct measurements at long time of flight, we apply a $350~\mu\mathrm{s}$ ODT pulse immediately before release to launch the molecular cloud upward. The pulse amplitude is adjusted for each ToF so that the cloud returns to the focal plane at the imaging time, allowing measurements from $6.5$ to $12.5~\mathrm{ms}$ [Fig.~\ref{fig:dropletproperties}(a)]. Details of the launch-and-refocus sequence are provided in Appendix~\ref{sec:ODTpulse}.

For each ToF, we fit the two-dimensional density profile to the sum of a broad thermal Gaussian and a narrow Gaussian component. Figure~\ref{fig:dropletproperties}(b) shows the extracted widths $\sigma_x$ and $\sigma_y$ of the compact component. Both remain near $1.4~\mu\mathrm{m}$ over the full ToF range, with no statistically significant dependence on expansion time. These widths are close to the independently calibrated effective detection resolution of approximately $1.2~\mu\mathrm{m}$ (see Appendix~\ref{sec:imagingresolution}), preventing a direct determination of the intrinsic droplet size. The absence of measurable expansion nevertheless shows that the compact component remains localized after removal of the trap, providing evidence for its self-bound character~\cite{zhang2026droplet, shi2026bec}.

The compact-component population exhibits a $1/e$ lifetime of $260(40)~\mathrm{ms}$, substantially shorter than the one-body lifetime measured for the dilute molecular gas. This accelerated decay is consistent with enhanced collisional loss in a high-density state. From the measured two-body loss rate of the droplet when transferred to an unshielded rotational state (see Appendix~\ref{sec:dropletdensity}), we measure a mean droplet density of
$6.3(1.3)\times10^{13}~\textrm{cm}^{-3}$ and a phase-space density of 18(4), putting the droplets deeply in the quantum-degenerate regime.

Finally, we also observe rare configurations containing two or three spatially separated compact components [Fig.~\ref{fig:dropletproperties}(c)]. These configurations are qualitatively distinct from an ordinary single gas-phase condensate and closely resemble the droplet arrays observed in other strongly dipolar gases~\cite{ferrier-barbut2016droplets,Schmitt2016Droplet,zhang2026droplet}. Taken together, the absence of measurable expansion, the short lifetime, the high density, and the observation of presumably metastable multi-droplet configurations provide complementary evidence that the final state is a self-bound dipolar droplet. 

\begin{figure*}[t]
    \centering
    \includegraphics[width=17.5cm]{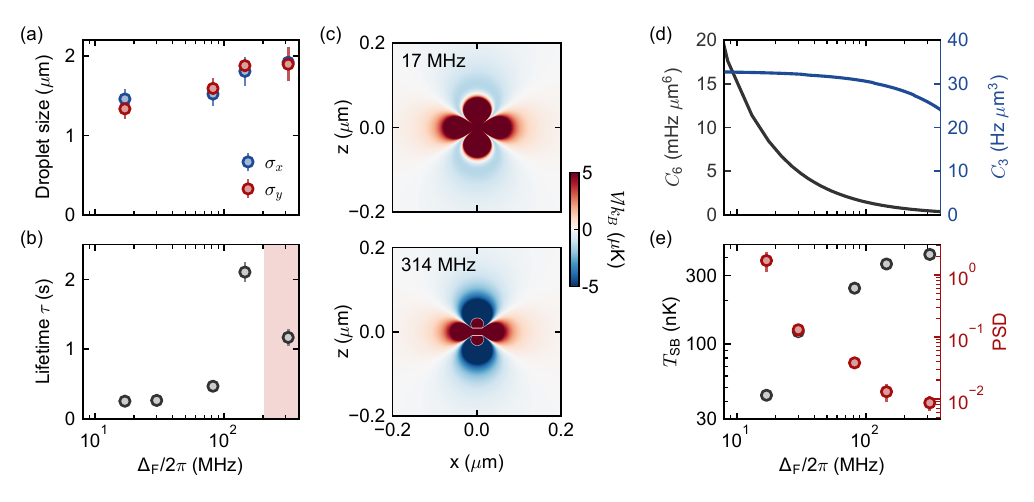}
    \caption{Droplet properties across various F\"orster defects. (a) Measured droplet sizes, $\sigma_x$ and $\sigma_y$. (b) Measured droplet lifetime, $\tau$. The red-shaded area represents the bound-state regime. (c) Effective interaction potential $V(\boldsymbol{r})$ from Eq.~(\ref{EffectivePotential}) in the $x-z$ plane. The upper and lower panels correspond to $\Delta_{\textrm{F}}/2\pi=17$~MHz and 314~MHz, respectively. The color scale is saturated for $\vert V/k_B\vert>5~\mu$K. (d) $C_6$ (left axis, black) and $C_3$ (right axis, blue) as functions of $\Delta_{\textrm{F}}$. (e) Measured self-bound droplet-formation temperature, $T_{\mathrm{SB}}$ (left axis, black), and the corresponding phase-space density (PSD) of the parent thermal cloud (right axis, red).}
    \label{fig:hightempdroplets}
\end{figure*}

\section{Formation of Droplets at Various F\"orster Defects} \label{sec:scanforster}
Having established the self-bound nature of the compact component at $\Delta_{\mathrm F}=2\pi\times17~\mathrm{MHz}$, we investigate whether droplets also form elsewhere in the F\"orster-shielded regime. At each selected $\Delta_{\mathrm F}$, we perform an evaporation at the corresponding electric field, held constant throughout. Compact components emerge at all five investigated defects, but at markedly different temperatures [Fig.~\ref{fig:hightempdroplets}].

To establish that these compact components are indeed self-bound, we repeat the ToF expansion measurement. At every investigated defect, we find that the transverse widths show no statistically significant dependence on ToF, similar to the expansion dynamics of Fig.~\ref{fig:dropletproperties} (see Appendix~\ref{sec:ODTpulse} for all plots). The absence of expansion confirms the self-bound character of the compact components at all five defects. However, the size of the droplets increases from $1.3$ to $1.9~\mu\mathrm{m}$ with increasing F\"orster defects [Fig.~\ref{fig:hightempdroplets}(a)]. 

We also characterize their stability by measuring the droplet lifetime at
each defect [Fig.~\ref{fig:hightempdroplets}(b)]. Its variation is qualitatively consistent with the two- and three-body loss landscape characterized above. The lifetime increases from $260(40)~\mathrm{ms}$ at $17~\mathrm{MHz}$ to $2.11(15)~\mathrm{s}$ at $146~\mathrm{MHz}$, where both two- and three-body loss coefficients are strongly suppressed. At $314~\mathrm{MHz}$, the lifetime decreases to $1.17(12)~\mathrm{s}$, coincident with the appearance of a field-linked bound state and enhanced collisional loss.

We next examine why the droplet-formation temperature varies so strongly with $\Delta_{\mathrm F}$. Figures~\ref{fig:hightempdroplets}(c) and \ref{fig:hightempdroplets}(d) show the evolution of the effective interaction introduced in Eq.~\eqref{EffectivePotential}. The coefficients governing the long-range dipolar interaction and the F\"orster-mediated repulsion are
\begin{equation}
\begin{aligned}
C_3&=\frac{d^2_{\tilde{1}\tilde{1}}}{4\pi\epsilon_0},\\
C_6&=
\left(
\frac{d_{\tilde{1}\tilde{2}}d_{\tilde{1}\tilde{0}}}
{4\pi\epsilon_0}
\right)^2
\frac{2}{\hbar\Delta_\mathrm{F}}.
\end{aligned}
\end{equation}
We define the well depth $E_{\mathrm{well}}>0$ as the magnitude of the global minimum of the effective interaction in Eq.~\eqref{EffectivePotential}. This minimum occurs along the attractive dipolar direction, $\theta=0$, at $r=(4C_6/C_3)^{1/3}$, giving $E_{\mathrm{well}}=C_3^2/4C_6$.

 Across $\Delta_{\mathrm F}/2\pi=17$ to $314~\mathrm{MHz}$, $C_3/h$ changes only modestly from $33$ to $26~\mathrm{Hz}\,\mu\mathrm{m}^3$, whereas $C_6/h$ decreases from $8.4$ to $0.4~\mathrm{mHz}\,\mu\mathrm{m}^6$ [Fig.~\ref{fig:hightempdroplets}(d)]. The weak variation of $C_3$ reflects the small change in the induced dipole moment, while $C_6$ decreases approximately as $1/\Delta_{\mathrm F}$ because of the energy denominator in the second-order coupling to the nearby pair channel. As a result, the dipolar interaction changes only weakly while the second-order repulsive contribution is strongly reduced. The repulsive core therefore shifts to smaller intermolecular separations, producing a deeper attractive well, as illustrated in Fig.~\ref{fig:hightempdroplets}(c).

We denote by $T_{\mathrm{SB}}$ the temperature of the thermal gas right before the self-bound droplet emerges. As shown in Fig.~\ref{fig:hightempdroplets}(e), $T_{\mathrm{SB}}$ increases from $44(4)~\mathrm{nK}$ at $\Delta_{\mathrm F}/2\pi=17~\mathrm{MHz}$ to $421(17)~\mathrm{nK}$ at $314~\mathrm{MHz}$. We find
\begin{equation}
\frac{k_{\mathrm B}T_{\mathrm{SB}}}{E_{\mathrm{well}}}\sim0.03
\end{equation}
at all five investigated F\"orster defects. This approximate scaling suggests that $E_{\mathrm{well}}$ sets the characteristic thermal energy scale for droplet formation. The rise in $T_{\mathrm{SB}}$ also reshapes the relation between droplet formation and quantum degeneracy of the parent gas. At $\Delta_{\mathrm F}/2\pi=17~\mathrm{MHz}$, the droplet emerges from a thermal gas in the quantum degenerate regime, characterized by a phase-space density of order unity. At $\Delta_{\mathrm F}/2\pi=146$ and $314~\mathrm{MHz}$, however, the droplets emerge at $360(30)$ and $421(17)~\mathrm{nK}$, respectively, from non-degenerate parent gases ($\mathrm{PSD} \sim 0.01$). A quantum-degenerate parent gas is therefore not a prerequisite for the formation of self-bound droplets. This contrasts with the conventional magnetic atom scenario, where droplets emerge from a Bose--Einstein condensate and are stabilized by repulsive Lee--Huang--Yang (LHY) quantum fluctuations~\cite{ferrier-barbut2016droplets,
Wachtler2016QuantumFilaments,Chomaz2016QuantumFluctuation}; here, the repulsive shielding core offers an alternative stabilization mechanism that operates even without a pre-existing condensate.


\begin{figure}[t]
    \centering
    \includegraphics[width=8.6cm]{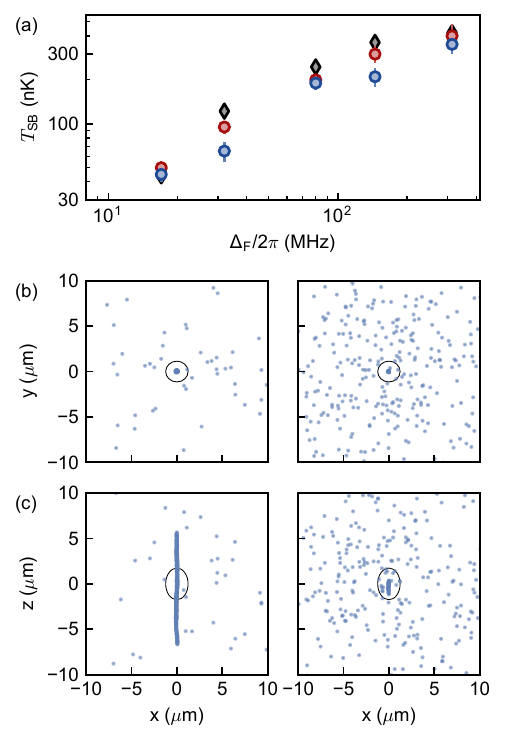}
    \caption{ Numerical simulations. (a) Temperatures at which the droplet-to-gas transition occurs, for $N=200$ molecules (blue) and $N=400$ molecules (red). Black diamonds mark the experimental values from Fig.~\ref{fig:hightempdroplets}. (b) Simulation snapshot, in the $x-y$ plane, of $N=400$ molecules at $\Delta_{\rm F} = 2\pi \times 80$ MHz, at $T = 150$ nK (left) and  $T = 250$ nK (right). Each blue dot corresponds to an imaginary-time particle image. The black ellipse represents the harmonic oscillator ground state. (c) Same as above, but in the $x-z$ plane.}
    \label{fig:simulations}
\end{figure}

In order to further investigate the physics of droplet formation observed in our measurements, we have performed Path Integral Monte Carlo simulations \cite{ceperley1995review} using the field-induced interaction potential Eq.~(\ref{EffectivePotential}) and the harmonic confinement of our experiments (see Appendix~\ref{sec:simulations}). We have simulated ensembles of $N=200$ and $N=400$ molecules, corresponding to the typical range of particle numbers in the experimentally observed droplets.

Upon varying $\Delta_{\rm F}$ and the temperature in the simulations, we observe the formation of bound structures. Remarkably, the clusters we find are highly anisotropic [see Fig.~\ref{fig:simulations}(c)]. Molecules align along the $z$-axis at typical intermolecular distances on the order of the separation corresponding to the potential minimum shown in Fig. \ref{fig:theory}(c). Such strongly correlated filaments are morphologically distinct from the self-bound droplets observed in dilute and weakly interacting dipolar Bose-Einstein condensates of magnetic atoms \cite{Schmitt2016Droplet, Chomaz2016QuantumFluctuation} with positive $s$-wave scattering length. Instead, similar filamentation processes have been found in previous PIMC studies in strongly interacting ensembles of dipolar bosons \cite{Cinti2017a,Cinti2017b}.

We can numerically detect the formation of filaments from the molecular gas phase as described in Appendix~\ref{sec:simulations}. Figure~\ref{fig:simulations}(a) shows the obtained critical temperature, $T_{\rm SB}$, for this gas-to-filament transition as a function of $\Delta_{\rm F}$ and for two different particle numbers $N=200$ and $N=400$. The calculated critical temperature depends only weakly on the number of particles  and agrees well with the results of our experimental measurements. It will be interesting in future experiments to explore the predicted filamentary nature of our observed droplets, for example by rotating the electric field direction so that the filaments align perpendicular to the high-resolution imaging axis.

\section{Conclusion}
In conclusion, we have identified a bound-state-free regime of dc F\"orster shielding that retains strong suppression of two-body loss without the enhanced three-body recombination associated with a field-linked bound state. This collisional stability enables efficient evaporation of bosonic NaRb molecules to quantum degeneracy with efficiency of $2.09(9)$. The emergence of dense, non-expanding components, even from parent clouds well outside the quantum degeneracy regime, provides evidence for self-bound dipolar molecular matter.

Direct evaporation into the self-bound regime, though it limits the droplet fraction in the gas due to the short droplet lifetime, provides an opportunity to map the finite-temperature phase diagram~\cite{Langen2025DropletPhase, Zhang2025DropletPhase}. The strength of the electric field in our shielding scheme simultaneously determines the elastic-to-inelastic collision ratio and the $s$-wave scattering length, leaving limited freedom to stabilize a gas-phase condensate. One approach to address this limitation is to combine dc F\"{o}rster shielding with a single microwave field, which could provide additional interaction tunability while avoiding the photon-changing collisions that dominate two-body loss in dual-microwave shielding~\cite{wang2026bound}. Another is to transfer the ultracold molecules to a quasi-two-dimensional geometry at temperatures above $T_{\mathrm{SB}}$. The trap geometry would provide further stability~\cite{de2011controlling,koch2008stabilization}, allowing the investigation of transitions between gaseous superfluids, crystalline phases, and potential supersolid phases even in a regime of dominant dipolar interactions~\cite{Buchler2007Strongly,Astrakharchik2007Quantum,Bombin2017Dipolar}. 

The quasi-two-dimensional geometry is also naturally compatible with our molecular quantum gas microscope~\cite{Rosenberg2022HBT,christakis2023probing}. Freezing the distribution of the continuum gas in a deep 2D lattice would enable site-resolved measurements of density fluctuations and spatial correlations~\cite{Xiang2025ContinuumQGM,Yao2025ContinuumQGM,deJongh2025ContinuumQGM}. Alternatively, adiabatically loading the molecular gas into a two-dimensional lattice would enable microscopic studies of the extended Bose--Hubbard model with long-range dipolar interactions~\cite{Goral2002EBH,Capogrosso-Sansone2010EBH}.

\section{Acknowledgments}

We would like to thank Jeremy Hutson for helpful discussions, and for sharing the results of his calculations on static-field shielding of NaRb, which we used to benchmark our own code. We also thank Lysander Christakis and Pascal Weckesser for earlier contributions to the experimental work. This work was supported by the NSF (grant
no. 2409375 and QLCI grant OMA-2120757), and the David and Lucile Packard Foundation (grant no. 2016-65128). L.F. was supported by the NSF Graduate Research Fellowship Program. T.P. and M.C. acknowledge support from the
Austrian Science Fund (FWF) through the SFB Qnnect (Grant No.\ 10.55776/F101200) and the cluster of excellence quantA (Grant No.\ 10.55776/COE1) and
the European Union (NextGenerationEU), from the SNSF
through the Swiss Quantum Initiative, and from the
European Research Council through the ERC Synergy
Grant ``SuperWave" (Grant No.\ 101071882). M.C. also acknowledges support from the European Union’s Horizon Europe research and innovation program under the Marie Sklodowska-Curie Grant Agreement No. 101205948 (Q-UDiM).

\appendix

\section{Experimental Upgrades}

Compared with our previous work~\cite{christakis2023probing}, several upgrades were made to the experimental apparatus to significantly increase the molecule number, including increasing the initial numbers of both atomic species before magnetoassociation, and switching from lattice-based molecule preparation to a bulk approach. To increase the atom numbers, we redesigned the magneto-optical trap (MOT) configuration. Previously, evaporation produced approximately $2.7\times10^5$ atoms in each of the Na and Rb Bose-Einstein condensates (BECs)~\cite{lysanderthesis}. While these atom numbers were sufficient for loading directly into an optical lattice, they limited the molecule number available for evaporative cooling toward quantum degeneracy. 

\begin{figure}[t]
    \centering
    \includegraphics[width=8.6cm]{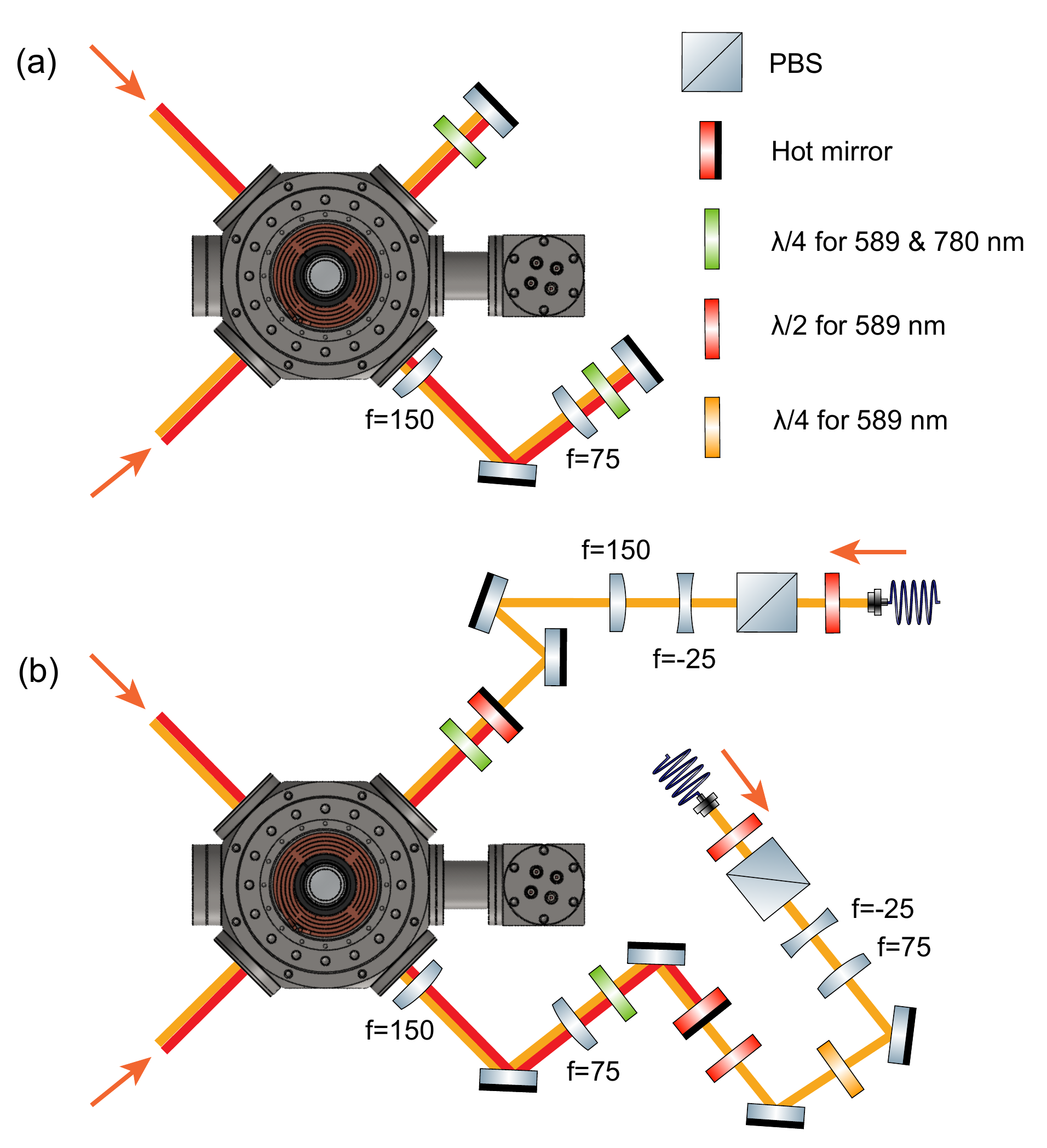}
    \caption{Three-dimensional MOT setups. The previous three-beam MOT setup is shown in (a) and the new five-beam MOT set-up is shown in (b). The yellow (red) lines represent the Na (Rb) cooling beams with wavelength $589$ nm ($780$ nm). Orange arrows indicate the incident beams. All the units of focal lengths are millimeters.}
    \label{fig:5beamMOT}
\end{figure}

Fig.~\ref{fig:5beamMOT} summarizes the modifications made to the system. In the previous configuration, we used a conventional three-beam retro-reflected MOT [Fig.~\ref{fig:5beamMOT}(a)]. Cooling beams enter along the $x$, $y$, and $z$ directions and are retro-reflected to generate the corresponding counter-propagating beams. While this geometry reduces the required optical power and simplifies the optical layout compared with six independently controlled beams, its performance in our vacuum chamber is limited by the available optical access. In particular, the electrodes used for our electric field clip part of the cooling light, casting shadows in the retro-reflected paths. Consequently, the retro-reflected beams have lower power than the incident beams, resulting in an imbalance between the counter-propagating beams. This imbalance degrades the optical-molasses stage and leads to additional heating during transfer into the magnetic trap. 

To solve this issue, we implemented a five-beam MOT configuration for Na [Fig.~\ref{fig:5beamMOT}(b)], in which the four horizontal ($\pm x$, $\pm y$) cooling beams are supplied independently. The $z$ axis beam remains retro-reflected because the high-resolution objective limits optical access through the top of the vacuum chamber. We implemented this modification only for Na because Na serves as the sympathetic coolant during dual-species evaporation. The Rb MOT retains the previous three-beam retro-reflected geometry, with the Na and Rb beam paths combined using dichroic mirrors that transmit Na light and reflect Rb light. Independently controlling the horizontal Na cooling beams allows the counter-propagating powers to be balanced, improving the number and temperature of Na atoms after loading into the magnetic trap.

Following optical molasses, the five-beam Na MOT yields $1.4\times10^9$ Na atoms at $96~\mu$K, of which $6.7\times10^8$ are subsequently captured in the magnetic trap at a temperature of $200~\mu$K. After loading into the optical dipole trap and evaporative cooling without Rb, we obtain $5.4\times10^6$ Na atoms at the BEC phase transition. For dual-species operation with a $30$~ms Rb MOT loading time, we can produce BECs containing $6.7 \times 10^5$ Na and $4.9 \times 10^5$ Rb atoms. These improvements were crucial for producing sufficiently large initial molecular samples for evaporation to quantum degeneracy.

\section{Electric Field}

\subsection{Fine-tuning electric field gradients}

A strong, tunable static electric field is generated using an octagonal arrangement of eight electrode rods~\cite{lysanderthesis}. Positive and negative voltages are set by two low-noise digital-to-analog converters (EVAL-AD5791SDZ) and amplified by high-voltage amplifiers (Trek-10-10). The set points for each electrode are generated from these voltages using precision voltage dividers~\cite{jasonthesis}. By tuning the voltage differences across the eight rods, electric field gradients can be fine-tuned to optimize evaporation efficiency and initial molecule conditions. 

We measured these electric field gradients by tracking the molecular cloud center-of-mass motion during ToF. Evaporation is optimized with a $3.51(5)~\textrm{m/s}^2$ acceleration in the $x$-$y$ plane. The initial cloud number and temperature are optimized with an acceleration of $8.44(16)~\textrm{m/s}^2$ in the $z$ direction, including gravitational acceleration $g$. The electric field is ramped up in the beginning of the sequence and remains fixed throughout atomic and molecular evaporation.

\subsection{STIRAP at finite electric field}
We perform STIRAP in the presence of the static electric field so that the molecules are shielded immediately after transfer to the electronic and vibrational ground-state manifold. This procedure is necessary because the bandwidth of our electric-field control system is approximately $1~\mathrm{kHz}$; if STIRAP were instead performed at zero field, the subsequent ramp into the shielding regime would be slow enough to lose a substantial fraction of the unshielded ground-state molecules before the target field was reached.

\begin{figure}[t]
    \centering
    \includegraphics[width=8.6cm]{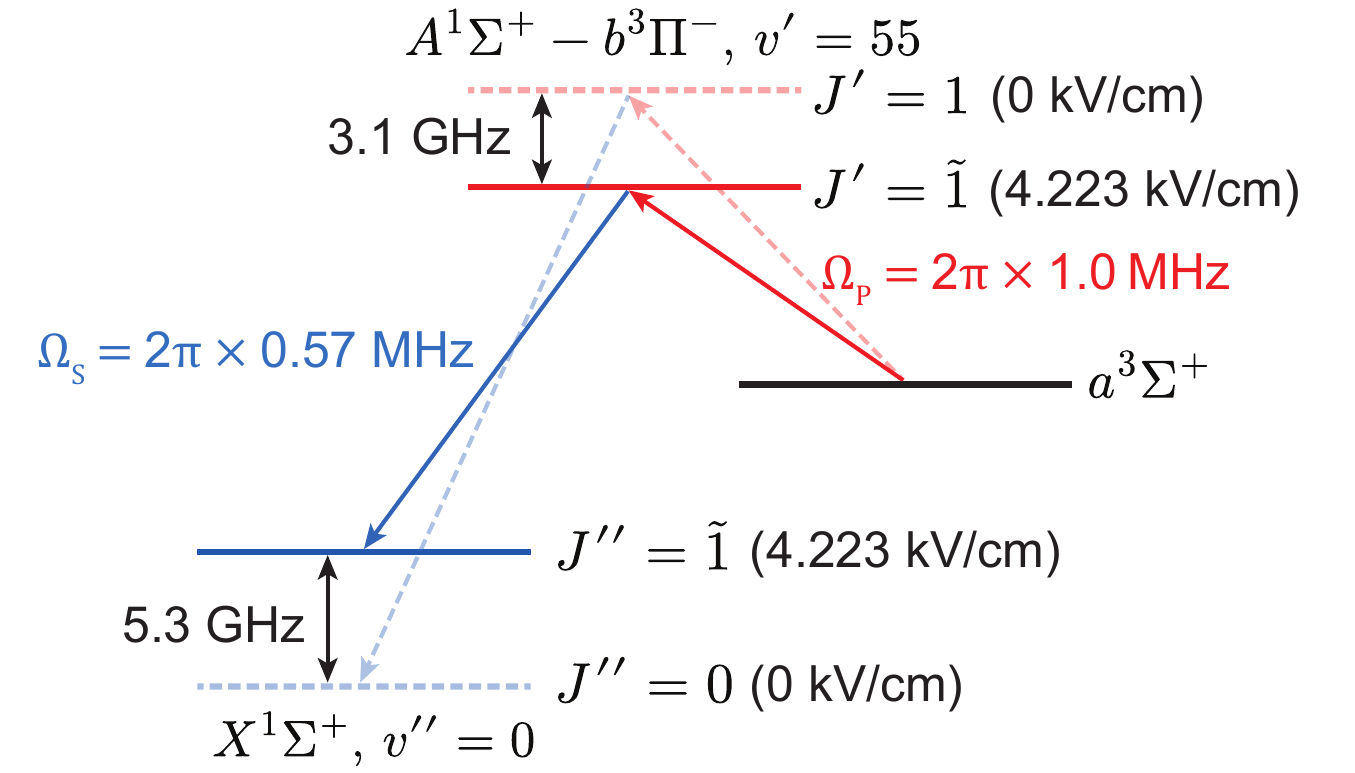}
    \caption{STIRAP diagram at electric fields of zero and 4.223 kV/cm. Dashed lines show energy levels of intermediate and ground states for STIRAP pulses implemented in previous studies at zero electric field~\cite{guo2018highresolution}. Energy levels of intermediate and ground states and STIRAP pulses we use at an electric field of 4.223 kV/cm are indicated by solid lines. Rabi frequencies for the pump and Stokes transitions are denoted as $\Omega_{\textrm{P}}$ and $\Omega_{\textrm{S}}$, respectively.}
    \label{fig:STIRAP}
\end{figure}

Our STIRAP scheme addresses a different intermediate hyperfine level and a different rotational target state from those used in previous zero-field experiments [Fig.~\ref{fig:STIRAP}]. The selected intermediate level lies $200~\mathrm{MHz}$ below the level used in Ref.~\cite{PhysRevA.96.052505}. Accounting for both the dc Stark shift at an electric field of $4.223$~kV/cm and the different choice of hyperfine state, the STIRAP frequency is shifted by $3.1$~GHz.

The target state is the field-dressed rotational state $\ket{\tilde{1},0}$ in the electronic and vibrational ground-state manifold, which is shielded above the F\"orster resonance. At an electric field of $4.223~\mathrm{kV/cm}$, this state lies $5.3~\mathrm{GHz}$ above the zero-field rovibrational ground state. The pump and Stokes Rabi frequencies are
$\Omega_{\textrm{P}}=2\pi\times1~\mathrm{MHz}$ and
$\Omega_{\textrm{S}}=2\pi\times0.57~\mathrm{MHz}$, respectively. The corresponding beam waists are $90$ and $150~\mu\mathrm{m}$, allowing reverse bound-to-free STIRAP to be performed after time-of-flight expansion. The one-way STIRAP efficiency at $346.7~\mathrm{G}$ is $94(2)\%$, while the reverse bound-to-free STIRAP efficiency near $347.6~\mathrm{G}$ is $87(2)\%$. 

Because STIRAP is performed at the same electric field used for each loss measurement, we measure the pump and Stokes spectra throughout the range $3~\mathrm{MHz}\leq\Delta_{\textrm{F}}/2\pi\leq314~\mathrm{MHz}$
and adjust the optical frequencies accordingly. 

\subsection{Electric-field stability}

The two-photon STIRAP linewidth provides a sensitive probe of transition-frequency fluctuations. We use its low-power limit to place an upper bound on electric-field-induced noise. All detunings and linewidths below are angular frequencies, and all linewidths denote Gaussian rms widths.

Near two-photon resonance, the intrinsic STIRAP response is approximated by~\cite{PhysRevA.65.043409}
\begin{align}
P_{\mathrm{STIRAP}}(\delta)
&\propto
\exp\left(
-\frac{\delta^2}{2\Delta_{\mathrm{Rabi}}^2}
\right),
\nonumber\\
\Delta_{\mathrm{Rabi}}^2
&=
\frac{\Omega_{\mathrm P}\Omega_{\mathrm S}}{\gamma\tau},
\label{eq:STIRAP_intrinsic_width}
\end{align}
where \(\Omega_{\mathrm P}\) and \(\Omega_{\mathrm S}\) are the pump and Stokes Rabi frequencies, \(\gamma\) is the intermediate-state decay rate, and \(\tau\) is the characteristic pulse duration.

Spatially uniform electric- and magnetic-field fluctuations are modeled as a Gaussian detuning offset,
\begin{equation}
P_{\mathrm{noise}}(\delta_0)
=
\frac{1}{\sqrt{2\pi}\Delta_{\mathrm{noise}}}
\exp\left(
-\frac{\delta_0^2}{2\Delta_{\mathrm{noise}}^2}
\right),
\end{equation}
with
\begin{equation}
\Delta_{\mathrm{noise}}^2=\Delta_{\mathrm{Electric}}^2+\Delta_{\mathrm{Magnetic}}^2,
\label{eq:field_noise}
\end{equation}
assuming that the two contributions are statistically independent. We note that for electric field the control-voltage noise is approximately white from \(10~\mathrm{Hz}\) while the bandwidth of the electrode assembly is $\sim1~\mathrm{kHz}$. We therefore approximate the resulting detuning as constant during each \(\sim30~\mu\mathrm{s}\) STIRAP pulse while allowing it to vary shot to shot. The same approximation also holds for magnetic field noise. The relative pump--Stokes frequency noise is estimated to contribute only at the kilohertz level and is neglected.

A spatial electric-field gradient produces an additional Gaussian broadening. For a thermal cloud confined harmonically along the gradient direction,
\begin{equation}
\Delta_{\mathrm{gradient}}=|\delta'|\sqrt{\frac{k_{\mathrm B}T}{m_{\mathrm{NaRb}}\omega_x^2}},
\qquad
\delta'=\frac{\partial\delta}{\partial x}.
\label{eq:gradient_broadening}
\end{equation}
Because the intrinsic response, spatially uniform field noise, and gradient broadening are independent and Gaussian, their variances add:
\begin{equation}
\Delta_{\mathrm{obs}}^2=\Delta_{\mathrm{Rabi}}^2+\Delta_{\mathrm{noise}}^2+\Delta_{\mathrm{gradient}}^2.
\label{eq:STIRAP_linewidth}
\end{equation}

As the pump and Stokes powers are reduced, the measured linewidth saturates at
\begin{equation}
\Delta_{\mathrm{sat}}=2\pi\times29(3)~\mathrm{kHz}.
\end{equation}

The measured two-photon STIRAP linewidth at full pump and Stokes laser power is $\Delta_{\textrm{obs}}=2\pi\times58(2)$~kHz. Therefore, spatially uniform field noise and the gradient are not expected to significantly affect STIRAP under full-power operating conditions.

An independent measurement gives
\begin{equation}
\delta'=2\pi\times0.97~\mathrm{kHz}/\mu\mathrm{m},
\end{equation}
which, together with the measured temperature and trap frequency, gives
\begin{equation}
\Delta_{\mathrm{gradient}}
=
2\pi\times12~\mathrm{kHz}.
\end{equation}
Subtracting the gradient contribution in quadrature yields
\begin{equation}
\Delta_{\mathrm{rem}}
=
\sqrt{
\Delta_{\mathrm{sat}}^2
-
\Delta_{\mathrm{gradient}}^2
}
=
2\pi\times26(3)~\mathrm{kHz}.
\label{eq:remaining_linewidth}
\end{equation}

Magnetic-field noise may also contribute to \(\Delta_{\mathrm{rem}}\). Conservatively attributing the entire remaining linewidth to electric-field noise therefore gives
\begin{equation}
\Delta_{\mathrm{Electric}}
\leq
\Delta_{\mathrm{rem}}
=
2\pi\times26(3)~\mathrm{kHz}.
\label{eq:electric_frequency_bound}
\end{equation}
The corresponding rms electric-field fluctuation satisfies
\begin{align}
\sigma_{\mathcal E}
\leq
\frac{\Delta_{\mathrm{rem}}}
{\left|\partial\delta/\partial\mathcal E\right|} = 1.11 (13)\times10^{-1}~\textrm{V/cm}, \nonumber \\
\sigma_{\mathcal E}/{\mathcal E} \leq 26(3)~\textrm{ppm}.
\end{align}
At the operating field, the ratio between the Stark sensitivities of the F\"orster defect \(\Delta_{\mathrm F}\) and the STIRAP resonance is \(5.0\). The resulting bound is therefore
\begin{equation}
\sigma_{\Delta_{\mathrm F}}\leq\left|
\frac{\partial\Delta_{\mathrm F}/\partial\mathcal E}{\partial\delta/\partial\mathcal E}\right|\Delta_{\mathrm{rem}}
=2\pi\times130(15)~\mathrm{kHz}.
\label{eq:forster_noise_bound}
\end{equation}
This upper bound is more than two orders of magnitude smaller than the F\"orster defect used for evaporative cooling, $\Delta_{\textrm{F}}/2\pi=17$~MHz, at which we reach quantum degeneracy. Thus, electric-field fluctuations are not expected to significantly perturb the shielding condition during evaporation.

\subsection{Measuring F\"orster defect $\Delta_{\textrm{F}}$}
We determine the F\"orster defect using microwave spectroscopy performed with an Agilent Technologies E8257C PSG analog signal generator. We first ramp our electric field to the desired point above the F\"orster resonance. After performing this ramp, we prepare the molecules in the field-dressed rotational state $\ket{\tilde{1},0}$. A microwave pulse then transfers the molecules to the unshielded $\ket{\tilde{2},0}$ or $\ket{\tilde{0},0}$ states. Molecules transferred out of the initial state are rapidly lost, producing dips in the detected molecule number as the microwave frequency is scanned across the resonance. 

From the centers of the loss features, we determine the transition frequencies
$\nu_{\tilde{1}\tilde{2}}$ and $\nu_{\tilde{1}\tilde{0}}$ for the
$\ket{\tilde{1},0}\rightarrow\ket{\tilde{2},0}$ and
$\ket{\tilde{1},0}\rightarrow\ket{\tilde{0},0}$ transitions,
respectively. Their difference gives the F\"orster defect,
\begin{equation}
    \hbar\Delta_{\mathrm{F}}
    =
    h\left(
    \nu_{\tilde{1}\tilde{0}}
    -
    \nu_{\tilde{1}\tilde{2}}
    \right)
    =
    E_{\tilde{1},\tilde{1}}
    -
    E_{\tilde{0},\tilde{2}},
\end{equation}
where $\Delta_{\mathrm{F}}$ is the F\"orster defect in angular-frequency units
and $E_{\tilde{N}_1,\tilde{N}_2}$ denotes the asymptotic energy of the dressed
rotational pair state
$\ket{\tilde{N}_1,0;\tilde{N}_2,0}$.

\section{Interaction potentials near the F\"orster resonance}
\label{sec:interactionpotentials}
We calculate the interaction potentials by constructing the two-molecule Hamiltonian $H(r)$ at fixed intermolecular separation $r$. The Hamiltonian includes the dc Stark shifts of the rotational states, the dipole--dipole interaction, the centrifugal energy associated with the relative orbital angular momentum, and the electronic van der Waals interaction. We use the basis
\[
\vert \tilde{N}_1,m_{N_1};\tilde{N}_2,m_{N_2}\rangle
\vert L,m_L\rangle,
\]
where $\vert L,m_L\rangle$ denotes the partial wave for the relative motion of the two molecules. The total projection
$M=m_{N_1}+m_{N_2}+m_L$
is conserved and is fixed within each symmetry block. We truncate the rotational basis to
$\tilde{N}_1,\tilde{N}_2\in\{0,1,2\}$
and include even partial waves up to $L=20$. The effects of higher rotational states are incorporated perturbatively through a Van Vleck transformation, which generates additional $1/r^6$ interaction terms~\cite{mukherjee2024controlling}. Couplings to hyperfine channels are neglected~\cite{mukherjee2024controlling}. With these truncations, the Hamiltonian has dimension $455\times455$. The adiabatic interaction potentials are obtained by diagonalizing $H(r)$ at each intermolecular separation.

For physical insight into the shielding mechanism, we also consider a reduced two-channel model. Near the F\"orster resonance, the entrance channel
$\vert\tilde{1},0;\tilde{1},0\rangle$
is resonantly coupled to the exchange-symmetric
$\vert\tilde{0},0;\tilde{2},0\rangle$
channel, while the remaining pair states are neglected in the reduced description~\cite{lassabliere2022model}. In this two-channel basis, the Hamiltonian is
\begin{eqnarray}
    E_{1}
    &=&
    \frac{d^2_{\tilde{1}\tilde{1}}}{4\pi\epsilon_0}
    \frac{1-3\cos^2\theta}{r^3},
    \nonumber\\
    E_{2}
    &=&
    \frac{
    d_{\tilde{0}\tilde{0}}d_{\tilde{2}\tilde{2}}
    +
    d_{\tilde{0}\tilde{2}}d_{\tilde{2}\tilde{0}}
    }{4\pi\epsilon_0}
    \frac{1-3\cos^2\theta}{r^3}
    -
    \hbar\Delta_{\textrm{F}},
    \nonumber\\
    W
    &=&
    \frac{
    d_{\tilde{1}\tilde{2}}d_{\tilde{1}\tilde{0}}
    }{4\pi\epsilon_0}
    \frac{\sqrt{2}(1-3\cos^2\theta)}{r^3},
    \nonumber\\
    H_{2\times2}
    &=&
    \begin{pmatrix}
        E_{1} & W\\
        W & E_{2}
    \end{pmatrix}.
    \label{EffectivePotentialHamiltonian}
\end{eqnarray}
Here,
$d_{\tilde{i}\tilde{j}}
=
\langle\tilde{i},0\vert d_z\vert\tilde{j},0\rangle$
is a dipole matrix element, and $\theta$ is the angle between the intermolecular axis and the electric-field direction. Taking the asymptotic energy of the entrance channel as the zero of energy, $E_{1}$ describes the diagonal dipole--dipole interaction in the entrance channel and approaches zero as $r\rightarrow\infty$. In contrast, $E_{2}$ describes the dipole--dipole interaction in the exchange-symmetric channel and approaches $-\hbar\Delta_{\textrm{F}}$ asymptotically. The off-diagonal matrix element $W$ describes the dipole--dipole coupling between the two channels.

For $\Delta_{\textrm{F}}>0$, the adiabatic potential correlated with the
$\vert\tilde{1},0;\tilde{1},0\rangle$
entrance channel is the upper eigenvalue of $H_{2\times2}$,
\begin{eqnarray}
    V(\boldsymbol{r})
    =
    \frac{1}{2}(E_{1}+E_{2})
    +
    \frac{1}{2}
    \sqrt{(E_{1}-E_{2})^2+4W^2}.
    \label{Eq:AppendixEffectivePotential}
\end{eqnarray}
This reduced model captures the avoided crossing responsible for the repulsive shielding interaction, while the full $455$-channel calculation provides the quantitative adiabatic potentials and bound-state spectrum used in the main text.

\section{Two-body coupled-channel calculation}\label{sec:coupled-channel}
\begin{figure}[t]
    \centering
    \includegraphics[width=8.6cm]{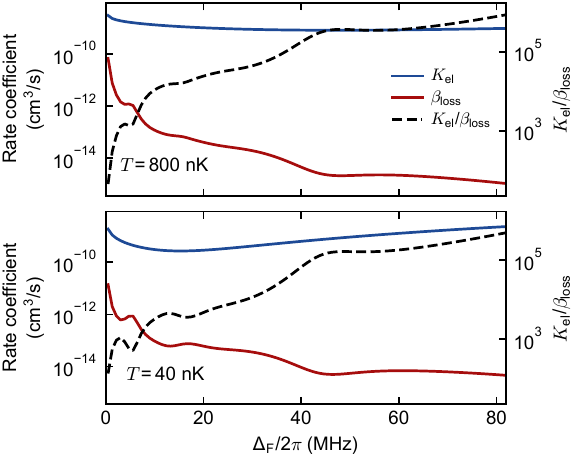}
    \caption{Thermally averaged elastic collision rate coefficient $K_{\textrm{el}}$ (blue) and loss rate coefficient $\beta_{\textrm{loss}}$ (red) as a function of the F\"orster defect at $T=800$~nK (upper panel) and $T=40$~nK (lower panel). The left axis shows the rate coefficients in units of cm$^3$/s. The right axis shows the elastic-to-inelastic ratio, $K_{\textrm{el}}/\beta_{\textrm{loss}}$, plotted as a black dashed line.}
    \label{fig:theory_elastic_to_inelastic}
\end{figure}

Two-body coupled-channel calculations are used to determine the two-body loss-rate coefficient $\beta_{\textrm{loss}}$ and the scattering length $a_s$ as functions of the F\"orster defect $\Delta_{\textrm{F}}$. The calculations describe collisions between two molecules that interact via the dipole--dipole interaction in the presence of a static electric field. The coupled radial equations generated by the Hamiltonian $H(r)$ described above are solved over the range of intermolecular separations $50a_0\leq r\leq2\times10^5a_0$ using the renormalized Numerov algorithm~\cite{johnson1978renormalized,janssen2012cold}. In this method, two linearly independent, real-valued solution matrices, $\mathbf{F}(r)$ and $\mathbf{G}(r)$, are propagated, allowing the physical short- and long-range boundary conditions to be imposed after propagation~\cite{karman2023resonances}. At the short-range matching radius, $r=50a_0$, we impose a universal-loss boundary condition. The resulting scattering matrix is used to obtain the cross sections for elastic scattering $\sigma_{\textrm{el}}$ and total loss $\sigma_{\textrm{loss}}$, and the scattering length $a_s$~\cite{mukherjee2023shielding}. We estimate $a_s$ at a collision energy corresponding to $E/k_B=100$~pK, below which its value is insensitive to further reductions in energy. Elastic collisional rate and two-body loss-rate coefficients at fixed collisional energy $E$ are given by
\begin{align}
K_{\textrm{el}} = \sqrt{\frac{2E}{\mu}}\sigma_{\textrm{el}}, \nonumber \\
\beta_{\textrm{loss}}(E)
=
\sqrt{\frac{2E}{\mu}}\sigma_{\textrm{loss}},
\end{align}
respectively. Here, $\mu=m_{\textrm{NaRb}}/2$ is the reduced mass. The results based on our two-body coupled-channel calculation are consistent with Ref.~\cite{mukherjee2024controlling} when $E/k_B=10$~nK. To compare with experimental results, we also calculate the thermally averaged elastic collisional rate and two-body loss-rate coefficients as

\begin{align}
K_{\textrm{el}}(T)
=
\sqrt{\frac{8k_BT}{\pi\mu}}
\frac{1}{(k_BT)^2}
\int_0^\infty
\sigma_{\textrm{el}}(E)e^{-E/k_BT}EdE, \nonumber \\
\beta_{\textrm{loss}}(T)
=
\sqrt{\frac{8k_BT}{\pi\mu}}
\frac{1}{(k_BT)^2}
\int_0^\infty
\sigma_{\textrm{loss}}(E)e^{-E/k_BT}EdE.
\end{align}
For the numerical integration, we calculate the cross sections at 32 logarithmically spaced collision energies between 3~nK and 30~$\mu$K.

Fig.~\ref{fig:theory_elastic_to_inelastic} shows the calculated thermally averaged $K_{\textrm{el}}$ and $\beta_{\textrm{loss}}$ as a function of $\Delta_{\textrm{F}}$. At the initial temperature, $T=800$~nK, the elastic-to-loss ratio is $K_{\textrm{el}}/\beta_{\textrm{loss}}\sim1.8\times10^4$, whereas at $T=40$~nK, where the gas reaches quantum degeneracy, it decreases to $\sim3.5\times10^3$.

\section{Kinetic model}\label{sec:kinetic model}
We use a kinetic model for the coupled evolution of the molecule number $N$ and total energy $E$ to extract the two- and three-body loss coefficients. During the loss measurements, the truncation parameter satisfies $\eta>10$, such that evaporation during the hold time is negligible.

For a thermal gas in a three-dimensional harmonic trap,
\begin{equation}
    E=3Nk_BT,
\end{equation}
and the peak density is
\begin{equation}
    n_0
    =
    N\omega_x\omega_y\omega_z
    \left(
        \frac{m_{\textrm{NaRb}}}{2\pi k_BT}
    \right)^{3/2}.
\end{equation}
The density moments relevant to two- and three-body loss are
\begin{equation}
\begin{aligned}
    \frac{1}{N}
    \int n^2(\boldsymbol r)\,d^3\boldsymbol r
    &=
    \frac{n_0}{2\sqrt{2}},\\
    \frac{1}{N}
    \int n^3(\boldsymbol r)\,d^3\boldsymbol r
    &=
    \frac{n_0^2}{3\sqrt{3}}.
\end{aligned}
\end{equation}
In addition to collisional loss, the model includes one-body decay with
lifetime $\tau$ and density-independent technical heating at a rate
$\dot{T}_c$.

\subsection{Measurement of one-body loss and background heating}

We determine the one-body lifetime using a low-density molecular sample for which collisional loss is negligible. To prepare this sample, we hold the molecules for more than $3~\mathrm{s}$ at $\Delta_{\textrm{F}}=2\pi\times82$ MHz after evaporation and subsequently recompress the optical dipole trap (ODT). An exponential fit to the molecule number gives $\tau=7.07(12)~\mathrm{s}$.

We independently determine the background heating rate at
$\Delta_{\mathrm{F}}=2\pi\times82~\mathrm{MHz}$, where the calculated two-body loss coefficient is below $10^{-14}~\mathrm{cm^3\,s^{-1}}$. Under these conditions, loss-induced anti-evaporation heating is negligible. The measured temperature increases approximately linearly with hold time, yielding $\dot{T}_c=70(3)~\mathrm{nK/s}$. 

\subsection{Two-body loss}

Including one- and two-body loss, the molecule number and total energy obey
\begin{equation}
\begin{aligned}
    \frac{dN}{dt}&= -\frac{N}{\tau} - \beta_{\mathrm{loss}} \frac{n_0}{2\sqrt{2}}N,\\
    \frac{dE}{dt}&= -\frac{E}{\tau} +(h_{\textrm{2B}}-\frac{3}{4})\beta_{\mathrm{loss}}\frac{n_0}{2\sqrt{2}}E+3Nk_B\dot{T}_c .
    \label{twobodyloss}
\end{aligned}
\end{equation}
Here, $\beta_{\mathrm{loss}}$ is the two-body loss coefficient. The dimensionless parameter $h_{\textrm{2B}}$ phenomenologically accounts for additional changes in the energy per particle associated with momentum dependence of the two-body loss and other loss-induced heating mechanisms~\cite{lin2023SingleMWShielding,ye2018collisions}. The factor $3/4$ in Eq.~(\ref{twobodyloss}) describes anti-evaporation
heating. Because two-body loss is weighted by $n^2(\boldsymbol r)$, molecules
are preferentially removed from the center of the trap. Their mean potential
energy is
\begin{equation}
    \langle U\rangle_{\mathrm{2B}}
    =
    \frac{
        \int U(\boldsymbol r)n^2(\boldsymbol r)\,d^3\boldsymbol r
    }{
        \int n^2(\boldsymbol r)\,d^3\boldsymbol r
    }
    =
    \frac{3}{4}k_BT.
\end{equation}
For a Maxwell--Boltzmann momentum distribution, the mean kinetic energy of a
lost molecule is $3k_BT/2$. Each lost molecule therefore removes
$9k_BT/4$, equal to $3/4$ of the cloud-averaged energy per particle
$E/N=3k_BT$. The corresponding two-body contribution to the energy equation
is
\begin{equation}
    \left.\frac{dE}{dt}\right|_{\mathrm{2B}}
    =
    -\frac{3}{4}
    \beta_{\mathrm{loss}}
    \frac{n_0}{2\sqrt{2}}E,
\end{equation}
before including the phenomenological correction $h_{\mathrm{2B}}$.

Combining the number and energy equations gives the temperature evolution
\begin{equation}
    \frac{dT}{dt}
    =
    \dot{T}_c
    +
    (h_{\textrm{2B}}+\frac{1}{4})
    \beta_{\mathrm{loss}}
    \frac{n_0}{2\sqrt{2}}T.
    \label{eq:twobody temp}
\end{equation}
Using the independently measured values of $\tau$ and $\dot{T}_c$, we fit
$N(t)$ and $T(t)$ simultaneously to Eq.~(\ref{twobodyloss}). All data sets obtained at different F\"orster defects are fit globally, with a separate value of $\beta_{\textrm{loss}}$ for each defect and a common value of $h_{\textrm{2B}}$. The global fit gives $h_{\textrm{2B}}=0.29(5)$. 

\subsection{Three-body loss}
For one- and three-body loss, the corresponding kinetic equations are
\begin{equation}
\begin{aligned}
    \frac{dN}{dt}
    &=
    -\frac{N}{\tau}
    -L_3\frac{n_0^2}{3\sqrt{3}}N,\\
    \frac{dE}{dt}
    &=
    -\frac{E}{\tau}
    +
    \left(h_{\mathrm{3B}}-\frac{2}{3}\right)L_3\frac{n_0^2}{3\sqrt{3}}E
    +3Nk_B\dot{T}_c .
    \label{threebodyloss}
\end{aligned}
\end{equation}
Here, $L_3$ is the three-body loss rate coefficient. Similar to $h_{\textrm{2B}}$, the dimensionless parameter $h_{\mathrm{3B}}$ accounts phenomenologically for additional possible energy changes associated with momentum correlations and other loss-induced heating mechanisms.

Three-body loss is weighted by $n^3(\boldsymbol r)$, giving a mean potential
energy per lost molecule of
\begin{equation}
    \langle U\rangle_{\mathrm{3B}}
    =
    \frac{
        \int U(\boldsymbol r)n^3(\boldsymbol r)\,d^3\boldsymbol r
    }{
        \int n^3(\boldsymbol r)\,d^3\boldsymbol r
    }
    =
    \frac{1}{2}k_BT.
\end{equation}
Together with the mean kinetic energy $3k_BT/2$, each lost molecule removes
$2k_BT$, equal to $2/3$ of the cloud-averaged energy per particle. The
anti-evaporation contribution to the energy equation is therefore
\begin{equation}
    \left.\frac{dE}{dt}\right|_{\mathrm{3B}}
    =
    -\frac{2}{3}L_3\frac{n_0^2}{3\sqrt{3}}E.
\end{equation}
The corresponding temperature equation is
\begin{equation}
    \frac{dT}{dt}
    =
    \dot{T}_c
    +
    \left(h_{\mathrm{3B}}+\frac{1}{3}\right)L_3\frac{n_0^2}{3\sqrt{3}}T.
    \label{eq:threebody_temp}
\end{equation}
When we estimate the distribution of $h_{\textrm{3B}}$ using a bootstrap method, we cannot detect $h_{\textrm{3B}}$ as the distribution includes $h_{\textrm{3B}}=0$. Therefore, in this data analysis, we fix $h_{\textrm{3B}}=0$ and fit the measured $N(t)$ and $T(t)$ simultaneously to Eq.~(\ref{threebodyloss}) to estimate $L_3$. Here, $\tau$ and $\dot{T}_c$ are determined from independent measurements. 

\subsection{Bootstrap uncertainty analysis}

We estimate the statistical uncertainties of the parameters obtained from the one-, two-, and three-body loss fits using bootstrap resampling. Within each hold-time group, the individual experimental realizations are resampled with replacement to generate $5000$ bootstrap data sets. The complete fitting procedure is repeated for each resampled data set, and the reported uncertainty of each fitted parameter is the standard deviation of the resulting bootstrap distribution. 

\subsection{Detection limits for two- and three-body loss}\label{sec:detection limit}

For the smallest measured loss coefficients, the fitted values become comparable to the sensitivity set by the finite observation time and the experimental uncertainties in the measured molecule number and temperature. We define the two- and three-body detection limits as the one-standard-deviation uncertainties of the corresponding fitted coefficients when the true coefficient is zero,

\begin{align}
\beta_{\mathrm{det}}
&\equiv
\left[
\operatorname{Var}
\left(
\widehat{\beta}_{\mathrm{loss}}
\mid
\beta_{\mathrm{loss}}=0
\right)
\right]^{1/2},
\nonumber \\
L_{\mathrm{det}}
&\equiv
\left[
\operatorname{Var}
\left(
\widehat{L}_3
\mid
L_3=0
\right)
\right]^{1/2}.
\label{eq:detection_limit_definition}
\end{align}

Although both detection limits are reported, they are evaluated using separate sensitivity analyses rather than a joint fit of $\beta_{\mathrm{loss}}$ and $L_3$. The two coefficients are extracted from different sets of loss measurements and in distinct interaction regimes. For the measurements used to determine $\beta_{\mathrm{loss}}$, we use the one- plus two-body kinetic model in Eqs.~(\ref{twobodyloss}) and~(\ref{eq:twobody temp}). For the measurements used to determine $L_3$, the calculated two-body loss coefficient is below the experimental two-body detection limit, and we therefore use the one- plus three-body kinetic model in Eqs.~(\ref{threebodyloss}) and~(\ref{eq:threebody_temp}). A joint Fisher analysis would instead quantify the ability of a single loss trace to distinguish simultaneous two- and three-body contributions, which is not the procedure used to extract the reported coefficients. We therefore determine the conditional sensitivity of each loss model separately.

For a given loss trace, we collect the measured number and temperature values in the vector
\begin{align}
\mathbf{y}
=
\left(
N_1,\ldots,N_M,
T_1,\ldots,T_M
\right)^{\mathrm T},
\end{align}
with covariance matrix
\begin{align}
\mathbf{C}
=
\operatorname{diag}
\left(
\sigma_{N,1}^{2},\ldots,\sigma_{N,M}^{2},
\sigma_{T,1}^{2},\ldots,\sigma_{T,M}^{2}
\right).
\end{align}
The nuisance parameters common to both analyses are
\begin{align}
\boldsymbol{\eta}
=
\left(
N_0,T_0,\Gamma_1,\dot{T}_c
\right),
\end{align}
where $\Gamma_1=1/\tau$ is the one-body loss rate and $\dot{T}_c$ is the independently measured background-heating rate.

\subsubsection{Two-body-loss sensitivity}

For the two-body analysis, the numerical solution of Eqs.~(\ref{twobodyloss}) and~(\ref{eq:twobody temp}) is denoted by
\begin{align}
\boldsymbol{\mu}_{2\mathrm B}
\left(
\beta_{\mathrm{loss}},
\boldsymbol{\eta}
\right).
\end{align}
Assuming Gaussian measurement uncertainties, the chi-squared function is specified by
\begin{align}
\chi_{2\mathrm B}^{2}
=
&\left[
\mathbf{y}
-
\boldsymbol{\mu}_{2\mathrm B}
\left(
\beta_{\mathrm{loss}},
\boldsymbol{\eta}
\right)
\right]^{\mathrm T}
\mathbf{C}^{-1}
\left[
\mathbf{y}
-
\boldsymbol{\mu}_{2\mathrm B}
\left(
\beta_{\mathrm{loss}},
\boldsymbol{\eta}
\right)
\right]
\nonumber\\
&+
\frac{
\left(
\Gamma_1-\Gamma_{1,\mathrm{cal}}
\right)^2
}{
\sigma_{\Gamma_1}^{2}
}
+
\frac{
\left(
\dot{T}_c-\dot{T}_{c,\mathrm{cal}}
\right)^2
}{
\sigma_{\dot{T}_c}^{2}
}.
\label{eq:chi2_two_body_detection}
\end{align}
The Fisher matrix is evaluated about the null trajectory obtained by setting $\beta_{\mathrm{loss}}=0$,
\begin{align}
N(t)&=N_0e^{-\Gamma_1t},
&
T(t)&=T_0+\dot{T}_ct.
\end{align}
Because the parameters span many orders of magnitude, we introduce the dimensionless parameter vector
\begin{align}
\boldsymbol{\theta}_{2\mathrm B}
=
\left(
b,\ln N_0,\ln T_0,u_{\Gamma},u_T
\right),
\qquad
b=\frac{\beta_{\mathrm{loss}}}{\beta_{\mathrm{ref}}},
\end{align}
for numerical conditioning, where
\begin{align}
u_{\Gamma}
&=
\frac{
\Gamma_1-\Gamma_{1,\mathrm{cal}}
}{
\sigma_{\Gamma_1}
},
&
u_T
&=
\frac{
\dot{T}_c-\dot{T}_{c,\mathrm{cal}}
}{
\sigma_{\dot{T}_c}
}.
\end{align}

Here, $\beta_{\mathrm{ref}}=10^{-14} \textrm{cm}^3/\textrm{s}$ is a fixed reference scale introduced only to improve numerical conditioning. The model Jacobian is

\begin{align}
J_{ia}^{(2\mathrm B)}
=
\left.
\frac{
\partial\mu_{2\mathrm B,i}
}{
\partial\theta_{2\mathrm B,a}
}
\right|_{\boldsymbol{\theta}_{2\mathrm B,0}},
\end{align}
where $\boldsymbol{\theta}_{2\mathrm B,0}$ denotes the best-fit null parameters. Including the independent calibration constraints, the Fisher matrix is
\begin{align}
\mathbf{F}_{2\mathrm B}
=
\left(
\mathbf{J}^{(2\mathrm B)}
\right)^{\mathrm T}
\mathbf{C}^{-1}
\mathbf{J}^{(2\mathrm B)}
+
\operatorname{diag}(0,0,0,1,1).
\end{align}
Within the local Gaussian approximation, the two-body detection limit is
\begin{align}
\beta_{\mathrm{det}}
=
\beta_{\mathrm{ref}}
\sqrt{
\left(
\mathbf{F}_{2\mathrm B}^{-1}
\right)_{bb}
}
.
\label{eq:beta_detection_fisher}
\end{align}
Because the complete Fisher matrix is inverted, this sensitivity includes correlations of $\beta_{\mathrm{loss}}$ with $N_0$, $T_0$, $\Gamma_1$, and $\dot{T}_c$. We note here that $h_{2B}$ is omitted from the nuisance vector because it only enters through the product $h_{2B}\beta_{\mathrm{loss}}$. Therefore, it has no first-order effect on the null trajectory ($\beta_\mathrm{loss}=0$) about which the Fisher matrix is evaluated and does not contribute to the detection limit.

\subsubsection{Three-body-loss sensitivity}

\begin{figure*}[t]
    \centering
    \includegraphics[width=17.5cm]{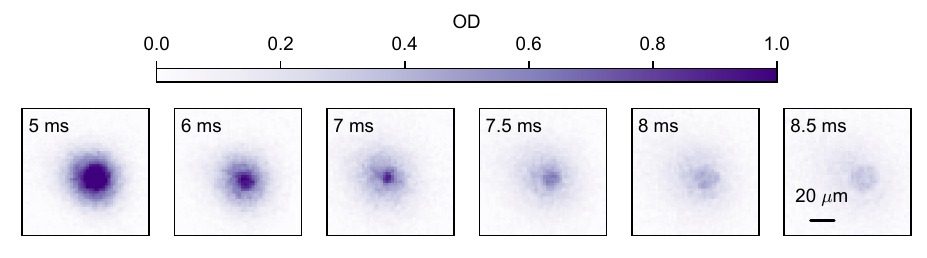}
    \caption{Droplet falling in and out of focus of the imaging system. ToFs are shown in the upper-left hand corner of the image.}
    \label{fig:dropletoutoffocus}
\end{figure*}

The three-body detection limit is determined independently using the numerical solution of Eqs.~(\ref{threebodyloss}) and~(\ref{eq:threebody_temp}), denoted by
\begin{align}
\boldsymbol{\mu}_{3\mathrm B}
\left(
L_3,\boldsymbol{\eta}
\right).
\end{align}
The corresponding objective function is
\begin{align}
\chi_{3\mathrm B}^{2}
=
&\left[
\mathbf{y}
-
\boldsymbol{\mu}_{3\mathrm B}
\left(
L_3,\boldsymbol{\eta}
\right)
\right]^{\mathrm T}
\mathbf{C}^{-1}
\left[
\mathbf{y}
-
\boldsymbol{\mu}_{3\mathrm B}
\left(
L_3,\boldsymbol{\eta}
\right)
\right]
\nonumber\\
&+
\frac{
\left(
\Gamma_1-\Gamma_{1,\mathrm{cal}}
\right)^2
}{
\sigma_{\Gamma_1}^{2}
}
+
\frac{
\left(
\dot{T}_c-\dot{T}_{c,\mathrm{cal}}
\right)^2
}{
\sigma_{\dot{T}_c}^{2}
}.
\label{eq:chi2_three_body_detection}
\end{align}
The Fisher matrix is again evaluated about the null trajectory, now with $L_3=0$. We introduce
\begin{align}
\boldsymbol{\theta}_{3\mathrm B}
=
\left(
\ell,\ln N_0,\ln T_0,u_{\Gamma},u_T
\right),
\qquad
\ell=\frac{L_3}{L_{\mathrm{ref}}},
\end{align}
where $L_{\mathrm{ref}}=10^{-25}\textrm{cm}^6/\textrm{s}$ is a fixed reference scale. The Jacobian and Fisher matrix are
\begin{align}
J_{ia}^{(3\mathrm B)}
&=
\left.
\frac{
\partial\mu_{3\mathrm B,i}
}{
\partial\theta_{3\mathrm B,a}
}
\right|_{\boldsymbol{\theta}_{3\mathrm B,0}},
\nonumber \\
\mathbf{F}_{3\mathrm B}
&=
\left(
\mathbf{J}^{(3\mathrm B)}
\right)^{\mathrm T}
\mathbf{C}^{-1}
\mathbf{J}^{(3\mathrm B)}
+
\operatorname{diag}(0,0,0,1,1).
\end{align}
The three-body detection limit is therefore
\begin{align}
L_{\mathrm{det}}
=
L_{\mathrm{ref}}
\sqrt{
\left(
\mathbf{F}_{3\mathrm B}^{-1}
\right)_{\ell\ell}
}
.
\label{eq:L3_detection_fisher}
\end{align}

This result includes the correlations of $L_3$ with the initial number and temperature, the one-body loss rate, and the background-heating rate. 

\section{Droplet Characterization}

\subsection{ODT pulse for longer time-of-flight}
\label{sec:ODTpulse}
Figure~\ref{fig:dropletoutoffocus} illustrates the sensitivity of the measured droplet width to defocusing. During time of flight, the molecules fall under gravity while the focal plane of the imaging system remains fixed. The object-side depth of field can be estimated as
\begin{equation}
d_{\mathrm{DOF}}=\frac{\lambda \cdot n}{\mathrm{NA}^2}+\frac{n\cdot e}{M \cdot \mathrm{NA}},
\end{equation}
where $\lambda$ is the imaging wavelength, $n$ is the refractive index, $M$ is the magnification, $\mathrm{NA}$ is the numerical aperture, and $e$ is the camera pixel pitch. For our imaging system, $M=30$, $\mathrm{NA}=0.5$, $e=6.5~\mu\mathrm{m}$, and $\lambda=780~\mathrm{nm}$. Taking $n\simeq 1$ gives $d_{\mathrm{DOF}}\simeq 3.5~\mu\mathrm{m}$.

The imaging system is focused on the molecular plane after $6.5~\mathrm{ms}$ of TOF. Increasing the TOF from $6.5$ to $7.5~\mathrm{ms}$ displaces the molecules vertically by
\begin{equation}
\Delta z=\frac{g}{2}\left[(7.5~\mathrm{ms})^2-(6.5~\mathrm{ms})^2\right]\simeq 69~\mu\mathrm{m}.
\end{equation}
This displacement is much larger than the depth of field, so the molecules are substantially out of focus at a ToF of 
7.5~ms and beyond. To study the ToF dependence of the droplet size, we apply a brief ODT pulse of duration 350~$\mu$s that imparts a sudden upward impulse on the cloud, compensating for the gravitational fall. By varying the pulse power, we adjust this impulse to return the droplet to the focal plane at the desired imaging time.

\begin{figure*}[t]
    \centering
    \includegraphics[width=17.4cm]{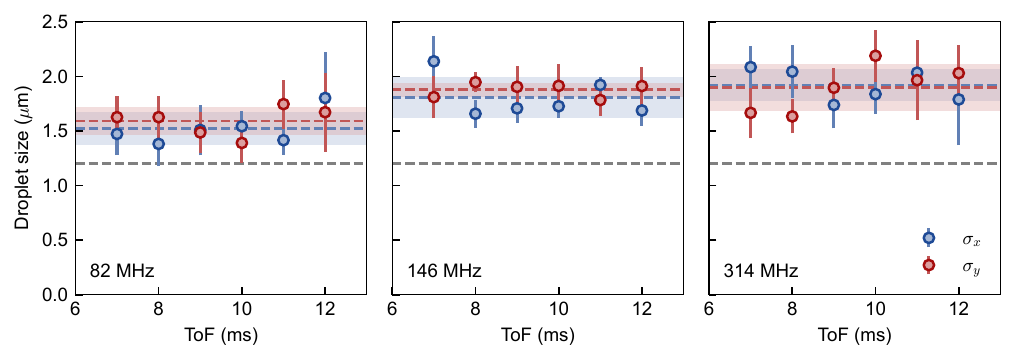 }
    \caption{Droplet sizes vs time of flight for various F\"orster defects.}
    \label{fig:dropletsizes}
\end{figure*}

As discussed in the main text, this method was used to explore the droplet expansion dynamics at various F\"orster defects. The expansion data for $\Delta_{\mathrm{F}}=2\pi \times 17$~MHz are shown in Fig.~\ref{fig:dropletproperties}, while the data for $\Delta_{\mathrm{F}}/2\pi=82$, $146$, and $314$~MHz are shown in Fig.~\ref{fig:dropletsizes}. 

\subsection{Double Gaussian bimodal fit} \label{subsection:doublegaussian}

In this paper, we use double Gaussian functions to fit the density distribution of a droplet with a thermal cloud. As mentioned in the main text, the size of a droplet is limited by imaging resolution, so we model the density distribution of a droplet from each image as a two-dimensional Gaussian function. The bimodal fitting function for the two-dimensional density distribution $n_{\textrm{2D}}(x,y)$ is
\begin{eqnarray}
    \sum_{i\in\{d,t\}}\frac{N_i}{2\pi\sigma_{x,i}\sigma_{y,i}}
    \exp\left[
    -\frac{(x-x_{0,i})^2}{2\sigma_{x,i}^2}
    -\frac{(y-y_{0,i})^2}{2\sigma_{y,i}^2}
    \right].
    \label{eq:droplet_gaussian}
\end{eqnarray}
Here, the subscript $i$ represents a thermal (droplet) component if $i=t$ ($i=d$). $N_i$ is the molecule number in component $i$; $\sigma_{x,i}$ and $\sigma_{y,i}$ are the Gaussian radii of the component $i$ along the $x$ and $y$ axes, respectively. Also, $(x_{0,i}, y_{0,i})$ is the center coordinate of the component $i$. 

As described in the main text, the position of the droplet varies from shot to shot. The pixel shifts are determined by computing the cross-correlation between each image and a reference image. Before averaging, the images are aligned by shifting each image so that the droplet centers are aligned. A double-Gaussian bimodal fit is then applied to both the individual images and the averaged image. The droplet size is extracted from the fit to the averaged image, while the corresponding error bars are estimated from the standard deviation of the droplet sizes obtained from the individual single-shot images.

\subsection{Imaging resolution}
\label{sec:imagingresolution}
We infer the density distribution of NaRb molecules by dissociating them through bound-to-free STIRAP and performing absorption imaging on the resulting Rb atoms. For the droplet-size measurement, a \(4~\mu\mathrm{s}\) repump pulse is followed by a \(30~\mu\mathrm{s}\) imaging pulse. The measured width is broadened by atomic motion during the detection sequence and by the finite point-spread function (PSF) of the imaging system. We consider three contributions: the relative recoil imparted during bound-to-free STIRAP, photon recoil during imaging, and the optical PSF. All widths quoted below are one-dimensional rms widths. For the motion-induced contributions, we use the rms displacement at the end of the corresponding detection interval, which provides a conservative estimate of the image broadening.

Bound-to-free STIRAP couples a bound molecular state to the continuum of two free atoms. When the STIRAP frequency \(f\) exceeds the dissociation threshold \(f_0\), the excess energy is converted into relative kinetic energy of the Na and Rb atoms. Conservation of energy in the center-of-mass frame gives
\begin{equation}
h(f-f_0)
=
\frac{p_{\mathrm{rel}}^2}{2m_{\mathrm{Na}}}
+
\frac{p_{\mathrm{rel}}^2}{2m_{\mathrm{Rb}}},
\end{equation}
and therefore
\begin{equation}
v_{\mathrm{Rb}}
=
\frac{p_{\mathrm{rel}}}{m_{\mathrm{Rb}}}
=
\sqrt{
\frac{2h m_{\mathrm{Na}}}
{m_{\mathrm{Rb}}
\left(m_{\mathrm{Na}}+m_{\mathrm{Rb}}\right)}
(f-f_0)
}.
\label{eq:kick_velocity}
\end{equation}

To determine \(v_{\mathrm{Rb}}\), we measure the two-dimensional Rb density distribution after variable TOF following bound-to-free STIRAP. Before dissociation, the molecular center-of-mass distribution is modeled as a Gaussian. Each molecule dissociates with the same relative momentum magnitude but a random direction. After a TOF \(t\), the Rb atom acquires an additional displacement \(v_{\mathrm{Rb}}t\,\hat{\boldsymbol{u}}\), where \(\hat{\boldsymbol{u}}\) is isotropically distributed. Averaging over \(\hat{\boldsymbol{u}}\) gives the two-dimensional density profile used to fit each absorption image.
\begin{widetext}
\begin{equation}
n_{\mathrm{2D}}(x,y;t)
=
\frac{N}{2\pi\sigma_x(t)\sigma_y(t)}
\int\frac{\mathrm{d}\Omega_{\boldsymbol{u}}}{4\pi}
\exp\Bigg[
-\frac{\left(x-x_0-R(t)u_x\right)^2}
{2\sigma_x^2(t)}
-\frac{\left(y-y_0-R(t)u_y\right)^2}
{2\sigma_y^2(t)}
\Bigg],
\label{eq:ToFSTIRAP}
\end{equation}
\end{widetext}
where
\begin{equation}
R(t)=v_{\mathrm{Rb}}t
\end{equation}
is the recoil radius. The Gaussian widths describe the center-of-mass distribution after TOF. For ballistic expansion of a thermal cloud,
\begin{equation}
\sigma_w^2(t)
=
\sigma_{w,0}^2
+
\frac{k_BT}{m_{\mathrm{NaRb}}}t^2,
\qquad
w\in\{x,y\}.
\label{eq:thermal_expansion}
\end{equation}

\begin{figure}[t]
\centering
\includegraphics[width=8.6cm]{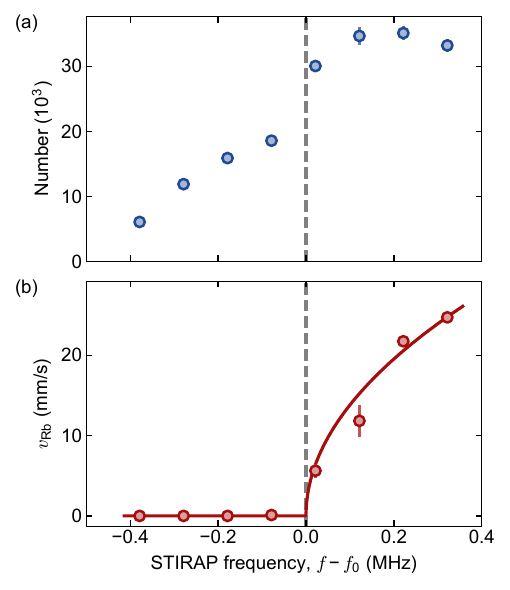}
\caption{Spectroscopy of the molecule number and the Rb recoil velocity \(v_{\mathrm{Rb}}\) as a function of the STIRAP frequency. The dashed line indicates the bound-to-free threshold \(f=f_0\). The threshold frequency \(f_0\) is obtained by fitting the measured recoil velocity to Eq.~\eqref{eq:kick_velocity}. The solid red line shows the fit.}
\label{fig:BtF}
\end{figure}

Equation~\eqref{eq:ToFSTIRAP} is fitted directly to each two-dimensional absorption image, with \(R(t)\), \(N\), \((x_0,y_0)\), and the Gaussian widths as fit parameters. We then determine \(v_{\mathrm{Rb}}\) from the linear dependence of the fitted recoil radius \(R(t)\) on \(t\). Figure~\ref{fig:BtF}(b) shows the resulting \(v_{\mathrm{Rb}}\) as a function of the STIRAP frequency. Above threshold, the measured velocities agree well with Eq.~\eqref{eq:kick_velocity}.

The relation between the fitted recoil radius and the one-dimensional rms broadening follows directly from isotropy:
\begin{equation}
\int\frac{\mathrm{d}\Omega_{\boldsymbol{u}}}{4\pi}
u_\alpha u_\beta
=
\frac{\delta_{\alpha\beta}}{3}.
\label{eq:isotropic_second_moment}
\end{equation}
Consequently, the second moments of Eq.~\eqref{eq:ToFSTIRAP} are
\begin{align}
\left\langle(x-x_0)^2\right\rangle
&=
\sigma_x^2(t)+\frac{R^2(t)}{3},
\nonumber\\
\left\langle(y-y_0)^2\right\rangle
&=
\sigma_y^2(t)+\frac{R^2(t)}{3}.
\end{align}
Thus, although the full two-dimensional distribution is used to determine \(R(t)\), its recoil contribution to the rms width along either transverse direction is \(R(t)/\sqrt{3}\).

For the droplet-size measurements, we choose
\[
f-f_0=0.32~\mathrm{MHz}.
\]
Using the displacement accumulated by the end of the \(34~\mu\mathrm{s}\) detection interval gives
\begin{equation}
\sigma_{\mathrm{STIRAP}}
=
\frac{v_{\mathrm{Rb}}\times34~\mu\mathrm{s}}{\sqrt{3}}
=
0.5~\mu\mathrm{m}.
\label{eq:STIRAP_broadening}
\end{equation}
A second contribution arises from photon recoil during the
\(30~\mu\mathrm{s}\) imaging pulse. Absorption transfers momentum along the imaging axis and therefore does not broaden the transverse
distribution, whereas spontaneous emission produces random transverse
recoil. For one transverse direction \(x\), the displacement at the end
of an imaging pulse of duration \(t\) is
\begin{equation}
\Delta x(t)
=
\frac{1}{m_{\mathrm{Rb}}}
\sum_{t_i<t}
p_{i,x}(t-t_i),
\end{equation}
where \(t_i\) and \(p_{i,x}\) are the time and transverse recoil momentum
of the \(i\)th emission event. Assuming independent, isotropically
distributed recoils,
\begin{equation}
\left\langle p_{i,x}p_{j,x}\right\rangle_{\Omega}
=
\delta_{ij}\frac{\hbar^2k_{780}^2}{3}.
\end{equation}
Averaging also over Poisson-distributed emission times with constant
scattering rate \(R_{\mathrm{sc}}\) gives
\begin{align}
\left\langle\Delta x^2(t)\right\rangle
&=
\frac{\hbar^2k_{780}^2}{3m_{\mathrm{Rb}}^2}
R_{\mathrm{sc}}
\int_0^t(t-t')^2\,\mathrm{d}t'
\nonumber\\
&=
\frac{R_{\mathrm{sc}}\hbar^2k_{780}^2t^3}
{9m_{\mathrm{Rb}}^2}.
\end{align}
Here, the average is taken over both the recoil directions and the
emission times. The resulting one-dimensional rms displacement is
\begin{equation}
\sigma_{\mathrm{photon}}
=
\frac{\hbar k_{780}}{3m_{\mathrm{Rb}}}
R_{\mathrm{sc}}^{1/2}t^{3/2}
=
0.98~\mu\mathrm{m}.
\end{equation}

Finally, the diffraction-limited point spread function of our imaging system at $780$~nm contributes an additional broadening of
\begin{align}
\sigma_{\textrm{PSF}}=
0.50~\mu\textrm{m},
\end{align}
as characterized independently in Ref.~\cite{Rosenberg2022HBT}.

Combining these three independent contributions in quadrature yields an overall imaging resolution of
\begin{align}
\sigma_{\textrm{resolution}}
=\sqrt{
\sigma_{\textrm{STIRAP}}^2
+
\sigma_{\textrm{photon}}^2
+
\sigma_{\textrm{PSF}}^2
}
=1.2~\mu\textrm{m}.
\end{align}

\subsection{Droplet lifetime}

\begin{figure*}[t]
    \centering
    \includegraphics[width=17.4cm]{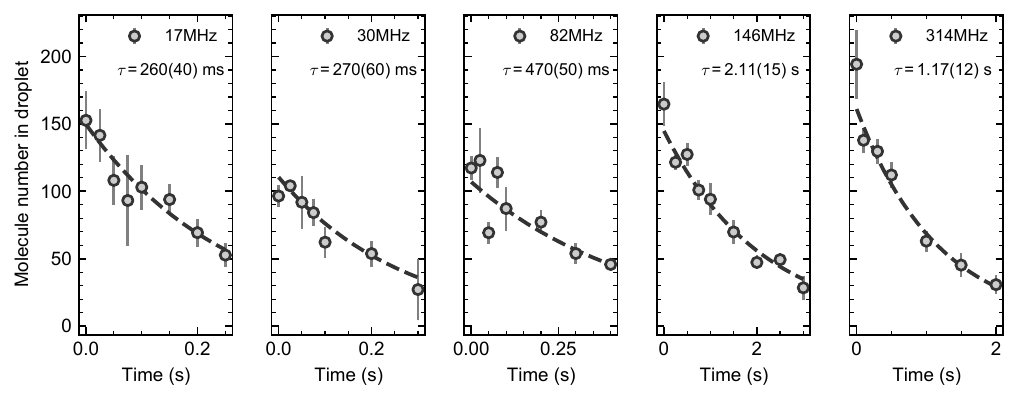}
    \caption{Droplet lifetime for various F\"orster defects $\Delta_{\mathrm{F}}$. The droplet number is extracted from a double Gaussian fit. The dashed black line is an exponential decay fit, from which we extract a droplet lifetime denoted for each F\"orster defect in the upper-right hand corner of the corresponding panel.}
    \label{fig:dropletlifetime}
\end{figure*}

The droplet lifetime is measured as a function of the hold time in the optical dipole trap (ODT), as shown in Fig.~\ref{fig:dropletlifetime}. At each hold time, the number of molecules in the droplet component is extracted using the double-Gaussian bimodal fit described in Section~\ref{subsection:doublegaussian}. There is large variation in the measured lifetimes as we scan F\"orster defects $\Delta_{\mathrm{F}}$. These values are given in Fig.~\ref{fig:dropletlifetime}. The uncertainties in the extracted molecule numbers are estimated by bootstrap resampling the absorption images over 5000 iterations. 

\subsection{Droplet density}
\label{sec:dropletdensity}
\begin{figure}[t]
    \centering
    \includegraphics[width=8.5cm]{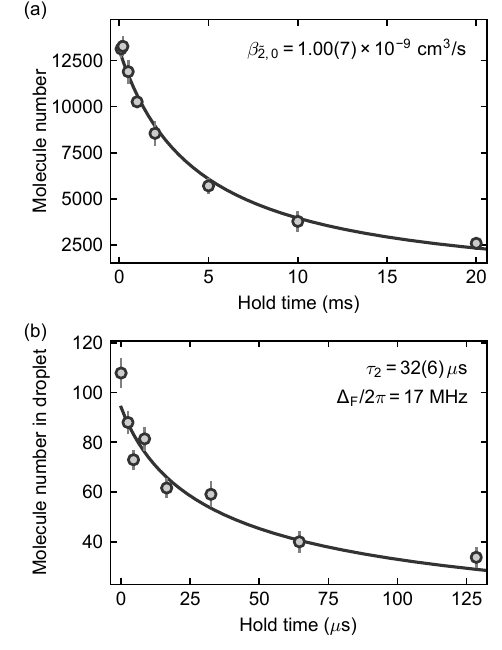}
    \caption{
    Density calibration from unshielded two-body loss at a F\"orster defect
    $\Delta_{\mathrm{F}}/2\pi=17$~MHz.
    (a) Two-body loss of a thermal molecular gas following transfer to
    $\vert\tilde{2},0\rangle$. The solid black line is a fit to the
    two-body loss model in Eq.~\eqref{unshieldkinetic}, from which we extract
    $\beta_{\tilde{2},0}$.
    (b) Decay of the droplet population following transfer to
    $\vert\tilde{2},0\rangle$. The droplet number is extracted from a
    double-Gaussian fit to the density distribution. The solid black line is
    a two-body decay fit from Eq.~\eqref{droplet_kinetic}, yielding the characteristic decay time $\tau_2$
    indicated in the upper-right corner.
    }
    \label{fig:dropletlifetime_unshielded}
\end{figure}

Direct determination of the droplet density from its spatial extent is limited by our imaging resolution. We therefore infer the density from the droplet decay. A potential concern with using the droplet lifetimes obtained in the previous section is the exchange of particles between the droplet and the surrounding thermal gas during the extended hold times. We therefore use a different approach in which the molecules are transferred to an unshielded rotational state, where the decay occurs on a much faster timescale~\cite{zhang2026droplet}. We transfer the molecules from $\vert\tilde{1},0\rangle$ to $\vert\tilde{2},0\rangle$ using a $1.5~\mu\mathrm{s}$ microwave $\pi$ pulse at $7.841$~GHz. Electric-field noise contributes a transition linewidth bounded by
\begin{equation}
\frac{
\left|
d_{\tilde{1}\tilde{1}}
-
d_{\tilde{2}\tilde{2}}
\right|
\sigma_{\mathcal{E}}
}{\hbar}
\leq
2\pi\times4.9(6)~\mathrm{kHz},
\end{equation}
which is small compared with the measured Rabi frequency
$\Omega=2\pi\times324(4)$~kHz.

We first calibrate the two-body loss coefficient
$\beta_{\tilde{2},0}$ using a thermal molecular gas. Molecules are prepared
in an optical dipole trap (ODT) with trapping frequencies
\begin{equation}
(\omega_x,\omega_y,\omega_z)
=
2\pi\times
(89.2(1.6),60.8(1.4),66.8(1.3))~\mathrm{Hz}
\end{equation}
at an initial temperature of $310(20)$~nK. We apply a $\pi$ pulse to
transfer the molecules to $\vert\tilde{2},0\rangle$, hold them for a
variable time, and subsequently transfer them back to
$\vert\tilde{1},0\rangle$ for detection. For this calibration measurement,
we use the conventional two-body loss model for a thermal gas and take the
temperature to remain constant during the loss measurement. For a thermal
gas, 
\begin{align}
\frac{dN(t)}{dt} = -\beta_{\tilde{2},0}\int n^2 (\boldsymbol{r},t) d^3 \boldsymbol{r} \nonumber \\
= -\beta_{\tilde{2},0}\omega_x\omega_y\omega_z
\left(
\frac{m_{\textrm{NaRb}}}{4\pi k_B T}
\right)^{3/2}N^2(t),
\label{unshieldkinetic}
\end{align}
so that the measured number evolution directly determines
$\beta_{\tilde{2},0}$.

We then perform the loss measurement on the droplet. The molecular gas is
evaporatively cooled at $\Delta_{\mathrm{F}}/2\pi=17$~MHz to
$T=63(3)$~nK, where the droplet forms (small differences in the trap alignment and evaporation curve lead to a slightly higher droplet formation temperature than in the main text). To minimize additional molecule loss
and reliably extract the droplet population in the presence of the
surrounding thermal background, the microwave transfer back to
$\vert\tilde{1},0\rangle$ is applied immediately before the bound-to-free
reverse STIRAP and imaging sequence, thereby preserving the bimodal density
distribution used in the fit. Specifically, we transfer the molecules to
$\vert\tilde{2},0\rangle$, hold them for a variable time, and then transfer
the remaining molecules back to $\vert\tilde{1},0\rangle$ immediately
before detection.

In contrast to the thermal-gas calibration, the loss of the droplet occurs
on a timescale short compared with density redistribution and
rethermalization. The inhomogeneous droplet therefore does not continuously
re-equilibrate during the decay, and its number evolution is not described
by the same global two-body loss curve used for the thermal gas. Instead, we
describe the droplet dynamics within a local-density approximation,
\begin{align}
\frac{\partial n_{\textrm{droplet}}(\boldsymbol{r},t)}{\partial t}
&=
-g\beta_{\tilde{2},0}n_{\textrm{droplet}}^2(\boldsymbol{r},t).
\end{align}
Here we take $g=1$ for the thermal gas and $g=1/2$ for the degenerate droplet, accounting for the reduction of local two-particle correlations relative to a thermal Bose gas.
Therefore, the number of molecules in a droplet can be expressed as
\begin{equation}
N_{\mathrm{droplet}}(t)
=
\int
\frac{
n_{\mathrm{droplet}}(\boldsymbol{r},0)
}{
1+
g\beta_{\tilde{2},0}
n_{\mathrm{droplet}}(\boldsymbol{r},0)t
}
d^3\boldsymbol{r}.
\end{equation}
Assuming that the initial droplet density follows a Gaussian distribution,
this expression becomes
\begin{align}
N_{\mathrm{droplet}}(t)
&=
\frac{2N_0}{\sqrt{\pi}}
\int_0^{\infty}
\frac{
\sqrt{u}e^{-u}
}{
1+2\sqrt{2}e^{-u}t/\tau_2
}du,
\nonumber\\
\frac{1}{\tau_2}
&=
g\beta_{\tilde{2},0}
\langle n_{\mathrm{droplet}}(t=0)\rangle.
\label{droplet_kinetic}
\end{align}
where
\begin{equation}
\langle n_{\textrm{droplet}}(t)\rangle
\equiv
\frac{1}{N_{\textrm{droplet}}(t)}
\int n_{\textrm{droplet}}^2(\boldsymbol{r},t)d^3\boldsymbol{r}
\end{equation}
is the mean density of a droplet at time $t$.

The measured loss dynamics are shown in
Fig.~\ref{fig:dropletlifetime_unshielded}. From the thermal-gas measurement
in Fig.~\ref{fig:dropletlifetime_unshielded}(a), using the conventional
two-body loss model at fixed temperature, we obtain
\begin{equation}
\beta_{\tilde{2},0}=1.00(7)\times10^{-9}~\mathrm{cm^3/s}.
\end{equation}
The fit in Fig.~\ref{fig:dropletlifetime_unshielded}(b) gives
$\tau_2=32(6)~\mu\mathrm{s}$, corresponding to
\begin{equation}
\langle n_{\mathrm{droplet}}(0)\rangle=6.3(1.3)\times10^{13}~\mathrm{cm^{-3}},
\end{equation}
for $g=1/2$. The resulting parameter satisfies
$\langle n_{\mathrm{droplet}}(0)\rangle a_{dd}^3 \ll 1$, indicating that
the characteristic kinetic-energy scale exceeds the dipolar interaction
energy scale.

Combining the measured density with the temperature, we estimate the
phase-space density of the droplet to be
\begin{align}
\mathrm{PSD}_{\mathrm{droplet}}
&=
\langle n_{\mathrm{droplet}}(0)\rangle
\lambda_{\mathrm{dB}}^3=18(4),
\nonumber \\
\lambda_{\mathrm{dB}}
&=
\frac{h}{\sqrt{2\pi m_{\textrm{NaRb}} k_B T}},
\end{align}
where $\lambda_{\mathrm{dB}}$ is the thermal de Broglie wavelength.
The resulting $\mathrm{PSD}_{\mathrm{droplet}}\gg1$ indicates that the droplet is deeply quantum degenerate and is consistent with the use of the bosonic bunching factor $g=1/2$ in extracting its density from the two-body decay.\\

\section{Numerical Simulations}
\label{sec:simulations}

For our theoretical calculations, we use Path Integral Monte Carlo (PIMC) \cite{ceperley1995review}, an \textit{ab initio} method which samples the equilibrium partition function of the system and thereby yields statistically exact results for observables in thermodynamic equilibrium at finite temperatures.
In PIMC simulations, each quantum particle is mapped to an imaginary-time trajectory at a discrete number of time slices, which are sampled with Monte Carlo techniques, whereby bosonic permutations can be sampled efficiently, e.g. with the worm algorithm \cite{boninsegni2006wormalgorithm}.
Continuous-space PIMC has been applied to dipolar particles in different contexts, and has recently emerged as a powerful instrument to study ultracold molecules numerically \cite{Langen2025DropletPhase, Ciardi2025crystal, ciardi2026metastable, Zhang2025DropletPhase, ArnoneCardinale2026, schindewolf2025few}.

We have performed simulations of $^{23}\mathrm{Na}^{87}\mathrm{Rb}$ molecules, in an anisotropic harmonic trap with frequencies $2\pi \times (60, 70, 30)$ Hz, close to the final trap depth at the end of the cooling process. 
For simplicity, we did not use different trapping frequencies in the simulations at the different temperatures, as implemented in the experimental protocol. The good agreement between the theoretical and experimental values of the critical temperature indicates that the droplet formation process is rather robust with respect to variations of the trap frequencies. 

We use the interaction potential given by equation \eqref{Eq:AppendixEffectivePotential}, with the known dipole moment of $d=3.3$ D and for varying values of  $\Delta_{\rm F}$ between $2\pi \times 17$ MHz and $2\pi \times 314$ MHz. Depending on the interaction strength, we have simulated temperatures down to 25 nK and up to 600 nK, choosing the number of time slices as large as 2048 at low temperatures, and as small as 16 at high temperatures. We do not attempt to simulate the large total particle numbers of the experiment, since equilibration between the small high-density droplet states and the dilute surrounding gas molecules may not be expected on the timescales of the experiment. Hence, we instead choose the particle number ($N=200$ and $N=400$) to match the typical numbers of molecules in the experimentally observed droplets.

We apply a simple and reliable procedure to detect the formation of bound clusters. Throughout the simulation, we mark molecules as neighbors based on their distance, setting a threshold on the order of the separation corresponding to the potential minimum. Then, we keep track of groups of neighboring particles, i.e. clusters. This allows us to determine, at a given $T$ and $\Delta_{\rm F}$, whether the molecules stabilize into a self-bound state or disperse into a gas, and thus estimate the critical temperature.


\begin{thebibliography}{94}%
\makeatletter
\providecommand \@ifxundefined [1]{%
 \@ifx{#1\undefined}
}%
\providecommand \@ifnum [1]{%
 \ifnum #1\expandafter \@firstoftwo
 \else \expandafter \@secondoftwo
 \fi
}%
\providecommand \@ifx [1]{%
 \ifx #1\expandafter \@firstoftwo
 \else \expandafter \@secondoftwo
 \fi
}%
\providecommand \natexlab [1]{#1}%
\providecommand \enquote  [1]{``#1''}%
\providecommand \bibnamefont  [1]{#1}%
\providecommand \bibfnamefont [1]{#1}%
\providecommand \citenamefont [1]{#1}%
\providecommand \href@noop [0]{\@secondoftwo}%
\providecommand \href [0]{\begingroup \@sanitize@url \@href}%
\providecommand \@href[1]{\@@startlink{#1}\@@href}%
\providecommand \@@href[1]{\endgroup#1\@@endlink}%
\providecommand \@sanitize@url [0]{\catcode `\\12\catcode `\$12\catcode `\&12\catcode `\#12\catcode `\^12\catcode `\_12\catcode `\%12\relax}%
\providecommand \@@startlink[1]{}%
\providecommand \@@endlink[0]{}%
\providecommand \url  [0]{\begingroup\@sanitize@url \@url }%
\providecommand \@url [1]{\endgroup\@href {#1}{\urlprefix }}%
\providecommand \urlprefix  [0]{URL }%
\providecommand \Eprint [0]{\href }%
\providecommand \doibase [0]{https://doi.org/}%
\providecommand \selectlanguage [0]{\@gobble}%
\providecommand \bibinfo  [0]{\@secondoftwo}%
\providecommand \bibfield  [0]{\@secondoftwo}%
\providecommand \translation [1]{[#1]}%
\providecommand \BibitemOpen [0]{}%
\providecommand \bibitemStop [0]{}%
\providecommand \bibitemNoStop [0]{.\EOS\space}%
\providecommand \EOS [0]{\spacefactor3000\relax}%
\providecommand \BibitemShut  [1]{\csname bibitem#1\endcsname}%
\let\auto@bib@innerbib\@empty
\bibitem [{\citenamefont {Micheli}\ \emph {et~al.}(2006)\citenamefont {Micheli}, \citenamefont {Brennen},\ and\ \citenamefont {Zoller}}]{micheli2006toolbox}%
  \BibitemOpen
  \bibfield  {author} {\bibinfo {author} {\bibfnamefont {A.}~\bibnamefont {Micheli}}, \bibinfo {author} {\bibfnamefont {G.~K.}\ \bibnamefont {Brennen}},\ and\ \bibinfo {author} {\bibfnamefont {P.}~\bibnamefont {Zoller}},\ }\bibfield  {title} {\bibinfo {title} {A toolbox for lattice-spin models with polar molecules},\ }\href {https://doi.org/10.1038/nphys287} {\bibfield  {journal} {\bibinfo  {journal} {Nat. Phys.}\ }\textbf {\bibinfo {volume} {2}},\ \bibinfo {pages} {341} (\bibinfo {year} {2006})}\BibitemShut {NoStop}%
\bibitem [{\citenamefont {Barnett}\ \emph {et~al.}(2006)\citenamefont {Barnett}, \citenamefont {Petrov}, \citenamefont {Lukin},\ and\ \citenamefont {Demler}}]{Barnett2006Magnetism}%
  \BibitemOpen
  \bibfield  {author} {\bibinfo {author} {\bibfnamefont {R.}~\bibnamefont {Barnett}}, \bibinfo {author} {\bibfnamefont {D.}~\bibnamefont {Petrov}}, \bibinfo {author} {\bibfnamefont {M.}~\bibnamefont {Lukin}},\ and\ \bibinfo {author} {\bibfnamefont {E.}~\bibnamefont {Demler}},\ }\bibfield  {title} {\bibinfo {title} {Quantum magnetism with multicomponent dipolar molecules in an optical lattice},\ }\href {https://doi.org/10.1103/PhysRevLett.96.190401} {\bibfield  {journal} {\bibinfo  {journal} {Phys. Rev. Lett.}\ }\textbf {\bibinfo {volume} {96}},\ \bibinfo {pages} {190401} (\bibinfo {year} {2006})}\BibitemShut {NoStop}%
\bibitem [{\citenamefont {Carr}\ \emph {et~al.}(2009)\citenamefont {Carr}, \citenamefont {DeMille}, \citenamefont {Krems},\ and\ \citenamefont {Ye}}]{carr2009cold}%
  \BibitemOpen
  \bibfield  {author} {\bibinfo {author} {\bibfnamefont {L.~D.}\ \bibnamefont {Carr}}, \bibinfo {author} {\bibfnamefont {D.}~\bibnamefont {DeMille}}, \bibinfo {author} {\bibfnamefont {R.~V.}\ \bibnamefont {Krems}},\ and\ \bibinfo {author} {\bibfnamefont {J.}~\bibnamefont {Ye}},\ }\bibfield  {title} {\bibinfo {title} {Cold and ultracold molecules: science, technology and applications},\ }\href {https://doi.org/10.1088/1367-2630/11/5/055049} {\bibfield  {journal} {\bibinfo  {journal} {New J. Phys.}\ }\textbf {\bibinfo {volume} {11}},\ \bibinfo {pages} {055049} (\bibinfo {year} {2009})}\BibitemShut {NoStop}%
\bibitem [{\citenamefont {Bohn}\ \emph {et~al.}(2017)\citenamefont {Bohn}, \citenamefont {Rey},\ and\ \citenamefont {Ye}}]{bohn2017cold}%
  \BibitemOpen
  \bibfield  {author} {\bibinfo {author} {\bibfnamefont {J.~L.}\ \bibnamefont {Bohn}}, \bibinfo {author} {\bibfnamefont {A.~M.}\ \bibnamefont {Rey}},\ and\ \bibinfo {author} {\bibfnamefont {J.}~\bibnamefont {Ye}},\ }\bibfield  {title} {\bibinfo {title} {Cold molecules: Progress in quantum engineering of chemistry and quantum matter},\ }\href {https://www.science.org/doi/10.1126/science.aam6299} {\bibfield  {journal} {\bibinfo  {journal} {Science}\ }\textbf {\bibinfo {volume} {357}},\ \bibinfo {pages} {1002} (\bibinfo {year} {2017})}\BibitemShut {NoStop}%
\bibitem [{\citenamefont {Yan}\ \emph {et~al.}(2013)\citenamefont {Yan}, \citenamefont {Moses}, \citenamefont {Gadway}, \citenamefont {Covey}, \citenamefont {Hazzard}, \citenamefont {Rey}, \citenamefont {Jin},\ and\ \citenamefont {Ye}}]{yan2013observation}%
  \BibitemOpen
  \bibfield  {author} {\bibinfo {author} {\bibfnamefont {B.}~\bibnamefont {Yan}}, \bibinfo {author} {\bibfnamefont {S.~A.}\ \bibnamefont {Moses}}, \bibinfo {author} {\bibfnamefont {B.}~\bibnamefont {Gadway}}, \bibinfo {author} {\bibfnamefont {J.~P.}\ \bibnamefont {Covey}}, \bibinfo {author} {\bibfnamefont {K.~R.}\ \bibnamefont {Hazzard}}, \bibinfo {author} {\bibfnamefont {A.~M.}\ \bibnamefont {Rey}}, \bibinfo {author} {\bibfnamefont {D.~S.}\ \bibnamefont {Jin}},\ and\ \bibinfo {author} {\bibfnamefont {J.}~\bibnamefont {Ye}},\ }\bibfield  {title} {\bibinfo {title} {Observation of dipolar spin-exchange interactions with lattice-confined polar molecules},\ }\href {https://doi.org/10.1038/nature12483} {\bibfield  {journal} {\bibinfo  {journal} {Nature}\ }\textbf {\bibinfo {volume} {501}},\ \bibinfo {pages} {521} (\bibinfo {year} {2013})}\BibitemShut {NoStop}%
\bibitem [{\citenamefont {Burchesky}\ \emph {et~al.}(2021)\citenamefont {Burchesky}, \citenamefont {Anderegg}, \citenamefont {Bao}, \citenamefont {Yu}, \citenamefont {Chae}, \citenamefont {Ketterle}, \citenamefont {Ni},\ and\ \citenamefont {Doyle}}]{Burchesky2021CaFCoherence}%
  \BibitemOpen
  \bibfield  {author} {\bibinfo {author} {\bibfnamefont {S.}~\bibnamefont {Burchesky}}, \bibinfo {author} {\bibfnamefont {L.}~\bibnamefont {Anderegg}}, \bibinfo {author} {\bibfnamefont {Y.}~\bibnamefont {Bao}}, \bibinfo {author} {\bibfnamefont {S.~S.}\ \bibnamefont {Yu}}, \bibinfo {author} {\bibfnamefont {E.}~\bibnamefont {Chae}}, \bibinfo {author} {\bibfnamefont {W.}~\bibnamefont {Ketterle}}, \bibinfo {author} {\bibfnamefont {K.-K.}\ \bibnamefont {Ni}},\ and\ \bibinfo {author} {\bibfnamefont {J.~M.}\ \bibnamefont {Doyle}},\ }\bibfield  {title} {\bibinfo {title} {Rotational coherence times of polar molecules in optical tweezers},\ }\href {https://doi.org/10.1103/PhysRevLett.127.123202} {\bibfield  {journal} {\bibinfo  {journal} {Phys. Rev. Lett.}\ }\textbf {\bibinfo {volume} {127}},\ \bibinfo {pages} {123202} (\bibinfo {year} {2021})}\BibitemShut {NoStop}%
\bibitem [{\citenamefont {Gregory}\ \emph {et~al.}(2024)\citenamefont {Gregory}, \citenamefont {Fernley}, \citenamefont {Tao}, \citenamefont {Bromley}, \citenamefont {Stepp}, \citenamefont {Zhang}, \citenamefont {Kotochigova}, \citenamefont {Hazzard},\ and\ \citenamefont {Cornish}}]{gregory2024second}%
  \BibitemOpen
  \bibfield  {author} {\bibinfo {author} {\bibfnamefont {P.~D.}\ \bibnamefont {Gregory}}, \bibinfo {author} {\bibfnamefont {L.~M.}\ \bibnamefont {Fernley}}, \bibinfo {author} {\bibfnamefont {A.~L.}\ \bibnamefont {Tao}}, \bibinfo {author} {\bibfnamefont {S.~L.}\ \bibnamefont {Bromley}}, \bibinfo {author} {\bibfnamefont {J.}~\bibnamefont {Stepp}}, \bibinfo {author} {\bibfnamefont {Z.}~\bibnamefont {Zhang}}, \bibinfo {author} {\bibfnamefont {S.}~\bibnamefont {Kotochigova}}, \bibinfo {author} {\bibfnamefont {K.~R.}\ \bibnamefont {Hazzard}},\ and\ \bibinfo {author} {\bibfnamefont {S.~L.}\ \bibnamefont {Cornish}},\ }\bibfield  {title} {\bibinfo {title} {Second-scale rotational coherence and dipolar interactions in a gas of ultracold polar molecules},\ }\href {https://doi.org/10.1038/s41567-023-02328-5} {\bibfield  {journal} {\bibinfo  {journal} {Nat. Phys.}\ }\textbf {\bibinfo {volume} {20}},\ \bibinfo {pages} {415} (\bibinfo {year} {2024})}\BibitemShut {NoStop}%
\bibitem [{\citenamefont {Holland}\ \emph {et~al.}(2023)\citenamefont {Holland}, \citenamefont {Lu},\ and\ \citenamefont {Cheuk}}]{Holland2023CaFEntanglement}%
  \BibitemOpen
  \bibfield  {author} {\bibinfo {author} {\bibfnamefont {C.~M.}\ \bibnamefont {Holland}}, \bibinfo {author} {\bibfnamefont {Y.}~\bibnamefont {Lu}},\ and\ \bibinfo {author} {\bibfnamefont {L.~W.}\ \bibnamefont {Cheuk}},\ }\bibfield  {title} {\bibinfo {title} {On-demand entanglement of molecules in a reconfigurable optical tweezer array},\ }\href {https://doi.org/10.1126/science.adf4272} {\bibfield  {journal} {\bibinfo  {journal} {Science}\ }\textbf {\bibinfo {volume} {382}},\ \bibinfo {pages} {1143} (\bibinfo {year} {2023})}\BibitemShut {NoStop}%
\bibitem [{\citenamefont {Bao}\ \emph {et~al.}(2023)\citenamefont {Bao}, \citenamefont {Yu}, \citenamefont {Anderegg}, \citenamefont {Chae}, \citenamefont {Ketterle}, \citenamefont {Ni},\ and\ \citenamefont {Doyle}}]{Bao2023SE}%
  \BibitemOpen
  \bibfield  {author} {\bibinfo {author} {\bibfnamefont {Y.}~\bibnamefont {Bao}}, \bibinfo {author} {\bibfnamefont {S.~S.}\ \bibnamefont {Yu}}, \bibinfo {author} {\bibfnamefont {L.}~\bibnamefont {Anderegg}}, \bibinfo {author} {\bibfnamefont {E.}~\bibnamefont {Chae}}, \bibinfo {author} {\bibfnamefont {W.}~\bibnamefont {Ketterle}}, \bibinfo {author} {\bibfnamefont {K.-K.}\ \bibnamefont {Ni}},\ and\ \bibinfo {author} {\bibfnamefont {J.~M.}\ \bibnamefont {Doyle}},\ }\bibfield  {title} {\bibinfo {title} {Dipolar spin-exchange and entanglement between molecules in an optical tweezer array},\ }\href {https://doi.org/10.1126/science.adf8999} {\bibfield  {journal} {\bibinfo  {journal} {Science}\ }\textbf {\bibinfo {volume} {382}},\ \bibinfo {pages} {1138} (\bibinfo {year} {2023})}\BibitemShut {NoStop}%
\bibitem [{\citenamefont {Picard}\ \emph {et~al.}(2025)\citenamefont {Picard}, \citenamefont {Park}, \citenamefont {Patenotte}, \citenamefont {Gebretsadkan}, \citenamefont {Wellnitz}, \citenamefont {Rey},\ and\ \citenamefont {Ni}}]{Picard2025SE}%
  \BibitemOpen
  \bibfield  {author} {\bibinfo {author} {\bibfnamefont {L.~R.~B.}\ \bibnamefont {Picard}}, \bibinfo {author} {\bibfnamefont {A.~J.}\ \bibnamefont {Park}}, \bibinfo {author} {\bibfnamefont {G.~E.}\ \bibnamefont {Patenotte}}, \bibinfo {author} {\bibfnamefont {S.}~\bibnamefont {Gebretsadkan}}, \bibinfo {author} {\bibfnamefont {D.}~\bibnamefont {Wellnitz}}, \bibinfo {author} {\bibfnamefont {A.~M.}\ \bibnamefont {Rey}},\ and\ \bibinfo {author} {\bibfnamefont {K.-K.}\ \bibnamefont {Ni}},\ }\bibfield  {title} {\bibinfo {title} {Entanglement and i{SWAP} gate between molecular qubits},\ }\href {https://doi.org/10.1038/s41586-024-08177-3} {\bibfield  {journal} {\bibinfo  {journal} {Nature}\ }\textbf {\bibinfo {volume} {637}},\ \bibinfo {pages} {821} (\bibinfo {year} {2025})}\BibitemShut {NoStop}%
\bibitem [{\citenamefont {Ruttley}\ \emph {et~al.}(2025)\citenamefont {Ruttley}, \citenamefont {Hepworth}, \citenamefont {Guttridge},\ and\ \citenamefont {Cornish}}]{Ruttley2025SE}%
  \BibitemOpen
  \bibfield  {author} {\bibinfo {author} {\bibfnamefont {D.~K.}\ \bibnamefont {Ruttley}}, \bibinfo {author} {\bibfnamefont {T.~R.}\ \bibnamefont {Hepworth}}, \bibinfo {author} {\bibfnamefont {A.}~\bibnamefont {Guttridge}},\ and\ \bibinfo {author} {\bibfnamefont {S.~L.}\ \bibnamefont {Cornish}},\ }\bibfield  {title} {\bibinfo {title} {Long-lived entanglement of molecules in magic-wavelength optical tweezers},\ }\href {https://doi.org/10.1038/s41586-024-08365-1} {\bibfield  {journal} {\bibinfo  {journal} {Nature}\ }\textbf {\bibinfo {volume} {637}},\ \bibinfo {pages} {827} (\bibinfo {year} {2025})}\BibitemShut {NoStop}%
\bibitem [{\citenamefont {Christakis}\ \emph {et~al.}(2023)\citenamefont {Christakis}, \citenamefont {Rosenberg}, \citenamefont {Raj}, \citenamefont {Chi}, \citenamefont {Morningstar}, \citenamefont {Huse}, \citenamefont {Yan},\ and\ \citenamefont {Bakr}}]{christakis2023probing}%
  \BibitemOpen
  \bibfield  {author} {\bibinfo {author} {\bibfnamefont {L.}~\bibnamefont {Christakis}}, \bibinfo {author} {\bibfnamefont {J.~S.}\ \bibnamefont {Rosenberg}}, \bibinfo {author} {\bibfnamefont {R.}~\bibnamefont {Raj}}, \bibinfo {author} {\bibfnamefont {S.}~\bibnamefont {Chi}}, \bibinfo {author} {\bibfnamefont {A.}~\bibnamefont {Morningstar}}, \bibinfo {author} {\bibfnamefont {D.~A.}\ \bibnamefont {Huse}}, \bibinfo {author} {\bibfnamefont {Z.~Z.}\ \bibnamefont {Yan}},\ and\ \bibinfo {author} {\bibfnamefont {W.~S.}\ \bibnamefont {Bakr}},\ }\bibfield  {title} {\bibinfo {title} {Probing site-resolved correlations in a spin system of ultracold molecules},\ }\href {https://doi.org/10.1038/s41586-022-05558-4} {\bibfield  {journal} {\bibinfo  {journal} {Nature}\ }\textbf {\bibinfo {volume} {614}},\ \bibinfo {pages} {64} (\bibinfo {year} {2023})}\BibitemShut {NoStop}%
\bibitem [{\citenamefont {Miller}\ \emph {et~al.}(2024)\citenamefont {Miller}, \citenamefont {Carroll}, \citenamefont {Lin}, \citenamefont {Hirzler}, \citenamefont {Gao}, \citenamefont {Zhou}, \citenamefont {Lukin},\ and\ \citenamefont {Ye}}]{Miller2024XYZ}%
  \BibitemOpen
  \bibfield  {author} {\bibinfo {author} {\bibfnamefont {C.}~\bibnamefont {Miller}}, \bibinfo {author} {\bibfnamefont {A.~N.}\ \bibnamefont {Carroll}}, \bibinfo {author} {\bibfnamefont {J.}~\bibnamefont {Lin}}, \bibinfo {author} {\bibfnamefont {H.}~\bibnamefont {Hirzler}}, \bibinfo {author} {\bibfnamefont {H.}~\bibnamefont {Gao}}, \bibinfo {author} {\bibfnamefont {H.}~\bibnamefont {Zhou}}, \bibinfo {author} {\bibfnamefont {M.~D.}\ \bibnamefont {Lukin}},\ and\ \bibinfo {author} {\bibfnamefont {J.}~\bibnamefont {Ye}},\ }\bibfield  {title} {\bibinfo {title} {Two-axis twisting using {Floquet}-engineered {XYZ} spin models with polar molecules},\ }\href {https://doi.org/10.1038/s41586-024-07883-2} {\bibfield  {journal} {\bibinfo  {journal} {Nature}\ }\textbf {\bibinfo {volume} {633}},\ \bibinfo {pages} {332} (\bibinfo {year} {2024})}\BibitemShut {NoStop}%
\bibitem [{\citenamefont {Lu}\ \emph {et~al.}(2026)\citenamefont {Lu}, \citenamefont {Holland}, \citenamefont {Welsh}, \citenamefont {Chen},\ and\ \citenamefont {Cheuk}}]{lu2026probing}%
  \BibitemOpen
  \bibfield  {author} {\bibinfo {author} {\bibfnamefont {Y.}~\bibnamefont {Lu}}, \bibinfo {author} {\bibfnamefont {C.~M.}\ \bibnamefont {Holland}}, \bibinfo {author} {\bibfnamefont {C.~L.}\ \bibnamefont {Welsh}}, \bibinfo {author} {\bibfnamefont {X.-Y.}\ \bibnamefont {Chen}},\ and\ \bibinfo {author} {\bibfnamefont {L.~W.}\ \bibnamefont {Cheuk}},\ }\bibfield  {title} {\bibinfo {title} {Probing coherent many-body spin dynamics in a molecular tweezer array quantum simulator},\ }\href {https://arxiv.org/abs/2603.19090} {\bibfield  {journal} {\bibinfo  {journal} {arXiv:2603.19090}\ } (\bibinfo {year} {2026})}\BibitemShut {NoStop}%
\bibitem [{\citenamefont {Li}\ \emph {et~al.}(2023)\citenamefont {Li}, \citenamefont {Matsuda}, \citenamefont {Miller}, \citenamefont {Carroll}, \citenamefont {Tobias}, \citenamefont {Higgins},\ and\ \citenamefont {Ye}}]{li2023tunable}%
  \BibitemOpen
  \bibfield  {author} {\bibinfo {author} {\bibfnamefont {J.-R.}\ \bibnamefont {Li}}, \bibinfo {author} {\bibfnamefont {K.}~\bibnamefont {Matsuda}}, \bibinfo {author} {\bibfnamefont {C.}~\bibnamefont {Miller}}, \bibinfo {author} {\bibfnamefont {A.~N.}\ \bibnamefont {Carroll}}, \bibinfo {author} {\bibfnamefont {W.~G.}\ \bibnamefont {Tobias}}, \bibinfo {author} {\bibfnamefont {J.~S.}\ \bibnamefont {Higgins}},\ and\ \bibinfo {author} {\bibfnamefont {J.}~\bibnamefont {Ye}},\ }\bibfield  {title} {\bibinfo {title} {Tunable itinerant spin dynamics with polar molecules},\ }\href {https://doi.org/10.1038/s41586-022-05479-2} {\bibfield  {journal} {\bibinfo  {journal} {Nature}\ }\textbf {\bibinfo {volume} {614}},\ \bibinfo {pages} {70} (\bibinfo {year} {2023})}\BibitemShut {NoStop}%
\bibitem [{\citenamefont {Carroll}\ \emph {et~al.}(2025)\citenamefont {Carroll}, \citenamefont {Hirzler}, \citenamefont {Miller}, \citenamefont {Wellnitz}, \citenamefont {Muleady}, \citenamefont {Lin}, \citenamefont {Zamarski}, \citenamefont {Wang}, \citenamefont {Bohn}, \citenamefont {Rey},\ and\ \citenamefont {Ye}}]{Carroll2025tJ}%
  \BibitemOpen
  \bibfield  {author} {\bibinfo {author} {\bibfnamefont {A.~N.}\ \bibnamefont {Carroll}}, \bibinfo {author} {\bibfnamefont {H.}~\bibnamefont {Hirzler}}, \bibinfo {author} {\bibfnamefont {C.}~\bibnamefont {Miller}}, \bibinfo {author} {\bibfnamefont {D.}~\bibnamefont {Wellnitz}}, \bibinfo {author} {\bibfnamefont {S.~R.}\ \bibnamefont {Muleady}}, \bibinfo {author} {\bibfnamefont {J.}~\bibnamefont {Lin}}, \bibinfo {author} {\bibfnamefont {K.~P.}\ \bibnamefont {Zamarski}}, \bibinfo {author} {\bibfnamefont {R.~R.~W.}\ \bibnamefont {Wang}}, \bibinfo {author} {\bibfnamefont {J.~L.}\ \bibnamefont {Bohn}}, \bibinfo {author} {\bibfnamefont {A.~M.}\ \bibnamefont {Rey}},\ and\ \bibinfo {author} {\bibfnamefont {J.}~\bibnamefont {Ye}},\ }\bibfield  {title} {\bibinfo {title} {Observation of generalized t-{J} spin dynamics with tunable dipolar interactions},\ }\href {https://doi.org/10.1126/science.adq0911} {\bibfield  {journal} {\bibinfo  {journal} {Science}\ }\textbf {\bibinfo {volume} {388}},\ \bibinfo {pages} {381}
  (\bibinfo {year} {2025})}\BibitemShut {NoStop}%
\bibitem [{\citenamefont {Langen}\ \emph {et~al.}(2024)\citenamefont {Langen}, \citenamefont {Valtolina}, \citenamefont {Wang},\ and\ \citenamefont {Ye}}]{langen2024quantum}%
  \BibitemOpen
  \bibfield  {author} {\bibinfo {author} {\bibfnamefont {T.}~\bibnamefont {Langen}}, \bibinfo {author} {\bibfnamefont {G.}~\bibnamefont {Valtolina}}, \bibinfo {author} {\bibfnamefont {D.}~\bibnamefont {Wang}},\ and\ \bibinfo {author} {\bibfnamefont {J.}~\bibnamefont {Ye}},\ }\bibfield  {title} {\bibinfo {title} {Quantum state manipulation and cooling of ultracold molecules},\ }\href {https://doi.org/10.1038/s41567-024-02423-1} {\bibfield  {journal} {\bibinfo  {journal} {Nat. Phys.}\ }\textbf {\bibinfo {volume} {20}},\ \bibinfo {pages} {702} (\bibinfo {year} {2024})}\BibitemShut {NoStop}%
\bibitem [{\citenamefont {Ni}\ \emph {et~al.}(2008)\citenamefont {Ni}, \citenamefont {Ospelkaus}, \citenamefont {De~Miranda}, \citenamefont {Pe'er}, \citenamefont {Neyenhuis}, \citenamefont {Zirbel}, \citenamefont {Kotochigova}, \citenamefont {Julienne}, \citenamefont {Jin},\ and\ \citenamefont {Ye}}]{ni2008high}%
  \BibitemOpen
  \bibfield  {author} {\bibinfo {author} {\bibfnamefont {K.-K.}\ \bibnamefont {Ni}}, \bibinfo {author} {\bibfnamefont {S.}~\bibnamefont {Ospelkaus}}, \bibinfo {author} {\bibfnamefont {M.}~\bibnamefont {De~Miranda}}, \bibinfo {author} {\bibfnamefont {A.}~\bibnamefont {Pe'er}}, \bibinfo {author} {\bibfnamefont {B.}~\bibnamefont {Neyenhuis}}, \bibinfo {author} {\bibfnamefont {J.}~\bibnamefont {Zirbel}}, \bibinfo {author} {\bibfnamefont {S.}~\bibnamefont {Kotochigova}}, \bibinfo {author} {\bibfnamefont {P.}~\bibnamefont {Julienne}}, \bibinfo {author} {\bibfnamefont {D.}~\bibnamefont {Jin}},\ and\ \bibinfo {author} {\bibfnamefont {J.}~\bibnamefont {Ye}},\ }\bibfield  {title} {\bibinfo {title} {A high phase-space-density gas of polar molecules},\ }\href {https://www.science.org/doi/10.1126/science.1163861} {\bibfield  {journal} {\bibinfo  {journal} {Science}\ }\textbf {\bibinfo {volume} {322}},\ \bibinfo {pages} {231} (\bibinfo {year} {2008})}\BibitemShut {NoStop}%
\bibitem [{\citenamefont {Shuman}\ \emph {et~al.}(2010)\citenamefont {Shuman}, \citenamefont {Barry},\ and\ \citenamefont {DeMille}}]{shuman2010laser}%
  \BibitemOpen
  \bibfield  {author} {\bibinfo {author} {\bibfnamefont {E.~S.}\ \bibnamefont {Shuman}}, \bibinfo {author} {\bibfnamefont {J.~F.}\ \bibnamefont {Barry}},\ and\ \bibinfo {author} {\bibfnamefont {D.}~\bibnamefont {DeMille}},\ }\bibfield  {title} {\bibinfo {title} {Laser cooling of a diatomic molecule},\ }\href {https://doi.org/10.1038/nature09443} {\bibfield  {journal} {\bibinfo  {journal} {Nature}\ }\textbf {\bibinfo {volume} {467}},\ \bibinfo {pages} {820} (\bibinfo {year} {2010})}\BibitemShut {NoStop}%
\bibitem [{\citenamefont {Anderegg}\ \emph {et~al.}(2019)\citenamefont {Anderegg}, \citenamefont {Cheuk}, \citenamefont {Bao}, \citenamefont {Burchesky}, \citenamefont {Ketterle}, \citenamefont {Ni},\ and\ \citenamefont {Doyle}}]{anderegg2019optical}%
  \BibitemOpen
  \bibfield  {author} {\bibinfo {author} {\bibfnamefont {L.}~\bibnamefont {Anderegg}}, \bibinfo {author} {\bibfnamefont {L.~W.}\ \bibnamefont {Cheuk}}, \bibinfo {author} {\bibfnamefont {Y.}~\bibnamefont {Bao}}, \bibinfo {author} {\bibfnamefont {S.}~\bibnamefont {Burchesky}}, \bibinfo {author} {\bibfnamefont {W.}~\bibnamefont {Ketterle}}, \bibinfo {author} {\bibfnamefont {K.-K.}\ \bibnamefont {Ni}},\ and\ \bibinfo {author} {\bibfnamefont {J.~M.}\ \bibnamefont {Doyle}},\ }\bibfield  {title} {\bibinfo {title} {An optical tweezer array of ultracold molecules},\ }\href {https://www.science.org/doi/10.1126/science.aax1265} {\bibfield  {journal} {\bibinfo  {journal} {Science}\ }\textbf {\bibinfo {volume} {365}},\ \bibinfo {pages} {1156} (\bibinfo {year} {2019})}\BibitemShut {NoStop}%
\bibitem [{\citenamefont {{De Marco}}\ \emph {et~al.}(2019)\citenamefont {{De Marco}}, \citenamefont {Valtolina}, \citenamefont {Matsuda}, \citenamefont {Tobias}, \citenamefont {Covey},\ and\ \citenamefont {Ye}}]{DeMarco2019KRbDegenerate}%
  \BibitemOpen
  \bibfield  {author} {\bibinfo {author} {\bibfnamefont {L.}~\bibnamefont {{De Marco}}}, \bibinfo {author} {\bibfnamefont {G.}~\bibnamefont {Valtolina}}, \bibinfo {author} {\bibfnamefont {K.}~\bibnamefont {Matsuda}}, \bibinfo {author} {\bibfnamefont {W.~G.}\ \bibnamefont {Tobias}}, \bibinfo {author} {\bibfnamefont {J.~P.}\ \bibnamefont {Covey}},\ and\ \bibinfo {author} {\bibfnamefont {J.}~\bibnamefont {Ye}},\ }\bibfield  {title} {\bibinfo {title} {A degenerate {Fermi} gas of polar molecules},\ }\href {https://doi.org/10.1126/science.aau7230} {\bibfield  {journal} {\bibinfo  {journal} {Science}\ }\textbf {\bibinfo {volume} {363}},\ \bibinfo {pages} {853} (\bibinfo {year} {2019})}\BibitemShut {NoStop}%
\bibitem [{\citenamefont {Schindewolf}\ \emph {et~al.}(2022)\citenamefont {Schindewolf}, \citenamefont {Bause}, \citenamefont {Chen}, \citenamefont {Duda}, \citenamefont {Karman}, \citenamefont {Bloch},\ and\ \citenamefont {Luo}}]{schindewolf2022evaporation}%
  \BibitemOpen
  \bibfield  {author} {\bibinfo {author} {\bibfnamefont {A.}~\bibnamefont {Schindewolf}}, \bibinfo {author} {\bibfnamefont {R.}~\bibnamefont {Bause}}, \bibinfo {author} {\bibfnamefont {X.-Y.}\ \bibnamefont {Chen}}, \bibinfo {author} {\bibfnamefont {M.}~\bibnamefont {Duda}}, \bibinfo {author} {\bibfnamefont {T.}~\bibnamefont {Karman}}, \bibinfo {author} {\bibfnamefont {I.}~\bibnamefont {Bloch}},\ and\ \bibinfo {author} {\bibfnamefont {X.-Y.}\ \bibnamefont {Luo}},\ }\bibfield  {title} {\bibinfo {title} {Evaporation of microwave-shielded polar molecules to quantum degeneracy},\ }\href {https://doi.org/10.1038/s41586-022-04900-0} {\bibfield  {journal} {\bibinfo  {journal} {Nature}\ }\textbf {\bibinfo {volume} {607}},\ \bibinfo {pages} {677} (\bibinfo {year} {2022})}\BibitemShut {NoStop}%
\bibitem [{\citenamefont {Bigagli}\ \emph {et~al.}(2024)\citenamefont {Bigagli}, \citenamefont {Yuan}, \citenamefont {Zhang}, \citenamefont {Bulatovic}, \citenamefont {Karman}, \citenamefont {Stevenson},\ and\ \citenamefont {Will}}]{bigagli2024bec}%
  \BibitemOpen
  \bibfield  {author} {\bibinfo {author} {\bibfnamefont {N.}~\bibnamefont {Bigagli}}, \bibinfo {author} {\bibfnamefont {W.}~\bibnamefont {Yuan}}, \bibinfo {author} {\bibfnamefont {S.}~\bibnamefont {Zhang}}, \bibinfo {author} {\bibfnamefont {B.}~\bibnamefont {Bulatovic}}, \bibinfo {author} {\bibfnamefont {T.}~\bibnamefont {Karman}}, \bibinfo {author} {\bibfnamefont {I.}~\bibnamefont {Stevenson}},\ and\ \bibinfo {author} {\bibfnamefont {S.}~\bibnamefont {Will}},\ }\bibfield  {title} {\bibinfo {title} {Observation of {Bose}--{Einstein} condensation of dipolar molecules},\ }\href {https://doi.org/10.1038/s41586-024-07492-z} {\bibfield  {journal} {\bibinfo  {journal} {Nature}\ }\textbf {\bibinfo {volume} {631}},\ \bibinfo {pages} {289} (\bibinfo {year} {2024})}\BibitemShut {NoStop}%
\bibitem [{\citenamefont {Shi}\ \emph {et~al.}(2026)\citenamefont {Shi}, \citenamefont {Huang}, \citenamefont {Deng}, \citenamefont {Jin}, \citenamefont {Yi}, \citenamefont {Shi},\ and\ \citenamefont {Wang}}]{shi2026bec}%
  \BibitemOpen
  \bibfield  {author} {\bibinfo {author} {\bibfnamefont {Z.}~\bibnamefont {Shi}}, \bibinfo {author} {\bibfnamefont {Z.}~\bibnamefont {Huang}}, \bibinfo {author} {\bibfnamefont {F.}~\bibnamefont {Deng}}, \bibinfo {author} {\bibfnamefont {W.-J.}\ \bibnamefont {Jin}}, \bibinfo {author} {\bibfnamefont {S.}~\bibnamefont {Yi}}, \bibinfo {author} {\bibfnamefont {T.}~\bibnamefont {Shi}},\ and\ \bibinfo {author} {\bibfnamefont {D.}~\bibnamefont {Wang}},\ }\bibfield  {title} {\bibinfo {title} {{Bose}-{Einstein} condensate of ultracold sodium-rubidium molecules with tunable dipolar interactions},\ }\href {https://doi.org/10.1038/s41567-026-03362-9} {\bibfield  {journal} {\bibinfo  {journal} {Nat. Phys.}\ } (\bibinfo {year} {2026})}\BibitemShut {NoStop}%
\bibitem [{\citenamefont {Zhang}\ \emph {et~al.}(2026)\citenamefont {Zhang}, \citenamefont {Yuan}, \citenamefont {Bigagli}, \citenamefont {Kwak}, \citenamefont {Karman}, \citenamefont {Stevenson},\ and\ \citenamefont {Will}}]{zhang2026droplet}%
  \BibitemOpen
  \bibfield  {author} {\bibinfo {author} {\bibfnamefont {S.}~\bibnamefont {Zhang}}, \bibinfo {author} {\bibfnamefont {W.}~\bibnamefont {Yuan}}, \bibinfo {author} {\bibfnamefont {N.}~\bibnamefont {Bigagli}}, \bibinfo {author} {\bibfnamefont {H.}~\bibnamefont {Kwak}}, \bibinfo {author} {\bibfnamefont {T.}~\bibnamefont {Karman}}, \bibinfo {author} {\bibfnamefont {I.}~\bibnamefont {Stevenson}},\ and\ \bibinfo {author} {\bibfnamefont {S.}~\bibnamefont {Will}},\ }\bibfield  {title} {\bibinfo {title} {Observation of self-bound droplets of ultracold dipolar molecules},\ }\href {https://doi.org/10.1038/s41586-026-10245-9} {\bibfield  {journal} {\bibinfo  {journal} {Nature}\ }\textbf {\bibinfo {volume} {651}},\ \bibinfo {pages} {601} (\bibinfo {year} {2026})}\BibitemShut {NoStop}%
\bibitem [{\citenamefont {Biswas}\ \emph {et~al.}(2026)\citenamefont {Biswas}, \citenamefont {Eppelt}, \citenamefont {Tian}, \citenamefont {Zhang}, \citenamefont {Deng}, \citenamefont {Frank}, \citenamefont {Shi}, \citenamefont {Bloch},\ and\ \citenamefont {Luo}}]{biswas2026controlled}%
  \BibitemOpen
  \bibfield  {author} {\bibinfo {author} {\bibfnamefont {S.}~\bibnamefont {Biswas}}, \bibinfo {author} {\bibfnamefont {S.}~\bibnamefont {Eppelt}}, \bibinfo {author} {\bibfnamefont {W.}~\bibnamefont {Tian}}, \bibinfo {author} {\bibfnamefont {W.}~\bibnamefont {Zhang}}, \bibinfo {author} {\bibfnamefont {F.}~\bibnamefont {Deng}}, \bibinfo {author} {\bibfnamefont {C.}~\bibnamefont {Frank}}, \bibinfo {author} {\bibfnamefont {T.}~\bibnamefont {Shi}}, \bibinfo {author} {\bibfnamefont {I.}~\bibnamefont {Bloch}},\ and\ \bibinfo {author} {\bibfnamefont {X.-Y.}\ \bibnamefont {Luo}},\ }\bibfield  {title} {\bibinfo {title} {Controlled symmetry breaking of the {Fermi} surface in ultracold polar molecules},\ }\href {https://arxiv.org/abs/2602.22447} {\bibfield  {journal} {\bibinfo  {journal} {arXiv:2602.22447}\ } (\bibinfo {year} {2026})}\BibitemShut {NoStop}%
\bibitem [{\citenamefont {Tang}\ \emph {et~al.}(2018)\citenamefont {Tang}, \citenamefont {Kao}, \citenamefont {Li}, \citenamefont {Seo}, \citenamefont {Mallayya}, \citenamefont {Rigol}, \citenamefont {Gopalakrishnan},\ and\ \citenamefont {Lev}}]{tang2018thermalization}%
  \BibitemOpen
  \bibfield  {author} {\bibinfo {author} {\bibfnamefont {Y.}~\bibnamefont {Tang}}, \bibinfo {author} {\bibfnamefont {W.}~\bibnamefont {Kao}}, \bibinfo {author} {\bibfnamefont {K.-Y.}\ \bibnamefont {Li}}, \bibinfo {author} {\bibfnamefont {S.}~\bibnamefont {Seo}}, \bibinfo {author} {\bibfnamefont {K.}~\bibnamefont {Mallayya}}, \bibinfo {author} {\bibfnamefont {M.}~\bibnamefont {Rigol}}, \bibinfo {author} {\bibfnamefont {S.}~\bibnamefont {Gopalakrishnan}},\ and\ \bibinfo {author} {\bibfnamefont {B.~L.}\ \bibnamefont {Lev}},\ }\bibfield  {title} {\bibinfo {title} {Thermalization near integrability in a dipolar quantum {N}ewton’s cradle},\ }\href {https://journals.aps.org/prx/abstract/10.1103/PhysRevX.8.021030} {\bibfield  {journal} {\bibinfo  {journal} {Phys. Rev. X}\ }\textbf {\bibinfo {volume} {8}},\ \bibinfo {pages} {021030} (\bibinfo {year} {2018})}\BibitemShut {NoStop}%
\bibitem [{\citenamefont {Chomaz}\ \emph {et~al.}(2019)\citenamefont {Chomaz}, \citenamefont {Petter}, \citenamefont {Ilzh{\"o}fer}, \citenamefont {Natale}, \citenamefont {Trautmann}, \citenamefont {Politi}, \citenamefont {Durastante}, \citenamefont {Van~Bijnen}, \citenamefont {Patscheider}, \citenamefont {Sohmen} \emph {et~al.}}]{chomaz2019long}%
  \BibitemOpen
  \bibfield  {author} {\bibinfo {author} {\bibfnamefont {L.}~\bibnamefont {Chomaz}}, \bibinfo {author} {\bibfnamefont {D.}~\bibnamefont {Petter}}, \bibinfo {author} {\bibfnamefont {P.}~\bibnamefont {Ilzh{\"o}fer}}, \bibinfo {author} {\bibfnamefont {G.}~\bibnamefont {Natale}}, \bibinfo {author} {\bibfnamefont {A.}~\bibnamefont {Trautmann}}, \bibinfo {author} {\bibfnamefont {C.}~\bibnamefont {Politi}}, \bibinfo {author} {\bibfnamefont {G.}~\bibnamefont {Durastante}}, \bibinfo {author} {\bibfnamefont {R.}~\bibnamefont {Van~Bijnen}}, \bibinfo {author} {\bibfnamefont {A.}~\bibnamefont {Patscheider}}, \bibinfo {author} {\bibfnamefont {M.}~\bibnamefont {Sohmen}}, \emph {et~al.},\ }\bibfield  {title} {\bibinfo {title} {Long-lived and transient supersolid behaviors in dipolar quantum gases},\ }\href {https://journals.aps.org/prx/abstract/10.1103/PhysRevX.9.021012} {\bibfield  {journal} {\bibinfo  {journal} {Phys. Rev. X}\ }\textbf {\bibinfo {volume} {9}},\ \bibinfo {pages} {021012} (\bibinfo {year}
  {2019})}\BibitemShut {NoStop}%
\bibitem [{\citenamefont {Su}\ \emph {et~al.}(2023)\citenamefont {Su}, \citenamefont {Douglas}, \citenamefont {Szurek}, \citenamefont {Groth}, \citenamefont {Ozturk}, \citenamefont {Krahn}, \citenamefont {H{\'e}bert}, \citenamefont {Phelps}, \citenamefont {Ebadi}, \citenamefont {Dickerson} \emph {et~al.}}]{su2023dipolar}%
  \BibitemOpen
  \bibfield  {author} {\bibinfo {author} {\bibfnamefont {L.}~\bibnamefont {Su}}, \bibinfo {author} {\bibfnamefont {A.}~\bibnamefont {Douglas}}, \bibinfo {author} {\bibfnamefont {M.}~\bibnamefont {Szurek}}, \bibinfo {author} {\bibfnamefont {R.}~\bibnamefont {Groth}}, \bibinfo {author} {\bibfnamefont {S.~F.}\ \bibnamefont {Ozturk}}, \bibinfo {author} {\bibfnamefont {A.}~\bibnamefont {Krahn}}, \bibinfo {author} {\bibfnamefont {A.~H.}\ \bibnamefont {H{\'e}bert}}, \bibinfo {author} {\bibfnamefont {G.~A.}\ \bibnamefont {Phelps}}, \bibinfo {author} {\bibfnamefont {S.}~\bibnamefont {Ebadi}}, \bibinfo {author} {\bibfnamefont {S.}~\bibnamefont {Dickerson}}, \emph {et~al.},\ }\bibfield  {title} {\bibinfo {title} {Dipolar quantum solids emerging in a {H}ubbard quantum simulator},\ }\href {https://www.nature.com/articles/s41586-023-06614-3} {\bibfield  {journal} {\bibinfo  {journal} {Nature}\ }\textbf {\bibinfo {volume} {622}},\ \bibinfo {pages} {724} (\bibinfo {year} {2023})}\BibitemShut {NoStop}%
\bibitem [{\citenamefont {He}\ \emph {et~al.}(2025)\citenamefont {He}, \citenamefont {Chen}, \citenamefont {Zhen}, \citenamefont {Huang}, \citenamefont {Parit},\ and\ \citenamefont {Jo}}]{he2025exploring}%
  \BibitemOpen
  \bibfield  {author} {\bibinfo {author} {\bibfnamefont {Y.}~\bibnamefont {He}}, \bibinfo {author} {\bibfnamefont {Z.}~\bibnamefont {Chen}}, \bibinfo {author} {\bibfnamefont {H.}~\bibnamefont {Zhen}}, \bibinfo {author} {\bibfnamefont {M.}~\bibnamefont {Huang}}, \bibinfo {author} {\bibfnamefont {M.~K.}\ \bibnamefont {Parit}},\ and\ \bibinfo {author} {\bibfnamefont {G.-B.}\ \bibnamefont {Jo}},\ }\bibfield  {title} {\bibinfo {title} {Exploring the {B}erezinskii-{K}osterlitz-{T}houless transition in a two-dimensional dipolar {B}ose gas},\ }\href {https://www.science.org/doi/10.1126/sciadv.adr2715} {\bibfield  {journal} {\bibinfo  {journal} {Sci. Adv.}\ }\textbf {\bibinfo {volume} {11}},\ \bibinfo {pages} {eadr2715} (\bibinfo {year} {2025})}\BibitemShut {NoStop}%
\bibitem [{\citenamefont {Chandrashekara}\ \emph {et~al.}(2026)\citenamefont {Chandrashekara}, \citenamefont {G{\"o}lzh{\"a}user}, \citenamefont {Platt}, \citenamefont {Gao}, \citenamefont {Kusch}, \citenamefont {Hoenen}, \citenamefont {Ballu}, \citenamefont {Kirkby},\ and\ \citenamefont {Chomaz}}]{chandrashekara2026competing}%
  \BibitemOpen
  \bibfield  {author} {\bibinfo {author} {\bibfnamefont {K.}~\bibnamefont {Chandrashekara}}, \bibinfo {author} {\bibfnamefont {C.}~\bibnamefont {G{\"o}lzh{\"a}user}}, \bibinfo {author} {\bibfnamefont {L.}~\bibnamefont {Platt}}, \bibinfo {author} {\bibfnamefont {J.}~\bibnamefont {Gao}}, \bibinfo {author} {\bibfnamefont {J.}~\bibnamefont {Kusch}}, \bibinfo {author} {\bibfnamefont {L.}~\bibnamefont {Hoenen}}, \bibinfo {author} {\bibfnamefont {M.}~\bibnamefont {Ballu}}, \bibinfo {author} {\bibfnamefont {W.}~\bibnamefont {Kirkby}},\ and\ \bibinfo {author} {\bibfnamefont {L.}~\bibnamefont {Chomaz}},\ }\bibfield  {title} {\bibinfo {title} {Competing triangular and stripe supersolid orders in a dipolar quantum gas},\ }\href {https://arxiv.org/abs/2608.20327} {\bibfield  {journal} {\bibinfo  {journal} {arXiv:2608.20327}\ } (\bibinfo {year} {2026})}\BibitemShut {NoStop}%
\bibitem [{\citenamefont {Kadau}\ \emph {et~al.}(2016)\citenamefont {Kadau}, \citenamefont {Schmitt}, \citenamefont {Wenzel}, \citenamefont {Wink}, \citenamefont {Maier}, \citenamefont {Ferrier-Barbut},\ and\ \citenamefont {Pfau}}]{kadau2016observing}%
  \BibitemOpen
  \bibfield  {author} {\bibinfo {author} {\bibfnamefont {H.}~\bibnamefont {Kadau}}, \bibinfo {author} {\bibfnamefont {M.}~\bibnamefont {Schmitt}}, \bibinfo {author} {\bibfnamefont {M.}~\bibnamefont {Wenzel}}, \bibinfo {author} {\bibfnamefont {C.}~\bibnamefont {Wink}}, \bibinfo {author} {\bibfnamefont {T.}~\bibnamefont {Maier}}, \bibinfo {author} {\bibfnamefont {I.}~\bibnamefont {Ferrier-Barbut}},\ and\ \bibinfo {author} {\bibfnamefont {T.}~\bibnamefont {Pfau}},\ }\bibfield  {title} {\bibinfo {title} {Observing the {R}osensweig instability of a quantum ferrofluid},\ }\href {https://www.nature.com/articles/nature16485} {\bibfield  {journal} {\bibinfo  {journal} {Nature}\ }\textbf {\bibinfo {volume} {530}},\ \bibinfo {pages} {194} (\bibinfo {year} {2016})}\BibitemShut {NoStop}%
\bibitem [{\citenamefont {Ferrier-Barbut}\ \emph {et~al.}(2016)\citenamefont {Ferrier-Barbut}, \citenamefont {Kadau}, \citenamefont {Schmitt}, \citenamefont {Wenzel},\ and\ \citenamefont {Pfau}}]{ferrier-barbut2016droplets}%
  \BibitemOpen
  \bibfield  {author} {\bibinfo {author} {\bibfnamefont {I.}~\bibnamefont {Ferrier-Barbut}}, \bibinfo {author} {\bibfnamefont {H.}~\bibnamefont {Kadau}}, \bibinfo {author} {\bibfnamefont {M.}~\bibnamefont {Schmitt}}, \bibinfo {author} {\bibfnamefont {M.}~\bibnamefont {Wenzel}},\ and\ \bibinfo {author} {\bibfnamefont {T.}~\bibnamefont {Pfau}},\ }\bibfield  {title} {\bibinfo {title} {Observation of quantum droplets in a strongly dipolar {Bose} gas},\ }\href {https://doi.org/10.1103/PhysRevLett.116.215301} {\bibfield  {journal} {\bibinfo  {journal} {Phys. Rev. Lett.}\ }\textbf {\bibinfo {volume} {116}},\ \bibinfo {pages} {215301} (\bibinfo {year} {2016})}\BibitemShut {NoStop}%
\bibitem [{\citenamefont {Schmitt}\ \emph {et~al.}(2016)\citenamefont {Schmitt}, \citenamefont {Wenzel}, \citenamefont {B{\"o}ttcher}, \citenamefont {Ferrier-Barbut},\ and\ \citenamefont {Pfau}}]{Schmitt2016Droplet}%
  \BibitemOpen
  \bibfield  {author} {\bibinfo {author} {\bibfnamefont {M.}~\bibnamefont {Schmitt}}, \bibinfo {author} {\bibfnamefont {M.}~\bibnamefont {Wenzel}}, \bibinfo {author} {\bibfnamefont {F.}~\bibnamefont {B{\"o}ttcher}}, \bibinfo {author} {\bibfnamefont {I.}~\bibnamefont {Ferrier-Barbut}},\ and\ \bibinfo {author} {\bibfnamefont {T.}~\bibnamefont {Pfau}},\ }\bibfield  {title} {\bibinfo {title} {Self-bound droplets of a dilute magnetic quantum liquid},\ }\href {https://doi.org/10.1038/nature20126} {\bibfield  {journal} {\bibinfo  {journal} {Nature}\ }\textbf {\bibinfo {volume} {539}},\ \bibinfo {pages} {259} (\bibinfo {year} {2016})}\BibitemShut {NoStop}%
\bibitem [{\citenamefont {Chomaz}\ \emph {et~al.}(2016)\citenamefont {Chomaz}, \citenamefont {Baier}, \citenamefont {Petter}, \citenamefont {Mark}, \citenamefont {W\"achtler}, \citenamefont {Santos},\ and\ \citenamefont {Ferlaino}}]{Chomaz2016QuantumFluctuation}%
  \BibitemOpen
  \bibfield  {author} {\bibinfo {author} {\bibfnamefont {L.}~\bibnamefont {Chomaz}}, \bibinfo {author} {\bibfnamefont {S.}~\bibnamefont {Baier}}, \bibinfo {author} {\bibfnamefont {D.}~\bibnamefont {Petter}}, \bibinfo {author} {\bibfnamefont {M.~J.}\ \bibnamefont {Mark}}, \bibinfo {author} {\bibfnamefont {F.}~\bibnamefont {W\"achtler}}, \bibinfo {author} {\bibfnamefont {L.}~\bibnamefont {Santos}},\ and\ \bibinfo {author} {\bibfnamefont {F.}~\bibnamefont {Ferlaino}},\ }\bibfield  {title} {\bibinfo {title} {Quantum-fluctuation-driven crossover from a dilute {Bose-Einstein} condensate to a macrodroplet in a dipolar quantum fluid},\ }\href {https://doi.org/10.1103/PhysRevX.6.041039} {\bibfield  {journal} {\bibinfo  {journal} {Phys. Rev. X}\ }\textbf {\bibinfo {volume} {6}},\ \bibinfo {pages} {041039} (\bibinfo {year} {2016})}\BibitemShut {NoStop}%
\bibitem [{\citenamefont {Cabrera}\ \emph {et~al.}(2018)\citenamefont {Cabrera}, \citenamefont {Tanzi}, \citenamefont {Sanz}, \citenamefont {Naylor}, \citenamefont {Thomas}, \citenamefont {Cheiney},\ and\ \citenamefont {Tarruell}}]{cabrera2018quantum}%
  \BibitemOpen
  \bibfield  {author} {\bibinfo {author} {\bibfnamefont {C.~R.}\ \bibnamefont {Cabrera}}, \bibinfo {author} {\bibfnamefont {L.}~\bibnamefont {Tanzi}}, \bibinfo {author} {\bibfnamefont {J.}~\bibnamefont {Sanz}}, \bibinfo {author} {\bibfnamefont {B.}~\bibnamefont {Naylor}}, \bibinfo {author} {\bibfnamefont {P.}~\bibnamefont {Thomas}}, \bibinfo {author} {\bibfnamefont {P.}~\bibnamefont {Cheiney}},\ and\ \bibinfo {author} {\bibfnamefont {L.}~\bibnamefont {Tarruell}},\ }\bibfield  {title} {\bibinfo {title} {Quantum liquid droplets in a mixture of {B}ose-{E}instein condensates},\ }\href {https://www.science.org/doi/10.1126/science.aao5686} {\bibfield  {journal} {\bibinfo  {journal} {Science}\ }\textbf {\bibinfo {volume} {359}},\ \bibinfo {pages} {301} (\bibinfo {year} {2018})}\BibitemShut {NoStop}%
\bibitem [{\citenamefont {Semeghini}\ \emph {et~al.}(2018)\citenamefont {Semeghini}, \citenamefont {Ferioli}, \citenamefont {Masi}, \citenamefont {Mazzinghi}, \citenamefont {Wolswijk}, \citenamefont {Minardi}, \citenamefont {Modugno}, \citenamefont {Modugno}, \citenamefont {Inguscio},\ and\ \citenamefont {Fattori}}]{semeghini2018self}%
  \BibitemOpen
  \bibfield  {author} {\bibinfo {author} {\bibfnamefont {G.}~\bibnamefont {Semeghini}}, \bibinfo {author} {\bibfnamefont {G.}~\bibnamefont {Ferioli}}, \bibinfo {author} {\bibfnamefont {L.}~\bibnamefont {Masi}}, \bibinfo {author} {\bibfnamefont {C.}~\bibnamefont {Mazzinghi}}, \bibinfo {author} {\bibfnamefont {L.}~\bibnamefont {Wolswijk}}, \bibinfo {author} {\bibfnamefont {F.}~\bibnamefont {Minardi}}, \bibinfo {author} {\bibfnamefont {M.}~\bibnamefont {Modugno}}, \bibinfo {author} {\bibfnamefont {G.}~\bibnamefont {Modugno}}, \bibinfo {author} {\bibfnamefont {M.}~\bibnamefont {Inguscio}},\ and\ \bibinfo {author} {\bibfnamefont {M.}~\bibnamefont {Fattori}},\ }\bibfield  {title} {\bibinfo {title} {Self-bound quantum droplets of atomic mixtures in free space},\ }\href {https://doi.org/10.1103/PhysRevLett.120.235301} {\bibfield  {journal} {\bibinfo  {journal} {Phys. Rev. Lett.}\ }\textbf {\bibinfo {volume} {120}},\ \bibinfo {pages} {235301} (\bibinfo {year} {2018})}\BibitemShut {NoStop}%
\bibitem [{\citenamefont {Ospelkaus}\ \emph {et~al.}(2010)\citenamefont {Ospelkaus}, \citenamefont {Ni}, \citenamefont {Wang}, \citenamefont {De~Miranda}, \citenamefont {Neyenhuis}, \citenamefont {Qu{\'e}m{\'e}ner}, \citenamefont {Julienne}, \citenamefont {Bohn}, \citenamefont {Jin},\ and\ \citenamefont {Ye}}]{ospelkaus2010quantum}%
  \BibitemOpen
  \bibfield  {author} {\bibinfo {author} {\bibfnamefont {S.}~\bibnamefont {Ospelkaus}}, \bibinfo {author} {\bibfnamefont {K.-K.}\ \bibnamefont {Ni}}, \bibinfo {author} {\bibfnamefont {D.}~\bibnamefont {Wang}}, \bibinfo {author} {\bibfnamefont {M.}~\bibnamefont {De~Miranda}}, \bibinfo {author} {\bibfnamefont {B.}~\bibnamefont {Neyenhuis}}, \bibinfo {author} {\bibfnamefont {G.}~\bibnamefont {Qu{\'e}m{\'e}ner}}, \bibinfo {author} {\bibfnamefont {P.}~\bibnamefont {Julienne}}, \bibinfo {author} {\bibfnamefont {J.}~\bibnamefont {Bohn}}, \bibinfo {author} {\bibfnamefont {D.}~\bibnamefont {Jin}},\ and\ \bibinfo {author} {\bibfnamefont {J.}~\bibnamefont {Ye}},\ }\bibfield  {title} {\bibinfo {title} {Quantum-state controlled chemical reactions of ultracold potassium-rubidium molecules},\ }\href {https://www.science.org/doi/10.1126/science.1184121} {\bibfield  {journal} {\bibinfo  {journal} {Science}\ }\textbf {\bibinfo {volume} {327}},\ \bibinfo {pages} {853} (\bibinfo {year} {2010})}\BibitemShut {NoStop}%
\bibitem [{\citenamefont {Hu}\ \emph {et~al.}(2019)\citenamefont {Hu}, \citenamefont {Liu}, \citenamefont {Grimes}, \citenamefont {Lin}, \citenamefont {Gheorghe}, \citenamefont {Vexiau}, \citenamefont {Bouloufa-Maafa}, \citenamefont {Dulieu}, \citenamefont {Rosenband},\ and\ \citenamefont {Ni}}]{Hu2019K2RB2}%
  \BibitemOpen
  \bibfield  {author} {\bibinfo {author} {\bibfnamefont {M.-G.}\ \bibnamefont {Hu}}, \bibinfo {author} {\bibfnamefont {Y.}~\bibnamefont {Liu}}, \bibinfo {author} {\bibfnamefont {D.~D.}\ \bibnamefont {Grimes}}, \bibinfo {author} {\bibfnamefont {Y.-W.}\ \bibnamefont {Lin}}, \bibinfo {author} {\bibfnamefont {A.~H.}\ \bibnamefont {Gheorghe}}, \bibinfo {author} {\bibfnamefont {R.}~\bibnamefont {Vexiau}}, \bibinfo {author} {\bibfnamefont {N.}~\bibnamefont {Bouloufa-Maafa}}, \bibinfo {author} {\bibfnamefont {O.}~\bibnamefont {Dulieu}}, \bibinfo {author} {\bibfnamefont {T.}~\bibnamefont {Rosenband}},\ and\ \bibinfo {author} {\bibfnamefont {K.-K.}\ \bibnamefont {Ni}},\ }\bibfield  {title} {\bibinfo {title} {Direct observation of bimolecular reactions of ultracold {KR}b molecules},\ }\href {https://doi.org/10.1126/science.aay9531} {\bibfield  {journal} {\bibinfo  {journal} {Science}\ }\textbf {\bibinfo {volume} {366}},\ \bibinfo {pages} {1111} (\bibinfo {year} {2019})}\BibitemShut {NoStop}%
\bibitem [{\citenamefont {Gregory}\ \emph {et~al.}(2019)\citenamefont {Gregory}, \citenamefont {Frye}, \citenamefont {Blackmore}, \citenamefont {Bridge}, \citenamefont {Sawant}, \citenamefont {Hutson},\ and\ \citenamefont {Cornish}}]{gregory2019sticky}%
  \BibitemOpen
  \bibfield  {author} {\bibinfo {author} {\bibfnamefont {P.~D.}\ \bibnamefont {Gregory}}, \bibinfo {author} {\bibfnamefont {M.~D.}\ \bibnamefont {Frye}}, \bibinfo {author} {\bibfnamefont {J.~A.}\ \bibnamefont {Blackmore}}, \bibinfo {author} {\bibfnamefont {E.~M.}\ \bibnamefont {Bridge}}, \bibinfo {author} {\bibfnamefont {R.}~\bibnamefont {Sawant}}, \bibinfo {author} {\bibfnamefont {J.~M.}\ \bibnamefont {Hutson}},\ and\ \bibinfo {author} {\bibfnamefont {S.~L.}\ \bibnamefont {Cornish}},\ }\bibfield  {title} {\bibinfo {title} {Sticky collisions of ultracold {R}b{C}s molecules},\ }\href {https://doi.org/10.1038/s41467-019-11033-y} {\bibfield  {journal} {\bibinfo  {journal} {Nat. Commun.}\ }\textbf {\bibinfo {volume} {10}},\ \bibinfo {pages} {3104} (\bibinfo {year} {2019})}\BibitemShut {NoStop}%
\bibitem [{\citenamefont {Schindewolf}\ \emph {et~al.}(2026)\citenamefont {Schindewolf}, \citenamefont {Hertkorn}, \citenamefont {Stevenson}, \citenamefont {Ciardi}, \citenamefont {Gro\ss{}}, \citenamefont {Wang}, \citenamefont {Karman}, \citenamefont {Qu\'em\'ener}, \citenamefont {Will}, \citenamefont {Pohl},\ and\ \citenamefont {Langen}}]{schindewolf2025few}%
  \BibitemOpen
  \bibfield  {author} {\bibinfo {author} {\bibfnamefont {A.}~\bibnamefont {Schindewolf}}, \bibinfo {author} {\bibfnamefont {J.}~\bibnamefont {Hertkorn}}, \bibinfo {author} {\bibfnamefont {I.}~\bibnamefont {Stevenson}}, \bibinfo {author} {\bibfnamefont {M.}~\bibnamefont {Ciardi}}, \bibinfo {author} {\bibfnamefont {P.}~\bibnamefont {Gro\ss{}}}, \bibinfo {author} {\bibfnamefont {D.}~\bibnamefont {Wang}}, \bibinfo {author} {\bibfnamefont {T.}~\bibnamefont {Karman}}, \bibinfo {author} {\bibfnamefont {G.}~\bibnamefont {Qu\'em\'ener}}, \bibinfo {author} {\bibfnamefont {S.}~\bibnamefont {Will}}, \bibinfo {author} {\bibfnamefont {T.}~\bibnamefont {Pohl}},\ and\ \bibinfo {author} {\bibfnamefont {T.}~\bibnamefont {Langen}},\ }\bibfield  {title} {\bibinfo {title} {Colloquium: {St}rongly dipolar molecular {Bo}se-{E}instein condensates: {F}rom few- to many-body physics},\ }\href {https://doi.org/10.1103/r55l-f93m} {\bibfield  {journal} {\bibinfo  {journal} {Rev. Mod. Phys.}\ }\textbf {\bibinfo {volume} {98}},\ \bibinfo
  {pages} {031002} (\bibinfo {year} {2026})}\BibitemShut {NoStop}%
\bibitem [{\citenamefont {Karman}\ and\ \citenamefont {Hutson}(2018)}]{Karman2018MWShielding}%
  \BibitemOpen
  \bibfield  {author} {\bibinfo {author} {\bibfnamefont {T.}~\bibnamefont {Karman}}\ and\ \bibinfo {author} {\bibfnamefont {J.~M.}\ \bibnamefont {Hutson}},\ }\bibfield  {title} {\bibinfo {title} {Microwave shielding of ultracold polar molecules},\ }\href {https://doi.org/10.1103/PhysRevLett.121.163401} {\bibfield  {journal} {\bibinfo  {journal} {Phys. Rev. Lett.}\ }\textbf {\bibinfo {volume} {121}},\ \bibinfo {pages} {163401} (\bibinfo {year} {2018})}\BibitemShut {NoStop}%
\bibitem [{\citenamefont {Avdeenkov}\ \emph {et~al.}(2006)\citenamefont {Avdeenkov}, \citenamefont {Kajita},\ and\ \citenamefont {Bohn}}]{Avdeenkov2006EfieldShielding}%
  \BibitemOpen
  \bibfield  {author} {\bibinfo {author} {\bibfnamefont {A.~V.}\ \bibnamefont {Avdeenkov}}, \bibinfo {author} {\bibfnamefont {M.}~\bibnamefont {Kajita}},\ and\ \bibinfo {author} {\bibfnamefont {J.~L.}\ \bibnamefont {Bohn}},\ }\bibfield  {title} {\bibinfo {title} {Suppression of inelastic collisions of polar $^{1}\ensuremath{\Sigma}$ state molecules in an electrostatic field},\ }\href {https://doi.org/10.1103/PhysRevA.73.022707} {\bibfield  {journal} {\bibinfo  {journal} {Phys. Rev. A}\ }\textbf {\bibinfo {volume} {73}},\ \bibinfo {pages} {022707} (\bibinfo {year} {2006})}\BibitemShut {NoStop}%
\bibitem [{\citenamefont {Anderegg}\ \emph {et~al.}(2021)\citenamefont {Anderegg}, \citenamefont {Burchesky}, \citenamefont {Bao}, \citenamefont {Yu}, \citenamefont {Karman}, \citenamefont {Chae}, \citenamefont {Ni}, \citenamefont {Ketterle},\ and\ \citenamefont {Doyle}}]{anderegg2021observation}%
  \BibitemOpen
  \bibfield  {author} {\bibinfo {author} {\bibfnamefont {L.}~\bibnamefont {Anderegg}}, \bibinfo {author} {\bibfnamefont {S.}~\bibnamefont {Burchesky}}, \bibinfo {author} {\bibfnamefont {Y.}~\bibnamefont {Bao}}, \bibinfo {author} {\bibfnamefont {S.~S.}\ \bibnamefont {Yu}}, \bibinfo {author} {\bibfnamefont {T.}~\bibnamefont {Karman}}, \bibinfo {author} {\bibfnamefont {E.}~\bibnamefont {Chae}}, \bibinfo {author} {\bibfnamefont {K.-K.}\ \bibnamefont {Ni}}, \bibinfo {author} {\bibfnamefont {W.}~\bibnamefont {Ketterle}},\ and\ \bibinfo {author} {\bibfnamefont {J.~M.}\ \bibnamefont {Doyle}},\ }\bibfield  {title} {\bibinfo {title} {Observation of microwave shielding of ultracold molecules},\ }\href {https://www.science.org/doi/10.1126/science.abg9502} {\bibfield  {journal} {\bibinfo  {journal} {Science}\ }\textbf {\bibinfo {volume} {373}},\ \bibinfo {pages} {779} (\bibinfo {year} {2021})}\BibitemShut {NoStop}%
\bibitem [{\citenamefont {De~Miranda}\ \emph {et~al.}(2011)\citenamefont {De~Miranda}, \citenamefont {Chotia}, \citenamefont {Neyenhuis}, \citenamefont {Wang}, \citenamefont {Qu{\'e}m{\'e}ner}, \citenamefont {Ospelkaus}, \citenamefont {Bohn}, \citenamefont {Ye},\ and\ \citenamefont {Jin}}]{de2011controlling}%
  \BibitemOpen
  \bibfield  {author} {\bibinfo {author} {\bibfnamefont {M.}~\bibnamefont {De~Miranda}}, \bibinfo {author} {\bibfnamefont {A.}~\bibnamefont {Chotia}}, \bibinfo {author} {\bibfnamefont {B.}~\bibnamefont {Neyenhuis}}, \bibinfo {author} {\bibfnamefont {D.}~\bibnamefont {Wang}}, \bibinfo {author} {\bibfnamefont {G.}~\bibnamefont {Qu{\'e}m{\'e}ner}}, \bibinfo {author} {\bibfnamefont {S.}~\bibnamefont {Ospelkaus}}, \bibinfo {author} {\bibfnamefont {J.}~\bibnamefont {Bohn}}, \bibinfo {author} {\bibfnamefont {J.}~\bibnamefont {Ye}},\ and\ \bibinfo {author} {\bibfnamefont {D.}~\bibnamefont {Jin}},\ }\bibfield  {title} {\bibinfo {title} {Controlling the quantum stereodynamics of ultracold bimolecular reactions},\ }\href {https://www.nature.com/articles/nphys1939} {\bibfield  {journal} {\bibinfo  {journal} {Nat. Phys.}\ }\textbf {\bibinfo {volume} {7}},\ \bibinfo {pages} {502} (\bibinfo {year} {2011})}\BibitemShut {NoStop}%
\bibitem [{\citenamefont {Qu{\'e}m{\'e}ner}\ and\ \citenamefont {Bohn}(2011)}]{quemener2011dynamics}%
  \BibitemOpen
  \bibfield  {author} {\bibinfo {author} {\bibfnamefont {G.}~\bibnamefont {Qu{\'e}m{\'e}ner}}\ and\ \bibinfo {author} {\bibfnamefont {J.~L.}\ \bibnamefont {Bohn}},\ }\bibfield  {title} {\bibinfo {title} {Dynamics of ultracold molecules in confined geometry and electric field},\ }\href {https://link.aps.org/doi/10.1103/PhysRevA.83.012705} {\bibfield  {journal} {\bibinfo  {journal} {Phys. Rev. A}\ }\textbf {\bibinfo {volume} {83}},\ \bibinfo {pages} {012705} (\bibinfo {year} {2011})}\BibitemShut {NoStop}%
\bibitem [{\citenamefont {Gonz{\'a}lez-Mart{\'\i}nez}\ \emph {et~al.}(2017)\citenamefont {Gonz{\'a}lez-Mart{\'\i}nez}, \citenamefont {Bohn},\ and\ \citenamefont {Qu{\'e}m{\'e}ner}}]{gonzalez2017adimensional}%
  \BibitemOpen
  \bibfield  {author} {\bibinfo {author} {\bibfnamefont {M.~L.}\ \bibnamefont {Gonz{\'a}lez-Mart{\'\i}nez}}, \bibinfo {author} {\bibfnamefont {J.~L.}\ \bibnamefont {Bohn}},\ and\ \bibinfo {author} {\bibfnamefont {G.}~\bibnamefont {Qu{\'e}m{\'e}ner}},\ }\bibfield  {title} {\bibinfo {title} {Adimensional theory of shielding in ultracold collisions of dipolar rotors},\ }\href {https://link.aps.org/doi/10.1103/PhysRevA.96.032718} {\bibfield  {journal} {\bibinfo  {journal} {Phys. Rev. A}\ }\textbf {\bibinfo {volume} {96}},\ \bibinfo {pages} {032718} (\bibinfo {year} {2017})}\BibitemShut {NoStop}%
\bibitem [{\citenamefont {Valtolina}\ \emph {et~al.}(2020)\citenamefont {Valtolina}, \citenamefont {Matsuda}, \citenamefont {Tobias}, \citenamefont {Li}, \citenamefont {De~Marco},\ and\ \citenamefont {Ye}}]{valtolina2020dipolar}%
  \BibitemOpen
  \bibfield  {author} {\bibinfo {author} {\bibfnamefont {G.}~\bibnamefont {Valtolina}}, \bibinfo {author} {\bibfnamefont {K.}~\bibnamefont {Matsuda}}, \bibinfo {author} {\bibfnamefont {W.~G.}\ \bibnamefont {Tobias}}, \bibinfo {author} {\bibfnamefont {J.-R.}\ \bibnamefont {Li}}, \bibinfo {author} {\bibfnamefont {L.}~\bibnamefont {De~Marco}},\ and\ \bibinfo {author} {\bibfnamefont {J.}~\bibnamefont {Ye}},\ }\bibfield  {title} {\bibinfo {title} {Dipolar evaporation of reactive molecules to below the {Fermi} temperature},\ }\href {https://doi.org/10.1038/s41586-020-2980-7} {\bibfield  {journal} {\bibinfo  {journal} {Nature}\ }\textbf {\bibinfo {volume} {588}},\ \bibinfo {pages} {239} (\bibinfo {year} {2020})}\BibitemShut {NoStop}%
\bibitem [{\citenamefont {Matsuda}\ \emph {et~al.}(2020)\citenamefont {Matsuda}, \citenamefont {De~Marco}, \citenamefont {Li}, \citenamefont {Tobias}, \citenamefont {Valtolina}, \citenamefont {Qu{\'e}m{\'e}ner},\ and\ \citenamefont {Ye}}]{matsuda2020resonant}%
  \BibitemOpen
  \bibfield  {author} {\bibinfo {author} {\bibfnamefont {K.}~\bibnamefont {Matsuda}}, \bibinfo {author} {\bibfnamefont {L.}~\bibnamefont {De~Marco}}, \bibinfo {author} {\bibfnamefont {J.-R.}\ \bibnamefont {Li}}, \bibinfo {author} {\bibfnamefont {W.~G.}\ \bibnamefont {Tobias}}, \bibinfo {author} {\bibfnamefont {G.}~\bibnamefont {Valtolina}}, \bibinfo {author} {\bibfnamefont {G.}~\bibnamefont {Qu{\'e}m{\'e}ner}},\ and\ \bibinfo {author} {\bibfnamefont {J.}~\bibnamefont {Ye}},\ }\bibfield  {title} {\bibinfo {title} {Resonant collisional shielding of reactive molecules using electric fields},\ }\href {https://www.science.org/doi/10.1126/science.abe7370} {\bibfield  {journal} {\bibinfo  {journal} {Science}\ }\textbf {\bibinfo {volume} {370}},\ \bibinfo {pages} {1324} (\bibinfo {year} {2020})}\BibitemShut {NoStop}%
\bibitem [{\citenamefont {Li}\ \emph {et~al.}(2021)\citenamefont {Li}, \citenamefont {Tobias}, \citenamefont {Matsuda}, \citenamefont {Miller}, \citenamefont {Valtolina}, \citenamefont {De~Marco}, \citenamefont {Wang}, \citenamefont {Lassabli{\`e}re}, \citenamefont {Qu{\'e}m{\'e}ner}, \citenamefont {Bohn},\ and\ \citenamefont {Ye}}]{li2021KRbEvap}%
  \BibitemOpen
  \bibfield  {author} {\bibinfo {author} {\bibfnamefont {J.-R.}\ \bibnamefont {Li}}, \bibinfo {author} {\bibfnamefont {W.~G.}\ \bibnamefont {Tobias}}, \bibinfo {author} {\bibfnamefont {K.}~\bibnamefont {Matsuda}}, \bibinfo {author} {\bibfnamefont {C.}~\bibnamefont {Miller}}, \bibinfo {author} {\bibfnamefont {G.}~\bibnamefont {Valtolina}}, \bibinfo {author} {\bibfnamefont {L.}~\bibnamefont {De~Marco}}, \bibinfo {author} {\bibfnamefont {R.~R.~W.}\ \bibnamefont {Wang}}, \bibinfo {author} {\bibfnamefont {L.}~\bibnamefont {Lassabli{\`e}re}}, \bibinfo {author} {\bibfnamefont {G.}~\bibnamefont {Qu{\'e}m{\'e}ner}}, \bibinfo {author} {\bibfnamefont {J.~L.}\ \bibnamefont {Bohn}},\ and\ \bibinfo {author} {\bibfnamefont {J.}~\bibnamefont {Ye}},\ }\bibfield  {title} {\bibinfo {title} {Tuning of dipolar interactions and evaporative cooling in a three-dimensional molecular quantum gas},\ }\href {https://doi.org/10.1038/s41567-021-01329-6} {\bibfield  {journal} {\bibinfo  {journal} {Nat. Phys.}\ }\textbf {\bibinfo {volume}
  {17}},\ \bibinfo {pages} {1144} (\bibinfo {year} {2021})}\BibitemShut {NoStop}%
\bibitem [{\citenamefont {Lin}\ \emph {et~al.}(2026)\citenamefont {Lin}, \citenamefont {Carroll}, \citenamefont {Martin}, \citenamefont {Miller}, \citenamefont {Wang}, \citenamefont {Xu}, \citenamefont {Bohn}, \citenamefont {de~Jongh},\ and\ \citenamefont {Ye}}]{lin2026KRbPauli}%
  \BibitemOpen
  \bibfield  {author} {\bibinfo {author} {\bibfnamefont {J.}~\bibnamefont {Lin}}, \bibinfo {author} {\bibfnamefont {A.~N.}\ \bibnamefont {Carroll}}, \bibinfo {author} {\bibfnamefont {P.}~\bibnamefont {Martin}}, \bibinfo {author} {\bibfnamefont {C.}~\bibnamefont {Miller}}, \bibinfo {author} {\bibfnamefont {R.~R.~W.}\ \bibnamefont {Wang}}, \bibinfo {author} {\bibfnamefont {K.}~\bibnamefont {Xu}}, \bibinfo {author} {\bibfnamefont {J.~L.}\ \bibnamefont {Bohn}}, \bibinfo {author} {\bibfnamefont {T.}~\bibnamefont {de~Jongh}},\ and\ \bibinfo {author} {\bibfnamefont {J.}~\bibnamefont {Ye}},\ }\bibfield  {title} {\bibinfo {title} {{Fermi} gas of polar molecules in the {P}auli-blocked regime},\ }\href {https://arxiv.org/abs/2606.14135} {\bibfield  {journal} {\bibinfo  {journal} {arXiv:2606.14135}\ } (\bibinfo {year} {2026})}\BibitemShut {NoStop}%
\bibitem [{\citenamefont {Stevenson}\ \emph {et~al.}(2024)\citenamefont {Stevenson}, \citenamefont {Singh}, \citenamefont {Elkamshishy}, \citenamefont {Bigagli}, \citenamefont {Yuan}, \citenamefont {Zhang}, \citenamefont {Greene},\ and\ \citenamefont {Will}}]{Stevenson2024ThreeBody}%
  \BibitemOpen
  \bibfield  {author} {\bibinfo {author} {\bibfnamefont {I.}~\bibnamefont {Stevenson}}, \bibinfo {author} {\bibfnamefont {S.}~\bibnamefont {Singh}}, \bibinfo {author} {\bibfnamefont {A.}~\bibnamefont {Elkamshishy}}, \bibinfo {author} {\bibfnamefont {N.}~\bibnamefont {Bigagli}}, \bibinfo {author} {\bibfnamefont {W.}~\bibnamefont {Yuan}}, \bibinfo {author} {\bibfnamefont {S.}~\bibnamefont {Zhang}}, \bibinfo {author} {\bibfnamefont {C.~H.}\ \bibnamefont {Greene}},\ and\ \bibinfo {author} {\bibfnamefont {S.}~\bibnamefont {Will}},\ }\bibfield  {title} {\bibinfo {title} {Three-body recombination of ultracold microwave-shielded polar molecules},\ }\href {https://doi.org/10.1103/PhysRevLett.133.263402} {\bibfield  {journal} {\bibinfo  {journal} {Phys. Rev. Lett.}\ }\textbf {\bibinfo {volume} {133}},\ \bibinfo {pages} {263402} (\bibinfo {year} {2024})}\BibitemShut {NoStop}%
\bibitem [{\citenamefont {Avdeenkov}\ and\ \citenamefont {Bohn}(2003)}]{Avdeenkov2003FieldLink}%
  \BibitemOpen
  \bibfield  {author} {\bibinfo {author} {\bibfnamefont {A.~V.}\ \bibnamefont {Avdeenkov}}\ and\ \bibinfo {author} {\bibfnamefont {J.~L.}\ \bibnamefont {Bohn}},\ }\bibfield  {title} {\bibinfo {title} {Linking ultracold polar molecules},\ }\href {https://doi.org/10.1103/PhysRevLett.90.043006} {\bibfield  {journal} {\bibinfo  {journal} {Phys. Rev. Lett.}\ }\textbf {\bibinfo {volume} {90}},\ \bibinfo {pages} {043006} (\bibinfo {year} {2003})}\BibitemShut {NoStop}%
\bibitem [{\citenamefont {Chen}\ \emph {et~al.}(2023)\citenamefont {Chen}, \citenamefont {Schindewolf}, \citenamefont {Eppelt}, \citenamefont {Bause}, \citenamefont {Duda}, \citenamefont {Biswas}, \citenamefont {Karman}, \citenamefont {Hilker}, \citenamefont {Bloch},\ and\ \citenamefont {Luo}}]{Chen2023FieldLinked}%
  \BibitemOpen
  \bibfield  {author} {\bibinfo {author} {\bibfnamefont {X.-Y.}\ \bibnamefont {Chen}}, \bibinfo {author} {\bibfnamefont {A.}~\bibnamefont {Schindewolf}}, \bibinfo {author} {\bibfnamefont {S.}~\bibnamefont {Eppelt}}, \bibinfo {author} {\bibfnamefont {R.}~\bibnamefont {Bause}}, \bibinfo {author} {\bibfnamefont {M.}~\bibnamefont {Duda}}, \bibinfo {author} {\bibfnamefont {S.}~\bibnamefont {Biswas}}, \bibinfo {author} {\bibfnamefont {T.}~\bibnamefont {Karman}}, \bibinfo {author} {\bibfnamefont {T.}~\bibnamefont {Hilker}}, \bibinfo {author} {\bibfnamefont {I.}~\bibnamefont {Bloch}},\ and\ \bibinfo {author} {\bibfnamefont {X.-Y.}\ \bibnamefont {Luo}},\ }\bibfield  {title} {\bibinfo {title} {Field-linked resonances of polar molecules},\ }\href {https://doi.org/10.1038/s41586-022-05651-8} {\bibfield  {journal} {\bibinfo  {journal} {Nature}\ }\textbf {\bibinfo {volume} {614}},\ \bibinfo {pages} {59} (\bibinfo {year} {2023})}\BibitemShut {NoStop}%
\bibitem [{\citenamefont {Chen}\ \emph {et~al.}(2024)\citenamefont {Chen}, \citenamefont {Biswas}, \citenamefont {Eppelt}, \citenamefont {Schindewolf}, \citenamefont {Deng}, \citenamefont {Shi}, \citenamefont {Yi}, \citenamefont {Hilker}, \citenamefont {Bloch},\ and\ \citenamefont {Luo}}]{Chen2024Tetramer}%
  \BibitemOpen
  \bibfield  {author} {\bibinfo {author} {\bibfnamefont {X.-Y.}\ \bibnamefont {Chen}}, \bibinfo {author} {\bibfnamefont {S.}~\bibnamefont {Biswas}}, \bibinfo {author} {\bibfnamefont {S.}~\bibnamefont {Eppelt}}, \bibinfo {author} {\bibfnamefont {A.}~\bibnamefont {Schindewolf}}, \bibinfo {author} {\bibfnamefont {F.}~\bibnamefont {Deng}}, \bibinfo {author} {\bibfnamefont {T.}~\bibnamefont {Shi}}, \bibinfo {author} {\bibfnamefont {S.}~\bibnamefont {Yi}}, \bibinfo {author} {\bibfnamefont {T.~A.}\ \bibnamefont {Hilker}}, \bibinfo {author} {\bibfnamefont {I.}~\bibnamefont {Bloch}},\ and\ \bibinfo {author} {\bibfnamefont {X.-Y.}\ \bibnamefont {Luo}},\ }\bibfield  {title} {\bibinfo {title} {Ultracold field-linked tetratomic molecules},\ }\href {https://doi.org/10.1038/s41586-023-06986-6} {\bibfield  {journal} {\bibinfo  {journal} {Nature}\ }\textbf {\bibinfo {volume} {626}},\ \bibinfo {pages} {283} (\bibinfo {year} {2024})}\BibitemShut {NoStop}%
\bibitem [{\citenamefont {Bigagli}\ \emph {et~al.}(2023)\citenamefont {Bigagli}, \citenamefont {Warner}, \citenamefont {Yuan}, \citenamefont {Zhang}, \citenamefont {Stevenson}, \citenamefont {Karman},\ and\ \citenamefont {Will}}]{bigagli2023collisionally}%
  \BibitemOpen
  \bibfield  {author} {\bibinfo {author} {\bibfnamefont {N.}~\bibnamefont {Bigagli}}, \bibinfo {author} {\bibfnamefont {C.}~\bibnamefont {Warner}}, \bibinfo {author} {\bibfnamefont {W.}~\bibnamefont {Yuan}}, \bibinfo {author} {\bibfnamefont {S.}~\bibnamefont {Zhang}}, \bibinfo {author} {\bibfnamefont {I.}~\bibnamefont {Stevenson}}, \bibinfo {author} {\bibfnamefont {T.}~\bibnamefont {Karman}},\ and\ \bibinfo {author} {\bibfnamefont {S.}~\bibnamefont {Will}},\ }\bibfield  {title} {\bibinfo {title} {Collisionally stable gas of bosonic dipolar ground-state molecules},\ }\href {https://doi.org/10.1038/s41567-023-02200-6} {\bibfield  {journal} {\bibinfo  {journal} {Nat. Phys.}\ }\textbf {\bibinfo {volume} {19}},\ \bibinfo {pages} {1579} (\bibinfo {year} {2023})}\BibitemShut {NoStop}%
\bibitem [{\citenamefont {Lin}\ \emph {et~al.}(2023)\citenamefont {Lin}, \citenamefont {Chen}, \citenamefont {Jin}, \citenamefont {Shi}, \citenamefont {Deng}, \citenamefont {Zhang}, \citenamefont {Qu\'em\'ener}, \citenamefont {Shi}, \citenamefont {Yi},\ and\ \citenamefont {Wang}}]{lin2023SingleMWShielding}%
  \BibitemOpen
  \bibfield  {author} {\bibinfo {author} {\bibfnamefont {J.}~\bibnamefont {Lin}}, \bibinfo {author} {\bibfnamefont {G.}~\bibnamefont {Chen}}, \bibinfo {author} {\bibfnamefont {M.}~\bibnamefont {Jin}}, \bibinfo {author} {\bibfnamefont {Z.}~\bibnamefont {Shi}}, \bibinfo {author} {\bibfnamefont {F.}~\bibnamefont {Deng}}, \bibinfo {author} {\bibfnamefont {W.}~\bibnamefont {Zhang}}, \bibinfo {author} {\bibfnamefont {G.}~\bibnamefont {Qu\'em\'ener}}, \bibinfo {author} {\bibfnamefont {T.}~\bibnamefont {Shi}}, \bibinfo {author} {\bibfnamefont {S.}~\bibnamefont {Yi}},\ and\ \bibinfo {author} {\bibfnamefont {D.}~\bibnamefont {Wang}},\ }\bibfield  {title} {\bibinfo {title} {Microwave shielding of bosonic {N}a{R}b molecules},\ }\href {https://doi.org/10.1103/PhysRevX.13.031032} {\bibfield  {journal} {\bibinfo  {journal} {Phys. Rev. X}\ }\textbf {\bibinfo {volume} {13}},\ \bibinfo {pages} {031032} (\bibinfo {year} {2023})}\BibitemShut {NoStop}%
\bibitem [{\citenamefont {Karman}\ \emph {et~al.}(2025)\citenamefont {Karman}, \citenamefont {Bigagli}, \citenamefont {Yuan}, \citenamefont {Zhang}, \citenamefont {Stevenson},\ and\ \citenamefont {Will}}]{Karman2025DoubleMWShielding}%
  \BibitemOpen
  \bibfield  {author} {\bibinfo {author} {\bibfnamefont {T.}~\bibnamefont {Karman}}, \bibinfo {author} {\bibfnamefont {N.}~\bibnamefont {Bigagli}}, \bibinfo {author} {\bibfnamefont {W.}~\bibnamefont {Yuan}}, \bibinfo {author} {\bibfnamefont {S.}~\bibnamefont {Zhang}}, \bibinfo {author} {\bibfnamefont {I.}~\bibnamefont {Stevenson}},\ and\ \bibinfo {author} {\bibfnamefont {S.}~\bibnamefont {Will}},\ }\bibfield  {title} {\bibinfo {title} {Double microwave shielding},\ }\href {https://doi.org/10.1103/b8pm-3prn} {\bibfield  {journal} {\bibinfo  {journal} {PRX Quantum}\ }\textbf {\bibinfo {volume} {6}},\ \bibinfo {pages} {020358} (\bibinfo {year} {2025})}\BibitemShut {NoStop}%
\bibitem [{\citenamefont {Mukherjee}\ and\ \citenamefont {Hutson}(2024)}]{mukherjee2024controlling}%
  \BibitemOpen
  \bibfield  {author} {\bibinfo {author} {\bibfnamefont {B.}~\bibnamefont {Mukherjee}}\ and\ \bibinfo {author} {\bibfnamefont {J.~M.}\ \bibnamefont {Hutson}},\ }\bibfield  {title} {\bibinfo {title} {Controlling collisional loss and scattering lengths of ultracold dipolar molecules with static electric fields},\ }\href {https://link.aps.org/doi/10.1103/PhysRevResearch.6.013145} {\bibfield  {journal} {\bibinfo  {journal} {Phys. Rev. Res.}\ }\textbf {\bibinfo {volume} {6}},\ \bibinfo {pages} {013145} (\bibinfo {year} {2024})}\BibitemShut {NoStop}%
\bibitem [{\citenamefont {Lassabli{\`e}re}\ and\ \citenamefont {Qu{\'e}m{\'e}ner}(2022)}]{lassabliere2022model}%
  \BibitemOpen
  \bibfield  {author} {\bibinfo {author} {\bibfnamefont {L.}~\bibnamefont {Lassabli{\`e}re}}\ and\ \bibinfo {author} {\bibfnamefont {G.}~\bibnamefont {Qu{\'e}m{\'e}ner}},\ }\bibfield  {title} {\bibinfo {title} {Model for two-body collisions between ultracold dipolar molecules around a {F}{\"o}rster resonance in an electric field},\ }\href {https://link.aps.org/doi/10.1103/PhysRevA.106.033311} {\bibfield  {journal} {\bibinfo  {journal} {Phys. Rev. A}\ }\textbf {\bibinfo {volume} {106}},\ \bibinfo {pages} {033311} (\bibinfo {year} {2022})}\BibitemShut {NoStop}%
\bibitem [{\citenamefont {Mukherjee}\ \emph {et~al.}(2025)\citenamefont {Mukherjee}, \citenamefont {Santos},\ and\ \citenamefont {Hutson}}]{Mukherjee2025EffectivePotential}%
  \BibitemOpen
  \bibfield  {author} {\bibinfo {author} {\bibfnamefont {B.}~\bibnamefont {Mukherjee}}, \bibinfo {author} {\bibfnamefont {L.}~\bibnamefont {Santos}},\ and\ \bibinfo {author} {\bibfnamefont {J.~M.}\ \bibnamefont {Hutson}},\ }\bibfield  {title} {\bibinfo {title} {Effective anisotropic interaction potentials for pairs of ultracold molecules shielded by a static electric field},\ }\href {https://doi.org/10.1088/1367-2630/ae05c1} {\bibfield  {journal} {\bibinfo  {journal} {New J. Phys.}\ }\textbf {\bibinfo {volume} {27}},\ \bibinfo {pages} {093204} (\bibinfo {year} {2025})}\BibitemShut {NoStop}%
\bibitem [{\citenamefont {Lam}\ \emph {et~al.}(2022)\citenamefont {Lam}, \citenamefont {Bigagli}, \citenamefont {Warner}, \citenamefont {Yuan}, \citenamefont {Zhang}, \citenamefont {Tiemann}, \citenamefont {Stevenson},\ and\ \citenamefont {Will}}]{Lam2022NaCsFeshbach}%
  \BibitemOpen
  \bibfield  {author} {\bibinfo {author} {\bibfnamefont {A.~Z.}\ \bibnamefont {Lam}}, \bibinfo {author} {\bibfnamefont {N.}~\bibnamefont {Bigagli}}, \bibinfo {author} {\bibfnamefont {C.}~\bibnamefont {Warner}}, \bibinfo {author} {\bibfnamefont {W.}~\bibnamefont {Yuan}}, \bibinfo {author} {\bibfnamefont {S.}~\bibnamefont {Zhang}}, \bibinfo {author} {\bibfnamefont {E.}~\bibnamefont {Tiemann}}, \bibinfo {author} {\bibfnamefont {I.}~\bibnamefont {Stevenson}},\ and\ \bibinfo {author} {\bibfnamefont {S.}~\bibnamefont {Will}},\ }\bibfield  {title} {\bibinfo {title} {High phase-space density gas of {N}a{C}s {Feshbach} molecules},\ }\href {https://doi.org/10.1103/PhysRevResearch.4.L022019} {\bibfield  {journal} {\bibinfo  {journal} {Phys. Rev. Res.}\ }\textbf {\bibinfo {volume} {4}},\ \bibinfo {pages} {L022019} (\bibinfo {year} {2022})}\BibitemShut {NoStop}%
\bibitem [{\citenamefont {Yuan}\ \emph {et~al.}(2025)\citenamefont {Yuan}, \citenamefont {Zhang}, \citenamefont {Bigagli}, \citenamefont {Kwak}, \citenamefont {Warner}, \citenamefont {Karman}, \citenamefont {Stevenson},\ and\ \citenamefont {Will}}]{yuan2025extreme}%
  \BibitemOpen
  \bibfield  {author} {\bibinfo {author} {\bibfnamefont {W.}~\bibnamefont {Yuan}}, \bibinfo {author} {\bibfnamefont {S.}~\bibnamefont {Zhang}}, \bibinfo {author} {\bibfnamefont {N.}~\bibnamefont {Bigagli}}, \bibinfo {author} {\bibfnamefont {H.}~\bibnamefont {Kwak}}, \bibinfo {author} {\bibfnamefont {C.}~\bibnamefont {Warner}}, \bibinfo {author} {\bibfnamefont {T.}~\bibnamefont {Karman}}, \bibinfo {author} {\bibfnamefont {I.}~\bibnamefont {Stevenson}},\ and\ \bibinfo {author} {\bibfnamefont {S.}~\bibnamefont {Will}},\ }\bibfield  {title} {\bibinfo {title} {Extreme loss suppression and wide tunability of dipolar interactions in an ultracold molecular gas},\ }\href {https://arxiv.org/abs/2505.08773} {\bibfield  {journal} {\bibinfo  {journal} {arXiv:2505.08773}\ } (\bibinfo {year} {2025})}\BibitemShut {NoStop}%
\bibitem [{\citenamefont {W\"achtler}\ and\ \citenamefont {Santos}(2016)}]{Wachtler2016QuantumFilaments}%
  \BibitemOpen
  \bibfield  {author} {\bibinfo {author} {\bibfnamefont {F.}~\bibnamefont {W\"achtler}}\ and\ \bibinfo {author} {\bibfnamefont {L.}~\bibnamefont {Santos}},\ }\bibfield  {title} {\bibinfo {title} {Quantum filaments in dipolar {B}ose-{E}instein condensates},\ }\href {https://doi.org/10.1103/PhysRevA.93.061603} {\bibfield  {journal} {\bibinfo  {journal} {Phys. Rev. A}\ }\textbf {\bibinfo {volume} {93}},\ \bibinfo {pages} {061603(R)} (\bibinfo {year} {2016})}\BibitemShut {NoStop}%
\bibitem [{\citenamefont {Ceperley}(1995)}]{ceperley1995review}%
  \BibitemOpen
  \bibfield  {author} {\bibinfo {author} {\bibfnamefont {D.~M.}\ \bibnamefont {Ceperley}},\ }\bibfield  {title} {\bibinfo {title} {Path integrals in the theory of condensed helium},\ }\href {https://doi.org/10.1103/RevModPhys.67.279} {\bibfield  {journal} {\bibinfo  {journal} {Rev. Mod. Phys.}\ }\textbf {\bibinfo {volume} {67}},\ \bibinfo {pages} {279} (\bibinfo {year} {1995})}\BibitemShut {NoStop}%
\bibitem [{\citenamefont {Cinti}\ and\ \citenamefont {Boninsegni}(2017)}]{Cinti2017a}%
  \BibitemOpen
  \bibfield  {author} {\bibinfo {author} {\bibfnamefont {F.}~\bibnamefont {Cinti}}\ and\ \bibinfo {author} {\bibfnamefont {M.}~\bibnamefont {Boninsegni}},\ }\bibfield  {title} {\bibinfo {title} {Classical and quantum filaments in the ground state of trapped dipolar {Bose} gases},\ }\href {https://doi.org/10.1103/PhysRevA.96.013627} {\bibfield  {journal} {\bibinfo  {journal} {Phys. Rev. A}\ }\textbf {\bibinfo {volume} {96}},\ \bibinfo {pages} {013627} (\bibinfo {year} {2017})}\BibitemShut {NoStop}%
\bibitem [{\citenamefont {Cinti}\ \emph {et~al.}(2017)\citenamefont {Cinti}, \citenamefont {Cappellaro}, \citenamefont {Salasnich},\ and\ \citenamefont {Macr\`{\i}}}]{Cinti2017b}%
  \BibitemOpen
  \bibfield  {author} {\bibinfo {author} {\bibfnamefont {F.}~\bibnamefont {Cinti}}, \bibinfo {author} {\bibfnamefont {A.}~\bibnamefont {Cappellaro}}, \bibinfo {author} {\bibfnamefont {L.}~\bibnamefont {Salasnich}},\ and\ \bibinfo {author} {\bibfnamefont {T.}~\bibnamefont {Macr\`{\i}}},\ }\bibfield  {title} {\bibinfo {title} {Superfluid filaments of dipolar bosons in free space},\ }\href {https://doi.org/10.1103/PhysRevLett.119.215302} {\bibfield  {journal} {\bibinfo  {journal} {Phys. Rev. Lett.}\ }\textbf {\bibinfo {volume} {119}},\ \bibinfo {pages} {215302} (\bibinfo {year} {2017})}\BibitemShut {NoStop}%
\bibitem [{\citenamefont {Langen}\ \emph {et~al.}(2025)\citenamefont {Langen}, \citenamefont {Boronat}, \citenamefont {S\'anchez-Baena}, \citenamefont {Bomb\'{\i}n}, \citenamefont {Karman},\ and\ \citenamefont {Mazzanti}}]{Langen2025DropletPhase}%
  \BibitemOpen
  \bibfield  {author} {\bibinfo {author} {\bibfnamefont {T.}~\bibnamefont {Langen}}, \bibinfo {author} {\bibfnamefont {J.}~\bibnamefont {Boronat}}, \bibinfo {author} {\bibfnamefont {J.}~\bibnamefont {S\'anchez-Baena}}, \bibinfo {author} {\bibfnamefont {R.}~\bibnamefont {Bomb\'{\i}n}}, \bibinfo {author} {\bibfnamefont {T.}~\bibnamefont {Karman}},\ and\ \bibinfo {author} {\bibfnamefont {F.}~\bibnamefont {Mazzanti}},\ }\bibfield  {title} {\bibinfo {title} {Dipolar droplets of strongly interacting molecules},\ }\href {https://doi.org/10.1103/PhysRevLett.134.053001} {\bibfield  {journal} {\bibinfo  {journal} {Phys. Rev. Lett.}\ }\textbf {\bibinfo {volume} {134}},\ \bibinfo {pages} {053001} (\bibinfo {year} {2025})}\BibitemShut {NoStop}%
\bibitem [{\citenamefont {Zhang}\ \emph {et~al.}(2025)\citenamefont {Zhang}, \citenamefont {Chen}, \citenamefont {Yi},\ and\ \citenamefont {Shi}}]{Zhang2025DropletPhase}%
  \BibitemOpen
  \bibfield  {author} {\bibinfo {author} {\bibfnamefont {W.}~\bibnamefont {Zhang}}, \bibinfo {author} {\bibfnamefont {K.}~\bibnamefont {Chen}}, \bibinfo {author} {\bibfnamefont {S.}~\bibnamefont {Yi}},\ and\ \bibinfo {author} {\bibfnamefont {T.}~\bibnamefont {Shi}},\ }\bibfield  {title} {\bibinfo {title} {Quantum phases for finite-temperature gases of bosonic polar molecules shielded by dual microwaves},\ }\href {https://doi.org/10.1103/9cxl-d9zg} {\bibfield  {journal} {\bibinfo  {journal} {PRX Quantum}\ }\textbf {\bibinfo {volume} {6}},\ \bibinfo {pages} {040307} (\bibinfo {year} {2025})}\BibitemShut {NoStop}%
\bibitem [{\citenamefont {Wang}(2026)}]{wang2026bound}%
  \BibitemOpen
  \bibfield  {author} {\bibinfo {author} {\bibfnamefont {R.~R.~W.}\ \bibnamefont {Wang}},\ }\bibfield  {title} {\bibinfo {title} {Bound-state-free {F}{\"o}rster resonant shielding of strongly dipolar ultracold molecules},\ }\href {https://arxiv.org/abs/2601.21928} {\bibfield  {journal} {\bibinfo  {journal} {arXiv:2601.21928}\ } (\bibinfo {year} {2026})}\BibitemShut {NoStop}%
\bibitem [{\citenamefont {Koch}\ \emph {et~al.}(2008)\citenamefont {Koch}, \citenamefont {Lahaye}, \citenamefont {Metz}, \citenamefont {Fr{\"o}hlich}, \citenamefont {Griesmaier},\ and\ \citenamefont {Pfau}}]{koch2008stabilization}%
  \BibitemOpen
  \bibfield  {author} {\bibinfo {author} {\bibfnamefont {T.}~\bibnamefont {Koch}}, \bibinfo {author} {\bibfnamefont {T.}~\bibnamefont {Lahaye}}, \bibinfo {author} {\bibfnamefont {J.}~\bibnamefont {Metz}}, \bibinfo {author} {\bibfnamefont {B.}~\bibnamefont {Fr{\"o}hlich}}, \bibinfo {author} {\bibfnamefont {A.}~\bibnamefont {Griesmaier}},\ and\ \bibinfo {author} {\bibfnamefont {T.}~\bibnamefont {Pfau}},\ }\bibfield  {title} {\bibinfo {title} {Stabilization of a purely dipolar quantum gas against collapse},\ }\href {https://www.nature.com/articles/nphys887} {\bibfield  {journal} {\bibinfo  {journal} {Nat. Phys.}\ }\textbf {\bibinfo {volume} {4}},\ \bibinfo {pages} {218} (\bibinfo {year} {2008})}\BibitemShut {NoStop}%
\bibitem [{\citenamefont {B\"uchler}\ \emph {et~al.}(2007)\citenamefont {B\"uchler}, \citenamefont {Demler}, \citenamefont {Lukin}, \citenamefont {Micheli}, \citenamefont {Prokof'ev}, \citenamefont {Pupillo},\ and\ \citenamefont {Zoller}}]{Buchler2007Strongly}%
  \BibitemOpen
  \bibfield  {author} {\bibinfo {author} {\bibfnamefont {H.~P.}\ \bibnamefont {B\"uchler}}, \bibinfo {author} {\bibfnamefont {E.}~\bibnamefont {Demler}}, \bibinfo {author} {\bibfnamefont {M.}~\bibnamefont {Lukin}}, \bibinfo {author} {\bibfnamefont {A.}~\bibnamefont {Micheli}}, \bibinfo {author} {\bibfnamefont {N.}~\bibnamefont {Prokof'ev}}, \bibinfo {author} {\bibfnamefont {G.}~\bibnamefont {Pupillo}},\ and\ \bibinfo {author} {\bibfnamefont {P.}~\bibnamefont {Zoller}},\ }\bibfield  {title} {\bibinfo {title} {Strongly correlated 2d quantum phases with cold polar molecules: Controlling the shape of the interaction potential},\ }\href {https://doi.org/10.1103/PhysRevLett.98.060404} {\bibfield  {journal} {\bibinfo  {journal} {Phys. Rev. Lett.}\ }\textbf {\bibinfo {volume} {98}},\ \bibinfo {pages} {060404} (\bibinfo {year} {2007})}\BibitemShut {NoStop}%
\bibitem [{\citenamefont {Astrakharchik}\ \emph {et~al.}(2007)\citenamefont {Astrakharchik}, \citenamefont {Boronat}, \citenamefont {Kurbakov},\ and\ \citenamefont {Lozovik}}]{Astrakharchik2007Quantum}%
  \BibitemOpen
  \bibfield  {author} {\bibinfo {author} {\bibfnamefont {G.~E.}\ \bibnamefont {Astrakharchik}}, \bibinfo {author} {\bibfnamefont {J.}~\bibnamefont {Boronat}}, \bibinfo {author} {\bibfnamefont {I.~L.}\ \bibnamefont {Kurbakov}},\ and\ \bibinfo {author} {\bibfnamefont {Y.~E.}\ \bibnamefont {Lozovik}},\ }\bibfield  {title} {\bibinfo {title} {Quantum phase transition in a two-dimensional system of dipoles},\ }\href {https://doi.org/10.1103/PhysRevLett.98.060405} {\bibfield  {journal} {\bibinfo  {journal} {Phys. Rev. Lett.}\ }\textbf {\bibinfo {volume} {98}},\ \bibinfo {pages} {060405} (\bibinfo {year} {2007})}\BibitemShut {NoStop}%
\bibitem [{\citenamefont {Bombin}\ \emph {et~al.}(2017)\citenamefont {Bombin}, \citenamefont {Boronat},\ and\ \citenamefont {Mazzanti}}]{Bombin2017Dipolar}%
  \BibitemOpen
  \bibfield  {author} {\bibinfo {author} {\bibfnamefont {R.}~\bibnamefont {Bombin}}, \bibinfo {author} {\bibfnamefont {J.}~\bibnamefont {Boronat}},\ and\ \bibinfo {author} {\bibfnamefont {F.}~\bibnamefont {Mazzanti}},\ }\bibfield  {title} {\bibinfo {title} {Dipolar {B}ose supersolid stripes},\ }\href {https://doi.org/10.1103/PhysRevLett.119.250402} {\bibfield  {journal} {\bibinfo  {journal} {Phys. Rev. Lett.}\ }\textbf {\bibinfo {volume} {119}},\ \bibinfo {pages} {250402} (\bibinfo {year} {2017})}\BibitemShut {NoStop}%
\bibitem [{\citenamefont {Rosenberg}\ \emph {et~al.}(2022)\citenamefont {Rosenberg}, \citenamefont {Christakis}, \citenamefont {Guardado-Sanchez}, \citenamefont {Yan},\ and\ \citenamefont {Bakr}}]{Rosenberg2022HBT}%
  \BibitemOpen
  \bibfield  {author} {\bibinfo {author} {\bibfnamefont {J.~S.}\ \bibnamefont {Rosenberg}}, \bibinfo {author} {\bibfnamefont {L.}~\bibnamefont {Christakis}}, \bibinfo {author} {\bibfnamefont {E.}~\bibnamefont {Guardado-Sanchez}}, \bibinfo {author} {\bibfnamefont {Z.~Z.}\ \bibnamefont {Yan}},\ and\ \bibinfo {author} {\bibfnamefont {W.~S.}\ \bibnamefont {Bakr}},\ }\bibfield  {title} {\bibinfo {title} {Observation of the {H}anbury {B}rown--{T}wiss effect with ultracold molecules},\ }\href {https://doi.org/10.1038/s41567-022-01695-9} {\bibfield  {journal} {\bibinfo  {journal} {Nat. Phys.}\ }\textbf {\bibinfo {volume} {18}},\ \bibinfo {pages} {1062} (\bibinfo {year} {2022})}\BibitemShut {NoStop}%
\bibitem [{\citenamefont {Xiang}\ \emph {et~al.}(2025)\citenamefont {Xiang}, \citenamefont {Cruz-Col\'on}, \citenamefont {Chua}, \citenamefont {Milner}, \citenamefont {de~Hond}, \citenamefont {Fricke},\ and\ \citenamefont {Ketterle}}]{Xiang2025ContinuumQGM}%
  \BibitemOpen
  \bibfield  {author} {\bibinfo {author} {\bibfnamefont {J.}~\bibnamefont {Xiang}}, \bibinfo {author} {\bibfnamefont {E.}~\bibnamefont {Cruz-Col\'on}}, \bibinfo {author} {\bibfnamefont {C.~C.}\ \bibnamefont {Chua}}, \bibinfo {author} {\bibfnamefont {W.~R.}\ \bibnamefont {Milner}}, \bibinfo {author} {\bibfnamefont {J.}~\bibnamefont {de~Hond}}, \bibinfo {author} {\bibfnamefont {J.~F.}\ \bibnamefont {Fricke}},\ and\ \bibinfo {author} {\bibfnamefont {W.}~\bibnamefont {Ketterle}},\ }\bibfield  {title} {\bibinfo {title} {In situ imaging of the thermal de {B}roglie wavelength in an ultracold {Bose} gas},\ }\href {https://doi.org/10.1103/PhysRevLett.134.183401} {\bibfield  {journal} {\bibinfo  {journal} {Phys. Rev. Lett.}\ }\textbf {\bibinfo {volume} {134}},\ \bibinfo {pages} {183401} (\bibinfo {year} {2025})}\BibitemShut {NoStop}%
\bibitem [{\citenamefont {Yao}\ \emph {et~al.}(2025)\citenamefont {Yao}, \citenamefont {Chi}, \citenamefont {Wang}, \citenamefont {Fletcher},\ and\ \citenamefont {Zwierlein}}]{Yao2025ContinuumQGM}%
  \BibitemOpen
  \bibfield  {author} {\bibinfo {author} {\bibfnamefont {R.}~\bibnamefont {Yao}}, \bibinfo {author} {\bibfnamefont {S.}~\bibnamefont {Chi}}, \bibinfo {author} {\bibfnamefont {M.}~\bibnamefont {Wang}}, \bibinfo {author} {\bibfnamefont {R.~J.}\ \bibnamefont {Fletcher}},\ and\ \bibinfo {author} {\bibfnamefont {M.}~\bibnamefont {Zwierlein}},\ }\bibfield  {title} {\bibinfo {title} {Measuring pair correlations in {Bose} and {Fermi} gases via atom-resolved microscopy},\ }\href {https://doi.org/10.1103/PhysRevLett.134.183402} {\bibfield  {journal} {\bibinfo  {journal} {Phys. Rev. Lett.}\ }\textbf {\bibinfo {volume} {134}},\ \bibinfo {pages} {183402} (\bibinfo {year} {2025})}\BibitemShut {NoStop}%
\bibitem [{\citenamefont {de~Jongh}\ \emph {et~al.}(2025)\citenamefont {de~Jongh}, \citenamefont {Verstraten}, \citenamefont {Dixmerias}, \citenamefont {Daix}, \citenamefont {Peaudecerf},\ and\ \citenamefont {Yefsah}}]{deJongh2025ContinuumQGM}%
  \BibitemOpen
  \bibfield  {author} {\bibinfo {author} {\bibfnamefont {T.}~\bibnamefont {de~Jongh}}, \bibinfo {author} {\bibfnamefont {J.}~\bibnamefont {Verstraten}}, \bibinfo {author} {\bibfnamefont {M.}~\bibnamefont {Dixmerias}}, \bibinfo {author} {\bibfnamefont {C.}~\bibnamefont {Daix}}, \bibinfo {author} {\bibfnamefont {B.}~\bibnamefont {Peaudecerf}},\ and\ \bibinfo {author} {\bibfnamefont {T.}~\bibnamefont {Yefsah}},\ }\bibfield  {title} {\bibinfo {title} {Quantum gas microscopy of fermions in the continuum},\ }\href {https://doi.org/10.1103/PhysRevLett.134.183403} {\bibfield  {journal} {\bibinfo  {journal} {Phys. Rev. Lett.}\ }\textbf {\bibinfo {volume} {134}},\ \bibinfo {pages} {183403} (\bibinfo {year} {2025})}\BibitemShut {NoStop}%
\bibitem [{\citenamefont {G\'oral}\ \emph {et~al.}(2002)\citenamefont {G\'oral}, \citenamefont {Santos},\ and\ \citenamefont {Lewenstein}}]{Goral2002EBH}%
  \BibitemOpen
  \bibfield  {author} {\bibinfo {author} {\bibfnamefont {K.}~\bibnamefont {G\'oral}}, \bibinfo {author} {\bibfnamefont {L.}~\bibnamefont {Santos}},\ and\ \bibinfo {author} {\bibfnamefont {M.}~\bibnamefont {Lewenstein}},\ }\bibfield  {title} {\bibinfo {title} {Quantum phases of dipolar bosons in optical lattices},\ }\href {https://doi.org/10.1103/PhysRevLett.88.170406} {\bibfield  {journal} {\bibinfo  {journal} {Phys. Rev. Lett.}\ }\textbf {\bibinfo {volume} {88}},\ \bibinfo {pages} {170406} (\bibinfo {year} {2002})}\BibitemShut {NoStop}%
\bibitem [{\citenamefont {Capogrosso-Sansone}\ \emph {et~al.}(2010)\citenamefont {Capogrosso-Sansone}, \citenamefont {Trefzger}, \citenamefont {Lewenstein}, \citenamefont {Zoller},\ and\ \citenamefont {Pupillo}}]{Capogrosso-Sansone2010EBH}%
  \BibitemOpen
  \bibfield  {author} {\bibinfo {author} {\bibfnamefont {B.}~\bibnamefont {Capogrosso-Sansone}}, \bibinfo {author} {\bibfnamefont {C.}~\bibnamefont {Trefzger}}, \bibinfo {author} {\bibfnamefont {M.}~\bibnamefont {Lewenstein}}, \bibinfo {author} {\bibfnamefont {P.}~\bibnamefont {Zoller}},\ and\ \bibinfo {author} {\bibfnamefont {G.}~\bibnamefont {Pupillo}},\ }\bibfield  {title} {\bibinfo {title} {Quantum phases of cold polar molecules in {2D} optical lattices},\ }\href {https://doi.org/10.1103/PhysRevLett.104.125301} {\bibfield  {journal} {\bibinfo  {journal} {Phys. Rev. Lett.}\ }\textbf {\bibinfo {volume} {104}},\ \bibinfo {pages} {125301} (\bibinfo {year} {2010})}\BibitemShut {NoStop}%
\bibitem [{\citenamefont {Christakis}(2023)}]{lysanderthesis}%
  \BibitemOpen
  \bibfield  {author} {\bibinfo {author} {\bibfnamefont {L.}~\bibnamefont {Christakis}},\ }\emph {\bibinfo {title} {Microscopy of quantum correlations in an ultracold molecular gas}},\ \href {https://dataspace.princeton.edu/handle/88435/dsp01th83m264v} {Ph.D. thesis},\ \bibinfo  {school} {Princeton University} (\bibinfo {year} {2023})\BibitemShut {NoStop}%
\bibitem [{\citenamefont {Rosenberg}(2025)}]{jasonthesis}%
  \BibitemOpen
  \bibfield  {author} {\bibinfo {author} {\bibfnamefont {J.}~\bibnamefont {Rosenberg}},\ }\emph {\bibinfo {title} {Probing and controlling ultracold polar molecules in a quantum gas microscope}},\ \href {https://dataspace.princeton.edu/handle/88435/dsp01ks65hg63b} {Ph.D. thesis},\ \bibinfo  {school} {Princeton University} (\bibinfo {year} {2025})\BibitemShut {NoStop}%
\bibitem [{\citenamefont {Guo}\ \emph {et~al.}(2017)\citenamefont {Guo}, \citenamefont {Vexiau}, \citenamefont {Zhu}, \citenamefont {Lu}, \citenamefont {Bouloufa-Maafa}, \citenamefont {Dulieu},\ and\ \citenamefont {Wang}}]{PhysRevA.96.052505}%
  \BibitemOpen
  \bibfield  {author} {\bibinfo {author} {\bibfnamefont {M.}~\bibnamefont {Guo}}, \bibinfo {author} {\bibfnamefont {R.}~\bibnamefont {Vexiau}}, \bibinfo {author} {\bibfnamefont {B.}~\bibnamefont {Zhu}}, \bibinfo {author} {\bibfnamefont {B.}~\bibnamefont {Lu}}, \bibinfo {author} {\bibfnamefont {N.}~\bibnamefont {Bouloufa-Maafa}}, \bibinfo {author} {\bibfnamefont {O.}~\bibnamefont {Dulieu}},\ and\ \bibinfo {author} {\bibfnamefont {D.}~\bibnamefont {Wang}},\ }\bibfield  {title} {\bibinfo {title} {High-resolution molecular spectroscopy for producing ultracold absolute-ground-state $^{23}${N}a$^{87}${R}b molecules},\ }\href {https://doi.org/10.1103/PhysRevA.96.052505} {\bibfield  {journal} {\bibinfo  {journal} {Phys. Rev. A}\ }\textbf {\bibinfo {volume} {96}},\ \bibinfo {pages} {052505} (\bibinfo {year} {2017})}\BibitemShut {NoStop}%
\bibitem [{\citenamefont {Guo}\ \emph {et~al.}(2018)\citenamefont {Guo}, \citenamefont {Ye}, \citenamefont {He}, \citenamefont {Qu\'em\'ener},\ and\ \citenamefont {Wang}}]{guo2018highresolution}%
  \BibitemOpen
  \bibfield  {author} {\bibinfo {author} {\bibfnamefont {M.}~\bibnamefont {Guo}}, \bibinfo {author} {\bibfnamefont {X.}~\bibnamefont {Ye}}, \bibinfo {author} {\bibfnamefont {J.}~\bibnamefont {He}}, \bibinfo {author} {\bibfnamefont {G.}~\bibnamefont {Qu\'em\'ener}},\ and\ \bibinfo {author} {\bibfnamefont {D.}~\bibnamefont {Wang}},\ }\bibfield  {title} {\bibinfo {title} {High-resolution internal state control of ultracold $^{23}${Na}$^{87}${Rb} molecules},\ }\href {https://doi.org/10.1103/PhysRevA.97.020501} {\bibfield  {journal} {\bibinfo  {journal} {Phys. Rev. A}\ }\textbf {\bibinfo {volume} {97}},\ \bibinfo {pages} {020501(R)} (\bibinfo {year} {2018})}\BibitemShut {NoStop}%
\bibitem [{\citenamefont {Yatsenko}\ \emph {et~al.}(2002)\citenamefont {Yatsenko}, \citenamefont {Romanenko}, \citenamefont {Shore},\ and\ \citenamefont {Bergmann}}]{PhysRevA.65.043409}%
  \BibitemOpen
  \bibfield  {author} {\bibinfo {author} {\bibfnamefont {L.~P.}\ \bibnamefont {Yatsenko}}, \bibinfo {author} {\bibfnamefont {V.~I.}\ \bibnamefont {Romanenko}}, \bibinfo {author} {\bibfnamefont {B.~W.}\ \bibnamefont {Shore}},\ and\ \bibinfo {author} {\bibfnamefont {K.}~\bibnamefont {Bergmann}},\ }\bibfield  {title} {\bibinfo {title} {Stimulated {R}aman adiabatic passage with partially coherent laser fields},\ }\href {https://doi.org/10.1103/PhysRevA.65.043409} {\bibfield  {journal} {\bibinfo  {journal} {Phys. Rev. A}\ }\textbf {\bibinfo {volume} {65}},\ \bibinfo {pages} {043409} (\bibinfo {year} {2002})}\BibitemShut {NoStop}%
\bibitem [{\citenamefont {Johnson}(1978)}]{johnson1978renormalized}%
  \BibitemOpen
  \bibfield  {author} {\bibinfo {author} {\bibfnamefont {B.~R.}\ \bibnamefont {Johnson}},\ }\bibfield  {title} {\bibinfo {title} {The renormalized {Numerov} method applied to calculating bound states of the coupled-channel {S}chr{\"o}dinger equation},\ }\href {https://doi.org/10.1063/1.436421} {\bibfield  {journal} {\bibinfo  {journal} {J. Chem. Phys.}\ }\textbf {\bibinfo {volume} {69}},\ \bibinfo {pages} {4678} (\bibinfo {year} {1978})}\BibitemShut {NoStop}%
\bibitem [{\citenamefont {Janssen}(2012)}]{janssen2012cold}%
  \BibitemOpen
  \bibfield  {author} {\bibinfo {author} {\bibfnamefont {L.~M.~C.}\ \bibnamefont {Janssen}},\ }\emph {\bibinfo {title} {Cold collision dynamics of {NH} radicals}},\ \href {https://repository.ubn.ru.nl/handle/2066/92746} {Ph.D. thesis},\ \bibinfo  {school} {Radboud University Nijmegen}, \bibinfo {address} {Nijmegen, The Netherlands} (\bibinfo {year} {2012})\BibitemShut {NoStop}%
\bibitem [{\citenamefont {Karman}(2023)}]{karman2023resonances}%
  \BibitemOpen
  \bibfield  {author} {\bibinfo {author} {\bibfnamefont {T.}~\bibnamefont {Karman}},\ }\bibfield  {title} {\bibinfo {title} {Resonances in non-universal dipolar collisions},\ }\href {https://doi.org/10.1021/acs.jpca.3c00797} {\bibfield  {journal} {\bibinfo  {journal} {J. Phys. Chem. A}\ }\textbf {\bibinfo {volume} {127}},\ \bibinfo {pages} {2194} (\bibinfo {year} {2023})}\BibitemShut {NoStop}%
\bibitem [{\citenamefont {Mukherjee}\ \emph {et~al.}(2023)\citenamefont {Mukherjee}, \citenamefont {Frye}, \citenamefont {Le~Sueur}, \citenamefont {Tarbutt},\ and\ \citenamefont {Hutson}}]{mukherjee2023shielding}%
  \BibitemOpen
  \bibfield  {author} {\bibinfo {author} {\bibfnamefont {B.}~\bibnamefont {Mukherjee}}, \bibinfo {author} {\bibfnamefont {M.~D.}\ \bibnamefont {Frye}}, \bibinfo {author} {\bibfnamefont {C.~R.}\ \bibnamefont {Le~Sueur}}, \bibinfo {author} {\bibfnamefont {M.~R.}\ \bibnamefont {Tarbutt}},\ and\ \bibinfo {author} {\bibfnamefont {J.~M.}\ \bibnamefont {Hutson}},\ }\bibfield  {title} {\bibinfo {title} {Shielding collisions of ultracold {CaF} molecules with static electric fields},\ }\href {https://doi.org/10.1103/PhysRevResearch.5.033097} {\bibfield  {journal} {\bibinfo  {journal} {Phys. Rev. Res.}\ }\textbf {\bibinfo {volume} {5}},\ \bibinfo {pages} {033097} (\bibinfo {year} {2023})}\BibitemShut {NoStop}%
\bibitem [{\citenamefont {Ye}\ \emph {et~al.}(2018)\citenamefont {Ye}, \citenamefont {Guo}, \citenamefont {González-Martínez}, \citenamefont {Quéméner},\ and\ \citenamefont {Wang}}]{ye2018collisions}%
  \BibitemOpen
  \bibfield  {author} {\bibinfo {author} {\bibfnamefont {X.}~\bibnamefont {Ye}}, \bibinfo {author} {\bibfnamefont {M.}~\bibnamefont {Guo}}, \bibinfo {author} {\bibfnamefont {M.~L.}\ \bibnamefont {González-Martínez}}, \bibinfo {author} {\bibfnamefont {G.}~\bibnamefont {Quéméner}},\ and\ \bibinfo {author} {\bibfnamefont {D.}~\bibnamefont {Wang}},\ }\bibfield  {title} {\bibinfo {title} {Collisions of ultracold $^{23}${N}a$^{87}${R}b molecules with controlled chemical reactivities},\ }\href {https://doi.org/10.1126/sciadv.aaq0083} {\bibfield  {journal} {\bibinfo  {journal} {Sci. Adv.}\ }\textbf {\bibinfo {volume} {4}},\ \bibinfo {pages} {eaaq0083} (\bibinfo {year} {2018})}\BibitemShut {NoStop}%
\bibitem [{\citenamefont {Boninsegni}\ \emph {et~al.}(2006)\citenamefont {Boninsegni}, \citenamefont {Prokof'ev},\ and\ \citenamefont {Svistunov}}]{boninsegni2006wormalgorithm}%
  \BibitemOpen
  \bibfield  {author} {\bibinfo {author} {\bibfnamefont {M.}~\bibnamefont {Boninsegni}}, \bibinfo {author} {\bibfnamefont {N.}~\bibnamefont {Prokof'ev}},\ and\ \bibinfo {author} {\bibfnamefont {B.}~\bibnamefont {Svistunov}},\ }\bibfield  {title} {\bibinfo {title} {Worm algorithm for continuous-space path integral {Monte Carlo} simulations},\ }\href {https://doi.org/10.1103/PhysRevLett.96.070601} {\bibfield  {journal} {\bibinfo  {journal} {Phys. Rev. Lett.}\ }\textbf {\bibinfo {volume} {96}},\ \bibinfo {pages} {070601} (\bibinfo {year} {2006})}\BibitemShut {NoStop}%
\bibitem [{\citenamefont {Ciardi}\ \emph {et~al.}(2025)\citenamefont {Ciardi}, \citenamefont {Pedersen}, \citenamefont {Langen},\ and\ \citenamefont {Pohl}}]{Ciardi2025crystal}%
  \BibitemOpen
  \bibfield  {author} {\bibinfo {author} {\bibfnamefont {M.}~\bibnamefont {Ciardi}}, \bibinfo {author} {\bibfnamefont {K.~R.}\ \bibnamefont {Pedersen}}, \bibinfo {author} {\bibfnamefont {T.}~\bibnamefont {Langen}},\ and\ \bibinfo {author} {\bibfnamefont {T.}~\bibnamefont {Pohl}},\ }\bibfield  {title} {\bibinfo {title} {Self-bound superfluid membranes and monolayer crystals of ultracold polar molecules},\ }\href {https://doi.org/10.1103/v7gw-xy36} {\bibfield  {journal} {\bibinfo  {journal} {Phys. Rev. Lett.}\ }\textbf {\bibinfo {volume} {135}},\ \bibinfo {pages} {153401} (\bibinfo {year} {2025})}\BibitemShut {NoStop}%
\bibitem [{\citenamefont {Ciardi}\ \emph {et~al.}(2026)\citenamefont {Ciardi}, \citenamefont {Schindewolf}, \citenamefont {Langen},\ and\ \citenamefont {Pohl}}]{ciardi2026metastable}%
  \BibitemOpen
  \bibfield  {author} {\bibinfo {author} {\bibfnamefont {M.}~\bibnamefont {Ciardi}}, \bibinfo {author} {\bibfnamefont {A.}~\bibnamefont {Schindewolf}}, \bibinfo {author} {\bibfnamefont {T.}~\bibnamefont {Langen}},\ and\ \bibinfo {author} {\bibfnamefont {T.}~\bibnamefont {Pohl}},\ }\bibfield  {title} {\bibinfo {title} {Equilibrium and non-equilibrium phases of microwave-dressed polar molecules beyond rotational symmetries},\ }\href {https://arxiv.org/abs/2606.30589} {\bibfield  {journal} {\bibinfo  {journal} {arXiv:2606.30589}\ } (\bibinfo {year} {2026})}\BibitemShut {NoStop}%
\bibitem [{\citenamefont {Arnone~Cardinale}\ \emph {et~al.}(2026)\citenamefont {Arnone~Cardinale}, \citenamefont {Bland},\ and\ \citenamefont {Reimann}}]{ArnoneCardinale2026}%
  \BibitemOpen
  \bibfield  {author} {\bibinfo {author} {\bibfnamefont {T.}~\bibnamefont {Arnone~Cardinale}}, \bibinfo {author} {\bibfnamefont {T.}~\bibnamefont {Bland}},\ and\ \bibinfo {author} {\bibfnamefont {S.~M.}\ \bibnamefont {Reimann}},\ }\bibfield  {title} {\bibinfo {title} {Exploring supersolids of single-microwave shielded molecules via exact and mean-field theories},\ }\href {https://doi.org/10.1038/s42005-026-02706-4} {\bibfield  {journal} {\bibinfo  {journal} {Commun. Phys.}\ }\textbf {\bibinfo {volume} {9}},\ \bibinfo {pages} {191} (\bibinfo {year} {2026})}\BibitemShut {NoStop}%
\end{thebibliography}
\providecommand{\noopsort}[1]{}\providecommand{\singleletter}[1]{#1}%

\end{document}